\documentclass[10pt,journal,letterpaper,onecolumn]{IE/IEEEtran}
\usepackage{cite}
\ifCLASSINFOpdf
  \usepackage[pdftex]{graphicx}
  \DeclareGraphicsExtensions{.pdf,.jpeg,.png}
\else
\fi
\usepackage{amsmath}
\usepackage{amssymb}
\allowdisplaybreaks[1]
\usepackage{algorithm}
\usepackage{algorithmicx}
\usepackage{algpseudocode}

\usepackage{array}

\usepackage{enumitem}
\SetLabelAlign{LeftAlignWithIndent}{\hspace*{1.0ex}\makebox[1.25em][l]{#1}}
\usepackage{epstopdf}

\usepackage{bm}
\usepackage{xcolor}
\usepackage{cleveref}
\usepackage[normalem]{ulem}
\usepackage{subfig}
\usepackage[font=small]{caption}
\usepackage{tikz}
\usepackage{pgfplots}
\usetikzlibrary{arrows}
\usetikzlibrary{fit,positioning}
\usetikzlibrary{decorations.pathreplacing}
\usetikzlibrary{shapes.misc}
\tikzstyle{block} = [rectangle, draw,  text centered, minimum height=2mm]
\tikzset{cross/.style={cross out, draw=black, minimum size=2mm, inner sep=0pt, outer sep=0pt},
  cross/.default={1pt}}
\tikzstyle{crosscircle} = [circle, draw=black,minimum size=4mm]
\tikzstyle{dottedblock} = [draw=blue,dashed,rounded corners=0.25cm,align=flush center,text width=12em,inner sep=4pt,minimum height=2.5cm]

\newtheorem{thm}{\protect\theoremname}
\newtheorem{lem}{\protect\lemmaname}
\newtheorem{rem}{\protect\remarkname}

\providecommand{\lemmaname}{Lemma}
\providecommand{\theoremname}{Theorem}
\providecommand{\remarkname}{Remark}
\providecommand{\corollaryname}{Corollary}
\providecommand{\definitionname}{Definition}

\makeatletter
\newcommand{\mathleft}{\@fleqntrue}
\newcommand{\mathcenter}{\@fleqnfalse}
\makeatother

\def\t{\bm{t}}

\begin{document}
%
\title{Generalized Compute-Compress-and-Forward}
%
%
%

\author{Hai~Cheng,
        Xiaojun~Yuan,~\IEEEmembership{Senior~Member,~IEEE,}
        and~Yihua~Tan
\thanks{H. Cheng and X. Yuan are with the School of Information Science and Technology, ShanghaiTech
University, Shanghai, China (e-mail: chenghai, yuanxj@shanghaitech.edu.cn)}
\thanks{Y. Tan is with the Department of Information Engineering, The Chinese University of Hong
Kong, Hong Kong, China (e-mail: tanyihua@cuhk.edu.hk).}}

\markboth{IEEE Transactions on Information Theory}%
{Cheng \MakeLowercase{\textit{et al.}}: Generalized Compute-Compress-and-Forward}

\maketitle

\begin{abstract}
Compute-and-forward (CF) harnesses interference in wireless communications by exploiting structured coding. The key idea of CF is to compute integer combinations of codewords from multiple source nodes, rather than to decode individual codewords by treating others as noise. Compute-compress-and-forward (CCF) can further enhance the network performance by introducing compression operations at receivers. In this paper, we develop a more general compression framework, termed generalized compute-compress-and-forward (GCCF), where the compression function involves multiple quantization-and-modulo lattice operations. We show that GCCF achieves a broader compression rate region than CCF. We also compare our compression rate region with the fundamental Slepian-Wolf (SW) region. We show that GCCF is optimal in the sense of achieving the minimum total compression rate. We also establish the criteria under which GCCF achieves the SW region. In addition, we consider a two-hop relay network employing the GCCF scheme. We formulate a sum-rate maximization problem and develop an approximate algorithm to solve the problem. Numerical results are presented to demonstrate the performance superiority of GCCF over CCF and other schemes.
\end{abstract}


%
\IEEEpeerreviewmaketitle

\section{Introduction}

\IEEEPARstart{C}{ompute-and-forward} (CF) is an advanced relay technique that exploits structured coding to harness interference in wireless communications \cite{nazer2011compute}. The key idea of CF is to suppress interference by computing integer combinations of source codewords, rather than decoding individual source codewords. CF employs nested lattice coding \cite{erez2004achieving} to ensure that the computed integer combinations in CF are still valid codewords. A nested lattice codebook is formed by the set of lattice points of a coding lattice confined within the fundamental Voronoi region of a coarser shaping lattice.

Since the advent of CF, many works followed up to enhance the throughput of CF-based relay networks \cite{ntranos2013asymmetric, zhu2014asymmetric, nazer2015expanding, nazer2012successive, Succ_integer_forcing, niesen2012degrees, tan2015compute, yihua2014asymmetric,zhan2014integer, TaoSSA, nam2010capacity, nam2011nested, osmane2011compute, sakzad2013integer}. In the original CF \cite{nazer2011compute}, all nested lattice codes share a common shaping lattice, and all transmitters are constrained by a same power budget. In \cite{ntranos2013asymmetric,zhu2014asymmetric,nazer2015expanding}, asymmetric CF allows asymmetric construction of shaping lattices and unequal power allocation across transmitters, so as to improve the computation performance. In \cite{nazer2012successive} and  \cite{Succ_integer_forcing}, the authors studied successive computation of multiple codeword combinations to enlarge the achievable rate region of a receiver. In \cite{niesen2012degrees}, the authors studied the degrees of freedom of CF to characterize the behavior of CF in the high signal-to-noise ratio (SNR) regime.

Recently, the authors in \cite{tan2015compute} pointed out that, as the computed codewords in a CF-based multi-hop relay network are in general correlated, the performance of the network can be enhanced if the codewords computed at relays are further compressed to reduce the information redundancy. The corresponding relaying strategy is referred to as compute-compress-and-forward (CCF). In CCF, each relay processes its computed message by taking quantization and modulo (QM) operation over a pair of carefully selected nested lattices. Significant performance gains of CCF over CF have been demonstrated by the numerical results in \cite{tan2015compute}. However, as one QM pair for CCF is not necessarily optimal, it is worth investigating the fundamental performance limits of CCF with multiple QM pairs.

In this paper, we consider the efficient transceiver design for an interference channel with $L$ transmitters and $L$ receivers, where $L$ is an arbitrary integer. We generalize CCF by allowing each receiver to compress its computed codeword by using multiple pairs of QM operations. We show that the generalized CCF (GCCF) scheme can achieve a broader compression rate region than the original CCF in \cite{tan2015compute}. We also show that the compression problem can be interpreted as a distributed source-coding problem. Based on that, we compare the compression rate region of GCCF with the Slepian-Wolf region, where the latter is the optimal rate region for distributed source coding \cite{Slepian1973Wolf}. We show that GCCF, though in general cannot achieve the entire SW region, is optimal in the sense of minimizing the total compression rate. Also, we prove that these two regions coincide in the following three cases: (i) the channel consists of only two transmitters and two receivers, i.e., $L = 2$; or (ii) all the transmitters share a common shaping lattice; or (iii) no dither is employed in nested lattice coding for the channel.

The proposed GCCF scheme, similar to CCF, can serve as a building block to construct a multi-hop relay network. In particular, we consider a two-hop relay network in which a single destination node is required to recover all the messages from the sources. We establish an achievable rate region of the relay network and then formulate a mixed-integer-programing problem for sum-rate maximization. We show that the problem can be approximately solved based on Lenstra-Lenstra-Lovsz (LLL) lattice basis reduction \cite{osmane2011compute}, \cite{sakzad2013integer} and differential evolution \cite{storn1997differential}. Numerical results are presented to demonstrate the superiority of our proposed GCCF scheme over the other benchmark schemes including CCF.

The remainder of this paper is organized as follows. In Section II, we introduce the system model and the background of CCF. In Section III, we describe the proposed GCCF scheme. In Section IV, we derive the compression rate region of GCCF, and discuss its relation with the Slepian-Wolf region. In Section V, we first establish an achievable rate region of a multi-hop relay network based on GCCF, and then present an approximate algorithm to solve the sum-rate maximization problem of a two-hop GCCF network. Numerical results are also provided to show the performance superiority of GCCF. Finally, conclusions are presented in Section VI.

\section{Preliminaries \label{sec:preliminary}}

  \subsection{System Model}
\begin{figure}
	\centering
	\includegraphics[width = 3.2in]{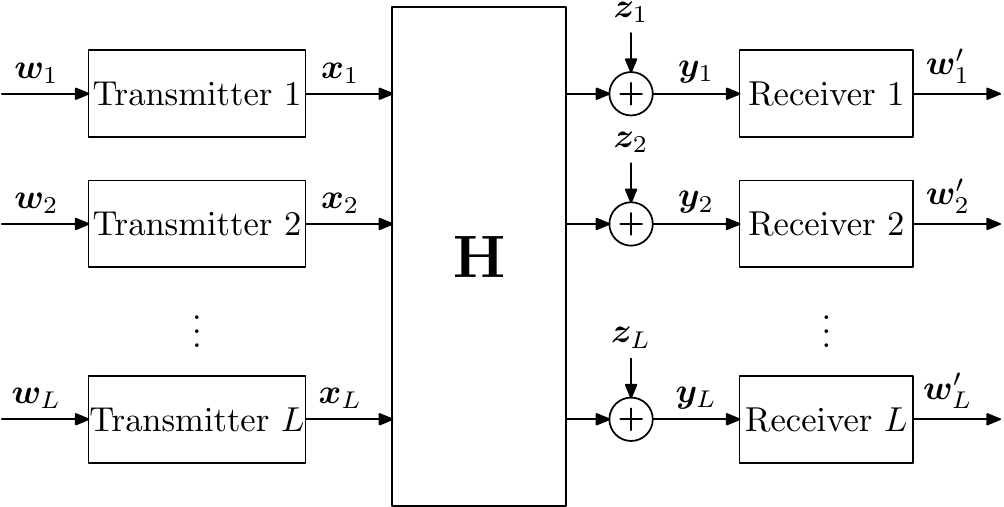}
	\caption{An interference channel with $L$ transmitters and $L$ receivers.}
	\label{fig:OneHop}
\end{figure}

  Consider an interference channel with $L$ transmitters and $L$ receivers, as illustrated in Fig. \ref{fig:OneHop}. Each transmitter or receiver node is equipped with a single antenna.
  The message of transmitter $l$ is a vector $\bm{\mathrm{w}}_l \in \mathbb{F}^{k_l}_{\gamma}$, where $\gamma$ is a prime number. Transmitter $l$ encodes message $\bm w_l$ into $\bm x_l \in \mathbb{R}^{n \times 1}$ and then transmits $\bm x_l$ to the receivers over an additive white Gaussian noise (AWGN) channel. Each receiver $m$ observes an output signal
  \begin{equation}\label{eq:channel model}
  \bm y_m=\sum_{l=1}^{L}{h_{ml}\bm x_l}+\bm z_m, \text{for}\ m \in \mathcal{I}_L
  \end{equation}
  where $h_{ml}\in \mathcal{N}(0,1)$ is the channel coefficient of the link from source $l$ to relay $m$, $\bm z_m\in \mathbb{R}^{n\times 1}$ is a Gaussian noise vector drawn from $\mathcal{N}(\bm 0,\bm I_{n})$ with $\bm I_{n}$ being the $n$-by-$n$ identity matrix, and $\mathcal{I}_l$ denotes the index set of integers from $1$ to $l$ and $\mathcal{I}_0 = \emptyset$. Denote by $\bm H = [\bm h_1,\bm h_2,\cdots,\bm h_L]^\mathrm{T}$ the channel matrix, where $\bm h_{m}=[h_{m1},h_{m2},\cdots,h_{mL}]^\mathrm T$ is the channel  vector seen by receiver $m$. The average power of transmitter $l$ is $p_l=\frac{1}{n}\text{E}\lVert\bm x_l\rVert^2$ satisfying $p_l \leq P_l$, where $P_l$ is the power budget of transmitter $l$. We assume full channel state information, i.e., all the channel coefficients are perfectly known. 

  The CCF scheme in \cite{tan2015compute} can be applied to the channel model in \eqref{eq:channel model}. In CCF, each transmitter encodes its message by a nested lattice code and then sends the codeword to the receivers. Each receiver computes an integer linear combination of the nested lattice codewords from the received signal, and then compresses the computed codeword. The goal of the compression operation is to reduce the forwarding rates of the receivers in relaying, so as to improve the spectrum efficiency of the relay network. In this paper, we aim to generalize the compression operation in CCF for more efficient forwarding.

  \subsection{Nested Lattice Codes \label{sec:lattice code}}

\begin{figure}
  	\centering
  	\includegraphics[width=3.2in]{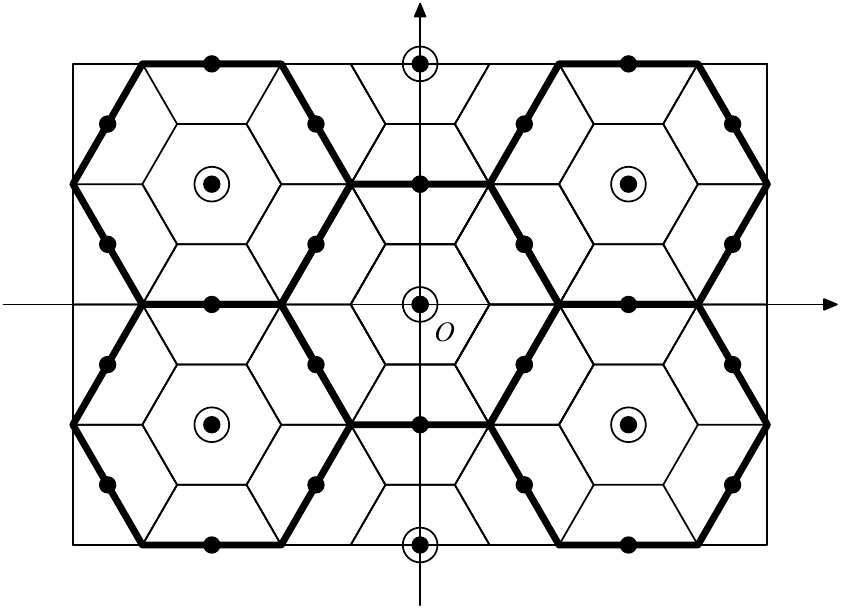}
  	\caption{An illustration of a nested lattice pair $(\Lambda_s, \Lambda_c)$ satisfying $\Lambda_s \subset \Lambda_c \subset \mathbb{R}^2$. Black points ``$\bullet$'' are elements of the coding lattice $\Lambda_c$ and the black circles ``$\bigcirc$'' are elements of the shaping lattice $\Lambda_s$. The Voronoi regions of $\Lambda_c$ and $\Lambda_s$ are hexagons with thin and thick edges, respectively. The nested lattice codebook consists of the set of all fine lattice points in the fundamental Voronoi region of the coarse lattice.}
  	\label{fig:lattice structure}
  \end{figure}

  We start with a brief introduction of nested lattice coding. A lattice $\Lambda \in \mathbb{R}^n$ is a discrete group with the following property. If $\bm t_1 \in \Lambda$ and $\bm t_2 \in \Lambda$, then $\bm t_1 + \bm t_2 \in \Lambda$; and if $\bm t_1 \in \Lambda$, then $-\bm t_1 \in \Lambda$. A lattice can be represented as 
  \begin{equation}
    \Lambda=\{\bm s=\bm{Gc}: \bm c \in \mathbb{Z}^{n \times 1}\}
  \end{equation}
  where $\bm G \in \mathbb{R}^{n\times n}$ is the generation matrix of $\Lambda$.
  The quantization of $\bm x\in \mathbb{R}^n$ on $\Lambda$ is the nearest lattice point to $\bm x$ in $\Lambda$, i.e.
  \begin{equation}
    \bm Q_{\Lambda}(\bm x)=\arg\min_{\bm{t}\in \Lambda} \lVert \bm{t-x} \rVert
  \end{equation}
  where $\lVert \cdot \rVert$ denotes the $l_2$ norm of a vector.
  The quantization error is given by 
  \begin{equation}\label{eq:modulo}
    \bm x\ \text{mod}\ \Lambda=\bm{x}-Q_{\Lambda}(\bm x)
  \end{equation}
  where ``mod'' represents the modulo operation. The fundamental Voronoi region of $\Lambda$ is defined by
  \begin{equation}
    \mathcal{V}=\{\bm x \in \mathbb{R}^n: Q_{\Lambda}(\bm x)=\bm 0\}.
  \end{equation}
  The second moment of $\Lambda$ is defined by
  \begin{equation}\label{eq:second moment}
	 \sigma_{\Lambda}^2 = \frac{1}{n}\int_{\mathcal{V}} \frac{\lVert \bm x \rVert^2}{\text{Vol}(\mathcal{V})}d \bm x
  \end{equation}
  where $\text{Vol}(\mathcal{V})$ is the volume of $\mathcal{V}$. The normalized second moment of $\Lambda$ is defined by
  \begin{equation}\label{eq:normalized second moment}
  	G(\Lambda) = \frac{\sigma_{\Lambda}^2}{(\text{Vol}(\mathcal{V}))^{\frac{2}{n}}}.
  \end{equation}
  If $ \Lambda_1 \subseteq \Lambda_2$, we say that $ \Lambda_1$ is nested in $\Lambda_2$ and that $ \Lambda_1$ is coarser than $ \Lambda_2$ (or alternatively, $\Lambda_2$ is finer than $\Lambda_1$). An example of a pair of nested lattices is given in Fig. \ref{fig:lattice structure}.
  We construct a lattice codebook $\mathcal{C}$ based on a nested lattice pair $(\Lambda_s, \Lambda_c)$ satisfying $\Lambda_s \subseteq \Lambda_c$, where $ \Lambda_s$ is the shaping lattice and $\Lambda_c$ is the coding lattice. Denote by $\mathcal{V}_s$ and $\mathcal{V}_c$ the fundamental Voronoi regions of $ \Lambda_s$ and $ \Lambda_c$, respectively. The lattice codebook $\mathcal{C}$ can be represented as 
  \begin{equation}
    \mathcal{C}= \Lambda_c\ \text{mod}\ \Lambda_s = \Lambda_c \cap \mathcal{V}_s.
  \end{equation}
  The rate of $\mathcal{C}$ is given by
  \begin{equation}
    R=\frac{1}{n}\log|\mathcal{C}|=\frac{1}{n}\log\frac{\text{Vol}(\mathcal{V}_{s})}{ \text{Vol}(\mathcal{V}_{c})}
  \end{equation}
  where $\log$ denotes logarithm with base $2$.

  \subsection{Encoding at Transmitters \label{sec: encoding at trans}}
  
  We now describe the encoding at the transmitters. We use the Construction A method \cite{nazer2011compute} to construct a chain of $2L$ nested lattices $\Lambda_1, \Lambda_2,\cdots, \Lambda_{2L}$ as follows. Let $i_1,i_2,\cdots,i_{2L}$ be integers satisfying $0 \leq i_1 \leq i_2 \leq \cdots \leq i_{2L} \leq K$, where $K$ is a sufficiently large integer. Consider a matrix $\bm{G} \in \mathbb{F}_{\gamma}^{n\times K}$ with i.i.d elements uniformly drawn over $\mathbb{F}_{\gamma}$. Let $\bm{G}_{k}$ be the matrix consisting of the first $i_k$ columns of $\bm{G}$, for $k = 1,2,\cdots,2L$. Denote $\mathcal{L}_{k} = \{\bm{G}_{k}\bm w : \bm w \in \mathbb{F}_{\gamma}^{i_k}\}$. Then construct the lattice $\Phi_k$ as $\Phi_k = \gamma^{-1} g(\mathcal{L}_{k}) + \mathbb{Z}^{(n)}$, where $g(\cdot)$ maps the elements of $\mathbb{F}_{\gamma}$ into the corresponding integers $\{0,1,\cdots,\gamma-1\}$.
  We are now ready to construct
  \begin{equation}
    \Lambda_{k} = \bm{B}\Phi_k = \bm{B}(\gamma^{-1} g(\mathcal{L}_{k}) + \mathbb{Z}^{(n)}),\ k \in \mathcal{I}_{2L}
  \end{equation}
  where $\bm{B} \in \mathbb{R}^{n\times n}$ is the generation matrix of lattice $\Lambda_{k}$. By construction, we have $i_1 \leq i_2 \leq \cdots \leq i_{2L}$. Hence the constructed lattices are nested as $\Lambda_1 \subseteq \Lambda_2 \subseteq \cdots \subseteq  \Lambda_{2L} $. Also, the constructed lattices $\{\Lambda_k\}_{k=1}^{2L}$ are simultaneously good for AWGN\cite{nazer2011compute} and good for MSE quantization\cite{erez2004achieving}. 
  
  For each source $l$, we choose a lattice pair $(\Lambda_{\mathrm{s},l}, \Lambda_{\mathrm{c},l})$ from the nested lattice chain to construct a lattice codebook 
  \begin{equation}\label{eq:lattice code}
    \mathcal{C}_l= \Lambda_{\mathrm{c},l} \cap \mathcal{V}_{\mathrm{s},l}.
  \end{equation}
 Let $\pi_s(\cdot)$ be a permutation satisfying $\Lambda_{s,\pi_s(1)} \subseteq \Lambda_{s,\pi_s(2)} \subseteq \cdots \subseteq \Lambda_{s,\pi_s(L)}$, which gives the nested order of $L$ shaping lattices. 
 Let lattice chain permutation $\pi(\cdot)$ be a bijective map from $\{1,\cdots,2L\}$ to $\{1,\cdots,2L\}$ satisfying
  \begin{equation}\label{eq:permutation nested}
    \Lambda_{\pi(2l-1)}\subseteq\Lambda_{\pi(2l)},\ l \in \mathcal{I}_L.
  \end{equation}
  With \eqref{eq:permutation nested}, we construct each $\mathcal{C}_{l}$ using the lattice pair
  $(\Lambda_{\pi(2l-1)},\Lambda_{\pi(2l)}),\ l\in \mathcal{I}_L.$
  This implies the following relation:
  \begin{equation}\label{eq:WhichLatticePair}
    l=\Bigl\lceil \frac{\pi^{-1}(k)}{2}\Bigr\rceil,\ 1\leq k\leq 2L.
  \end{equation}
  From \eqref{eq:WhichLatticePair}, if $ \pi^{-1}(k) $ is even, then $\Lambda_k = \Lambda_{\mathrm{c},l}$ with $l = \frac{\pi^{-1}(k)}{2}$; otherwise, $\Lambda_k = \Lambda_{\mathrm{s},l}$ with $l = \frac{\pi^{-1}(k) + 1}{2}$.
  
  The message $\bm{w}_l$ of transmitter $l$ is uniformly drawn from $\mathbb{F}_{\gamma}^{k_l}$ with $k_l= i_{\pi(2l)} - i_{\pi(2l-1)}$. 
  We zero-pad $\bm w_l$ with $i_{\pi(2l-1)}$ leading zeros and $K - i_{\pi(2l)}$ trailing zeros, yielding
  \begin{equation}\label{eq:zero-pad}
   [\bm{0}_{i_{\pi(2l-1)}}^\mathrm{T},\ \bm w_l^\mathrm{T},\ \bm{0}_{K - i_{\pi(2l)}}^\mathrm{T}]^\mathrm{T} \in \mathbb{F}_{\gamma}^{K},\ l \in \mathcal{I}_L.
  \end{equation}
  With some abuse of notation, we henceforth denote by $\bm{w}_l$ the zero-padded vector in \eqref{eq:zero-pad}. We construct a function $\phi_l(\cdot)$ that maps message $\bm w_l$ into $\mathcal{C}_l$ using a common matrix $\bm G$:
  \begin{equation}\label{eq:encoding mapping}
	  \bm t_l = \phi_l(\bm w_l) = (\bm{B}\gamma^{-1} g(\bm{G}\bm w_l)) \bmod \Lambda_{\mathrm{s},l}\in \mathcal{C}_l,\ l \in \mathcal{I}_L.
  \end{equation}
  The function $\phi_l(\cdot)$ gives an isomorphism between $\mathbb{F}_{\gamma}^{k_l}$ and $\mathcal{C}_l$ \cite{nazer2011compute,feng2013algebraic}. Thus, $\bm t_l$ is uniformly distributed over $\mathcal{C}_l$. The inverse of $\phi_{l}(\cdot)$ is denoted by $\phi_{l}^{-1}(\cdot)$.
  The rate of source $l$ is given by 
  \begin{equation}\label{eq:source rate}
  	r_l=\frac{1}{n}\log|\mathcal{C}_l|=\frac{1}{n}\log\frac{\text{Vol}(\mathcal{V}_{\mathrm{s},l})}{\text{Vol}(\mathcal{V}_{\mathrm{c},l})}.
  \end{equation}
  Following the approach in \cite{zhu2014asymmetric}, we construct the channel input vector of transmitter $l$ as
  \begin{equation}\label{eq:channel input}
    \bm x_l=(\bm t_l/\beta_l-\bm d_l)\ \text{mod}\ \Lambda_{\mathrm{s},l}/\beta_l
  \end{equation}
  where $\beta_l \in \mathbb{R}$ is a scaling factor, and $\bm d_l$ is a random dithering signal uniformly distributed over the scaled Voronoi region $\mathcal{V}_{\mathrm{s},l}/\beta_l$. From the Crypto lemma\cite{erez2004achieving}, $\bm x_l$ is independent of $\bm t_l$ and is uniformly distributed over $\mathcal{V}_{\mathrm{s},l}/\beta_l$ \cite{erez2004achieving}. 
  We note that $\{\beta_l\}$ are set as $\beta_l = 1, l\in \mathcal{I}_L$ in the original CCF scheme in \cite{tan2015compute}. Here we treat $\{\beta_l\}$ as system variables to be optimized.
  
  From \eqref{eq:second moment} and \eqref{eq:normalized second moment}, the average power of $\bm x_l$ is given by
  \begin{equation}\label{eq:power&beta}
  \begin{split}
  p_l &= \frac{1}{n}\text{E}\lVert \bm x_l \rVert^2 = \frac{1}{n}\int_{\mathcal{V}_{\mathrm{s},l}/\beta_l} \frac{\lVert \bm x \rVert^2}{\text{Vol}(\mathcal{V}_{\mathrm{s},l}/\beta_l)}d \bm x \\
  &= G(\Lambda_{\mathrm{s},l}/\beta_l)\left(\text{Vol}(\mathcal{V}_{\mathrm{s},l}/\beta_l)\right)^{\frac{2}{n}}. 
  \end{split}
  \end{equation}
  As $\Lambda_{\mathrm{s},l}/\beta_l$ is good for MSE quantization, we obtain $\lim_{n \to \infty} G(\Lambda_{\mathrm{s},l}/\beta_l) = \frac{1}{2\pi e}$\cite{erez2004achieving}. Then, we have the following relation: 
  \begin{equation}\label{volumn_shaping_lattice}
  \text{Vol}(\mathcal{V}_{\mathrm{s},l})\! =\! \text{Vol}(\mathcal{V}_{\mathrm{s},l}/\beta_l)\beta_l^n\! =\! \left(\!\frac{p_l \beta_l^2}{G(\Lambda_{\mathrm{s},l}/\beta_l)}\!\right)^{\frac{n}{2}}\!\! \approx\!  (2\pi e p_l \beta_l^2)^{\frac{n}{2}}.
  \end{equation}
  This implies that the nesting order of $\{\Lambda_{\mathrm{s},l}\}$ is determined by the order of $\{p_l \beta_l^2\}$.

  \subsection{Computation at Receivers}

  Upon receiving $\bm y_m$ in \eqref{eq:channel model}, 
  receiver $m$ decodes a linear combination of lattice codewords from $\bm y_m$, denoted by
  \begin{equation}\label{eq:computation}
  	\hat{\bm v}_m = \varphi_m(\bm y_m)
  \end{equation}
  where $\varphi_m(\cdot)$ is the decoding function of receiver $m$. We now give more details of $\varphi_m(\cdot)$ following the CF approach in \cite{nazer2011compute}. Receiver $m$ first multiplies $\bm y_m$ by $\alpha_m$ and then cancels the dithering signals, yielding 
  \begin{equation}
    \begin{split}
    \bm{s}_m&=\alpha_m \bm y_m+\sum_{l=1}^{L}{a_{ml}\beta_l\bm d_l}\\
    &\stackrel{(a)}{=}\sum_{l=1}^{L}{a_{ml}\beta_l(\bm{t_l}/\beta_l-Q_{\Lambda_{\mathrm{s},l}/\beta_l}(\bm{t_l}/\beta_l-\bm d_l))}+\bm z_m'\\
    &\stackrel{(b)}{=}\sum_{l=1}^{L}{a_{ml}(\bm{t_l}-Q_{\Lambda_{\mathrm{s},l}}(\bm{t_l}-\beta_l\bm d_l))}+\bm z_m'
    \end{split}
  \end{equation}
  where $a_{ml}$ is an integer coefficient, and $\bm z_m' \triangleq \sum_{l=1}^{L}{(\alpha_m h_{ml}-a_{ml}\beta_l)\bm x_l}+\alpha_m \bm z_m$. In the above, step (a) follows from \eqref{eq:channel model} and \eqref{eq:channel input}, and step (b) follows by noting 
  $Q_{\Lambda}(\beta \bm t)=\beta Q_{\Lambda/\beta}(\bm t)\ \text{for}\ \beta >0.$
  Then, receiver $m$ quantizes $\bm{s}_m$ over $\Lambda_{f,m}$ and takes modulo over the coarsest lattice $\Lambda_{1}$, yielding
  \begin{equation}\label{eq:decoded codeword}
  \begin{split}
	\hat{\bm v}_m &= Q_{\Lambda_{f,m}}(\bm{s}_m)\bmod\Lambda_{1} \\
	&= Q_{\Lambda_{f,m}}\!\left(\!\sum_{l=1}^{L}{a_{ml}\!\left(\bm{t_l}-Q_{\Lambda_{\mathrm{s},l}}(\bm{t_l}-\beta_l\bm d_l)\right)}\! +\! \bm z_m' \!\right)\!\bmod\!\Lambda_{1}
  \end{split}
  \end{equation}
  where $\Lambda_{f,m}$ is the finest lattice in $\{\Lambda_{\mathrm{c},l}\}_{1}^{L}$ with $a_{ml}\neq 0$. Ignoring the equivalent noise $\bm z_m'$, we obtain the signal desired by receiver $m$:
  \begin{equation}\label{eq:computed codeword}
  \bm v_m = \left(\sum_{l=1}^{L}a_{ml}(\bm t_l-Q_{\Lambda_{\mathrm{s},l}}(\bm t_l-\beta_l \bm d_l)) \right)\bmod \Lambda_{1}.
  \end{equation}
  We say that a rate tuple $(r_1,r_2,\cdots,r_L)$ is achievable if 
  \begin{equation}
    \lim\limits_{n\rightarrow \infty}\text{Pr}\{\hat{\bm v}_m\neq \bm v_m\}=0,\ \text{for}\ m \in \mathcal{I}_L。
  \end{equation}
  Based on the results in \cite{zhu2014asymmetric}, the rate tuple $(r_1,r_2,\cdots,r_L)$ is achievable if for $l \in \mathcal{I}_L$,
  \begin{equation}\label{eq:computation rate constraint}
    r_l < \frac{1}{2}\log^{+} \left\{
    \min_{m:a_{ml}\neq 0}{\frac{p_l \beta_l^2}{\lVert \bm{P}^{\frac{1}{2}}\tilde{\bm{a}}_m\rVert^2 - \frac{(\bm h_m^\mathrm T \bm P \tilde{\bm{a}}_m)^2}{1+\lVert\bm{P}^{\frac{1}{2}} \bm h_m\rVert^2}}}
    \right\} \triangleq \hat{r}_l
  \end{equation}
   where $\bm{P} = \text{diag}\{p_1, p_2, \cdots, p_L\}$ and  $\tilde{\bm{a}}_m = [\beta_1 a_{m1},\beta_2 a_{m2},\cdots,\beta_L a_{mL}]^\mathrm{T}$. Denote by $\bm A$ the integer coefficient matrix with the $(m,l)$-th element given by $a_{ml}$. Following  \cite{nazer2011compute}, we always assume $\bm A$ is invertible over $\mathbb{F}_{\gamma}^{L\times L}$, so that $\{\t_l\}_1^L$ can be recovered from $\{\bm v_m\}_1^L$. An achievable rate region is then given by
    \begin{equation}\label{computation rate region}
      \mathcal{R}_{\mathrm{cpu}} = \left\{(r_1,r_2,\cdots,r_L)\in\mathbb{R}_{+}^{L} | r_l \leq \hat{r}_l, l \in \mathcal{I}_L \right\}.
    \end{equation}
    We refer to the rate region in \eqref{computation rate region} as the computation rate region. Here we use shorthand ``cpu'' for computation.

  \subsection{Compression at Receivers \label{sec: compression at receivers}}
  From \eqref{eq:computed codeword}, $\{\hat{\bm v}_m\}$ computed at receivers are generally correlated, as they are constructed by the same set of $\{\t_l\}$. Recall that each receiver in the channel model \eqref{eq:channel model} serves as a relay node. Forwarding $\{\hat{\bm v}_m\}$ directly at the receivers may lead to spectral inefficiency. That is, each receiver $m$ needs to compress $\hat{\bm v}_m$, so as to reduce the forwarding rate. Specifically, the compression at receiver $m$ is to generate
  \begin{equation}
  \hat{\bm \delta}_m=\psi_m(\hat{\bm v}_m)
  \end{equation}
  at a reduced rate $R_m(\leq r_m)$, where $\psi_m(\cdot)$ is referred to as the compression function of receiver $m$. For the overall scheme, the compression is required to be information lossless, i.e., $\{\hat{\bm v}_m\}_1^L $ can be exactly recovered from $\{\hat{\bm \delta}_m\}_1^L$. We say that a compression rate tuple $(R_1,R_2,\cdots,R_L)$ is achievable if $\{\hat{\bm v}_m\}_1^L$ can be recovered from $\{\hat{\bm \delta}_m\}_1^L$  without distortion. The convex hull of all achievable compression rate tuples gives the \textit{compression rate region}, denoted by $\mathcal{R}_{\mathrm{cpr}}$. For convenience of discussion, we henceforth assume that there is no error in receiver computations, i.e., $\hat{\bm v}_m = \bm{v}_m\ \text{for}\ m \in \mathcal{I}_L$. Correspondingly, the error-free version of $\hat{\bm \delta}_m$ is denoted by $ \bm \delta_m $. Before presenting the next section, we list some frequently used notations in Table \ref{tab:Notation}.
\begin{table}[h]
	\textcolor{black}{
		\caption{Frequently used notation\label{tab:Notation}}
		\centering{}
		\begin{tabular}{|>{\centering}p{0.2\columnwidth}|>{\centering}p{0.68\columnwidth}|}
			\hline 
			Notation & Definition\tabularnewline
			\hline 
			\hline 
			$\bm w_{l}$ & The message  of the $l$-th source\tabularnewline
			\hline
			$\bm t_{l}$ & Lattice codeword of the $l$-th source\tabularnewline
			\hline 
			$\bm{x}_{l}$ & Transmitted signal of the $l$-th source\tabularnewline
			\hline 
			$\bm \Lambda_{k}$ & The $k$-th lattice on the lattice chain \tabularnewline
			\hline 
			$\Lambda_{\mathrm{c},l}$ & Coding lattice  of the $l$-th source\tabularnewline
			\hline
			$\Lambda_{\mathrm{s},l}$ & Shaping lattice of the $l$-th source\tabularnewline
			\hline 
			$ \pi(\cdot) $ & A permutation of lattice chain satisfying $ \Lambda_{\pi(2l-1)} = \Lambda_{\mathrm{s},l} $ and $\Lambda_{\pi(2l)} = \Lambda_{\mathrm{c},l}$\tabularnewline
			\hline
			$\pi_{\mathrm{s}}(\cdot)$ & A permutation satisfying $\Lambda_{\mathrm{s},\pi_{\mathrm{s}}(1)} \subseteq \Lambda_{\mathrm{s},\pi_{\mathrm{s}}(2)} \subseteq \cdots \subseteq \Lambda_{\mathrm{s},\pi_{\mathrm{s}}(L)}$\tabularnewline
			\hline
			$\pi_{\mathrm{c}}(\cdot)$ &  A permutation satisfying $\Lambda_{\mathrm{c},\pi_{\mathrm{c}}(1)} \subseteq \Lambda_{\mathrm{c},\pi_{\mathrm{c}}(2)} \subseteq \cdots \subseteq \Lambda_{\mathrm{c},\pi_{\mathrm{c}}(L)}$\tabularnewline
			\hline 
			$Q_{\Lambda}\left(\bm{x}\right)$ & Quantization of $\bm{x}$ over lattice $\Lambda$\tabularnewline
			\hline 
			$\bm{x}\bmod\Lambda$ & $\bm{x}$ modulo $\Lambda$ \tabularnewline
			\hline 
			$\mathcal{V}$ & Fundamental Voronoi region of a lattice  \tabularnewline
			\hline
			$\textrm{Vol}\left(\mathcal{V}\right)$ & Volume of $\mathcal{V}$\tabularnewline
			\hline 
			$\beta_{l}$ & Precoding coefficient at the $l$-th source. \tabularnewline
			\hline 
			$\mathbf{A}$& Linear combination matrices $\bm A \in \mathbb{R}^{L\times M}$ \tabularnewline
			\hline
			$\mathbf{A}(\mathcal{S},\mathcal{S}')$ & Submatrix of $\bm A$ with rows indexed by $\mathcal{S}$ and the columns indexed by $\mathcal{S}'$\tabularnewline
			\hline
			$|\mathcal{S}|$ & Cardinality of $\mathcal{S}$\tabularnewline
			\hline
			$\text{rank}(\bm A)$ & Rank of matrix $\bm A$\tabularnewline
			\hline 
			$r_{l}$ & Transmission rate of the $l$-th source \tabularnewline
			\hline
			$R_{m}$ & Forwarding rate of the $m$-th receiver \tabularnewline
			\hline
			$\bm v_m$ & Computed codeword at receiver $m$ \tabularnewline
			\hline
			$\bm \delta_m$ & Compressed codeword at receiver $m$ \tabularnewline
			\hline
			$\mathcal{R}_{\mathrm{cpu}}$ & Computation rate region in \eqref{computation rate region}\tabularnewline
			\hline
			$\mathcal{R}_{\mathrm{cpr}}$ & Compression rate region in \eqref{eq:Origin Slepian-Wolf Bound}\tabularnewline
			\hline
			$\mathcal{C}_{\mathrm{v},k}$ & The $k$-th virtual lattice codebook  \tabularnewline
			\hline
			$r_{\mathrm{v},k}$ & Rate of $\mathcal{C}_{\mathrm{v},k}$ \tabularnewline
			\hline
			$\t_{l,k}$ & Codeword component of $\t_l$ in $\mathcal{C}_{v,k}$, defined in \eqref{eq:codeword component}\tabularnewline
			\hline
			$ \mathcal{I}_l$ & Index set $\{1,2,\cdots,l\}$\tabularnewline
			\hline
			$ \mathcal{L}_k$ & The index set that consists of all $l$ satisfying $\mathcal{C}_{\mathrm{v},k} \subseteq \mathcal{C}_l$ for given $k$ \tabularnewline		
			\hline
			$ \mathcal{K}_l $ & The index set that consists of all $k$ satisfying $\mathcal{C}_{\mathrm{v},k} \subseteq \mathcal{C}_{l}$ for given $l$\tabularnewline
			\hline
			$ \alpha_m$ & Weight coefficient of $R_m$ in \eqref{eq: oringinal_proble_a}  \tabularnewline
			\hline
			$\pi_{\alpha}(\cdot)$ &  Permutation defined by the order of $\{\alpha_m\}$\tabularnewline
			\hline 
		\end{tabular}
	}
\end{table}

\section{Proposed Compression Scheme \label{sec: compression}}

In this section, we propose a compression scheme based on the basic lattice operation introduced in Section \ref{sec:lattice code}: quantization and modulo operations. We first describe a technique termed \textit{lattice codeword splitting}. We then present the proposed compression scheme by utilizing the codeword splitting. Finally, we discuss the relation between the proposed scheme and the original CCF in \cite{tan2015compute}.

  \subsection{Lattice Codeword Splitting \label{sec:codeword splitting}}

  \begin{figure*}
	\centering
	\includegraphics[width=5.0in]{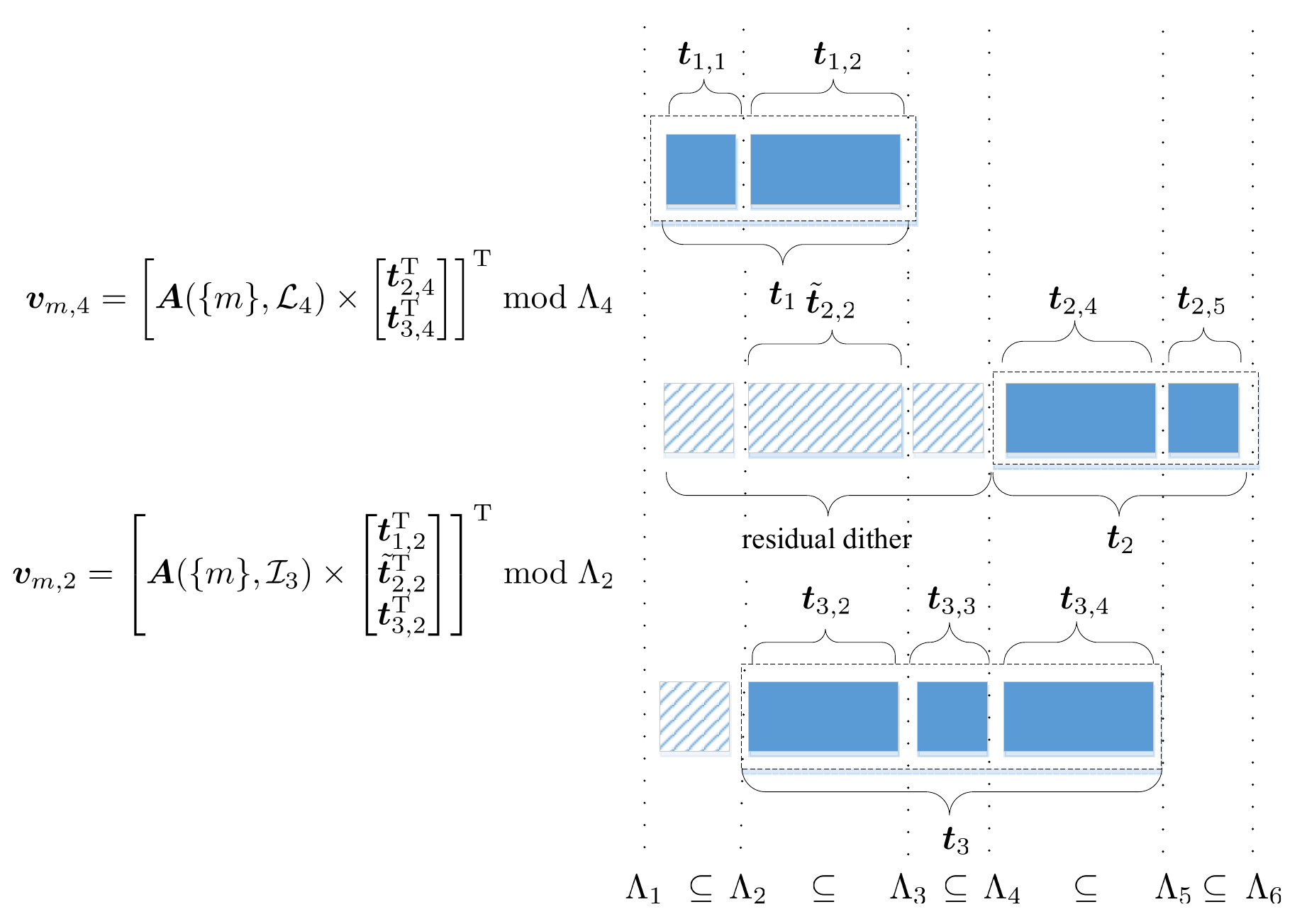}
	\caption{An illustration of codeword splitting with the nested lattice chain $\Lambda_1 (\Lambda_{\mathrm{s},1}) \subseteq \Lambda_2 (\Lambda_{\mathrm{s},3}) \subseteq \Lambda_3 (\Lambda_{\mathrm{c},1}) \subseteq \Lambda_4 (\Lambda_{\mathrm{s},2}) \subseteq \Lambda_5 (\Lambda_{\mathrm{c},3}) \subseteq \Lambda_6 (\Lambda_{\mathrm{c},2})$. The source codewords are given by $\t_1 \in \Lambda_{c,3}\cap \mathcal{V}_1$, $\t_2 \in \Lambda_{c,6}\cap \mathcal{V}_4$, and $\t_3 \in \Lambda_{c,5}\cap \mathcal{V}_2$. Each vertical dotted lines corresponds to a lattice in the nested lattice chain. Each solid rectangle represents a codeword component. For example, the codeword component $\t_{1,1}$ is obtained by quantizing $\t_1$ over $\Lambda_2$ and then taking modulo over $\Lambda_1$. Each rectangle filled with oblique lines represents a dithered codeword component. For example,  $\tilde{\t}_{2,2}$ in the  figure is defined as $\tilde{\t}_{2,2} =  Q_{\Lambda_3}(\bm t_2 - Q_{\Lambda_4}(\bm t_2 - \beta_2 \bm d_2))\bmod \Lambda_{2}$. Also, since \eqref{eq_zero_dither} holds for $k=4$, $\bm v_{m,4}$ (i.e., the fourth codeword component of receiver $m$) is a combination of $\t_{2,4}$ and $\t_{3,4}$ with no residual dither.}
	\label{fig:illus-splitting}
\end{figure*}

  To start with, we describe lattice codeword splitting by quantization and modulo operations. These operations will be frequently used in the proposed compression scheme. Construct a virtual nested lattice codebook $\mathcal{C}_{\mathrm{v},k}$ with lattice pair $(\Lambda_k,\Lambda_{k+1})$, i.e.
  \begin{equation}\label{eq:virtual lattice code}
  \mathcal{C}_{\mathrm{v},k}=\Lambda_{k+1}\cap \mathcal{V}_k,\ k=1,\cdots,2L-1,
  \end{equation}
  where the subscript $_\mathrm{v}$ means ``virtual''. We call $\mathcal{C}_{\mathrm{v},k}$ the $k$-th virtual lattice codebook.
  The rate of $\mathcal{C}_{\mathrm{v},k}$ is 
  \begin{equation}\label{eq:rate of virtual lattice code}
  r_{\mathrm{v},k}=\frac{1}{n}\log|\mathcal{C}_{\mathrm{v},k}| = \frac{1}{n}\log\frac{\text{Vol}(\mathcal{V}_{k})}{\text{Vol}(\mathcal{V}_{k+1})}.
  \end{equation}
  Define the $k$-th \textit{codeword component} of $\t_{l}$ as
  \begin{equation}\label{eq:codeword component}
  \t_{l,k} = Q_{\Lambda_{k+1}}(\t_l)\bmod\Lambda_{k}.
  \end{equation}
  Clearly, $ \t_{l,k}$ is a codeword in $\mathcal{C}_{\mathrm{v},k}$. Note that the map from $\t_{l}$ to $\t_{l,k}$ is a homomorphism and $\t_{l}$ is uniformly distributed $\mathcal{C}_{\mathrm{v}}$, then $\t_{l,k}$ is uniformly distributed over $\mathcal{C}_{\mathrm{v},k}$. Define 
  \mathleft
  \begin{equation}\label{eq:Index set I_k}
  \mathcal{L}_k=\{l|\Lambda_{\mathrm{s},l}\subseteq \Lambda_k\subseteq\Lambda_{k+1}\subseteq\Lambda_{\mathrm{c},l},l\in \mathcal{I}_L\},k\in \mathcal{I}_{2L-1} ,
  \end{equation}
  \begin{equation}\label{eq:Index set J_l}
  \mathcal{K}_l=\{k|\Lambda_{\mathrm{s},l}\subseteq \Lambda_k\subseteq\Lambda_{k+1}\subseteq\Lambda_{\mathrm{c},l},k\in\mathcal{I}_{2L-1}\}, l\in \mathcal{I}_L.
  \end{equation}
  \mathcenter
  Note that $\Lambda_{\mathrm{s},l}\subseteq \Lambda_k\subseteq\Lambda_{k+1}\subseteq\Lambda_{\mathrm{c},l}$ implies that
  $\mathcal{C}_{\mathrm{v},k} \subseteq \mathcal{C}_{l}.$
  Thus, $\mathcal{L}_k$ is the index set that consists of all $l$ satisfying $\mathcal{C}_{\mathrm{v},k} \subseteq \mathcal{C}_l$; $\mathcal{K}_l$ is the index set that consists of all $k$ satisfying $\mathcal{C}_{\mathrm{v},k} \subseteq \mathcal{C}_{l}$. 
  
  We are now ready to split codeword $\t_l$ into a combination of the codeword components in \eqref{eq:codeword component}:
  \begin{subequations}
	  \begin{align}
	  \t_l &= \t_l \bmod \Lambda_1\\
	  &= \left( Q_{\Lambda_2}(\t_l) + \t_l\bmod \Lambda_2 \right)\bmod \Lambda_1 \label{eq_split_temp0}\\
	  &= \left( Q_{\Lambda_2}(\t_l)\bmod \Lambda_1 + \t_l\bmod \Lambda_2 \right)\bmod \Lambda_1\\
	  &= \left( \t_{l,1} + \t_l\bmod \Lambda_2 \right)\bmod \Lambda_1
	  \end{align}
  \end{subequations}
  where \eqref{eq_split_temp0} follows from \eqref{eq:modulo}. By induction, $\t_l$ can be split as
  \begin{equation}\label{eq:codeword_splitting_0}
  	\t_l = \left( \sum_{k=1}^{2L-1} \t_{l,k} \right)\bmod \Lambda_1.
  \end{equation}
  Taking modulo over $\Lambda_{\mathrm{s},l}$ on both sides of \eqref{eq:codeword_splitting_0}, we obtain
  \begin{subequations}\label{eq:codeword splitting}
  	\begin{align}
  		\t_l &= \t_l\bmod \Lambda_{\mathrm{s},l} \label{eq:codeword splitting_0}\\
  		&= \left(  \sum_{k=1}^{2L-1} \t_{l,k} \right)\bmod \Lambda_{\mathrm{s},l} \label{eq:codeword splitting_1}\\
  		&= \left(\sum_{k \in \mathcal{K}_l} \t_{l,k} \right)\bmod \Lambda_{\mathrm{s},l} \label{eq:codeword splitting_2}
  	\end{align}
  \end{subequations}
  where \eqref{eq:codeword splitting_0} follows from $\t_l \in \mathcal{V}_{\mathrm{s},l}$, \eqref{eq:codeword splitting_1} from \eqref{eq:codeword_splitting_0}, and \eqref{eq:codeword splitting_2} from the fact that $\t_{l,k} = \bm 0$ for $k\notin \mathcal{K}_l$.

  From the splitting operation above, $\t_l$ is uniquely mapped to the codeword component tuple $ (\t_{l,1},\t_{l,2},\cdots,\t_{l,2L-1})$;  for any given $ (\t_{l,1},\t_{l,2},\cdots,\t_{l,2L-1})$, $\t_l$ can be constructed by using \eqref{eq:codeword splitting}. Thus, the map between $\t_l$ and $(\t_{l,1},\t_{l,2},\cdots,\t_{l,2L-1})$ is bijective. 
  Recall that $\t_{l}$ is uniformly distributed over $\mathcal{C}_{\mathrm{v}}$ and $\t_{l,k}$ is uniformly distributed over $\mathcal{C}_{\mathrm{v},k}$. Thus, $\t_{l,k}$ is independent of $\t_{l,k'}$ for $k \neq k'$. Also, 
   the rate of $\t_l$ can be represented by
  \begin{equation}\label{eq:Source rate splitting}
  r_l=\sum_{k\in \mathcal{K}_l}{r_{\mathrm{v},k}},\ \text{for}\ l\in \mathcal{I}_L.
  \end{equation}

We next split each $\bm v_m$ as follows.
Define the $k$-th codeword component of $\bm v_k$ as 
\begin{equation}\label{eq_v_splitting}
	\bm v_{m,k} = Q_{\Lambda_{k+1}}(\bm v_m) \bmod \Lambda_k. 
\end{equation}
Similar to \eqref{eq:codeword_splitting_0}, we obtain 
\begin{equation}
	\bm v_m = \left(\sum_{k=1}^{2L-1} \bm v_{m,k} \right)\bmod \Lambda_{1}.
\end{equation}

Furthermore, we can represent $\bm v_{m,k}$ as a function of $\t_l$:
\begin{equation}\label{eq:component of v_m}
	\bm v_{m,k} = \left(\sum_{l=1}^{L}a_{ml} \tilde{\t}_{l,k} \right) \bmod \Lambda_{k}
\end{equation}
where 
\begin{align}
	\tilde{\t}_{l,k} &= Q_{\Lambda_{k+1}}(\tilde{\t}_{l})\bmod \Lambda_k, \label{eq_tilde_t_l_k}\\
	\tilde{\t}_{l} &= \t_l - Q_{\Lambda_{\mathrm{s},l}}(\t_l - \beta_l \bm d_l).\label{eq_tilde_t_l}
\end{align}
The derivation of \eqref{eq:component of v_m} is given in Appendix \ref{app:derivation}.
Following from $Q_{\Lambda_{k+1}}(\bm x)\bmod \Lambda_k = Q_{\Lambda_{k+1}}(\bm x \bmod \Lambda_k)\bmod \Lambda_k$,
we represented $\tilde{\t}_{l,k}$ as 
\begin{equation}\label{eq:component of v_m_2}
	\tilde{\t}_{l,k} =  Q_{\Lambda_{k+1}}\left( \t_l - Q_{\Lambda_{\mathrm{s},l}}(\t_l - \beta_l \bm d_l)\bmod \Lambda_k \right)\bmod \Lambda_k.
\end{equation}
In \eqref{eq:component of v_m_2}, $Q_{\Lambda_{\mathrm{s},l}}(\t_l - \beta_l \bm d_l)\bmod \Lambda_{k}$ is a residual dithering signal. Since $Q_{\Lambda_{\mathrm{s},l}}(\cdot) \in \Lambda_{\mathrm{s},l}$, we obtain  $Q_{\Lambda_{\mathrm{s},l}}(\t_l - \beta_l \bm d_l)\bmod \Lambda_{k} = \bm 0$ and 
\begin{equation}\label{eq_t_tild_t}
	\tilde{\t}_{l,k} = \t_{l,k} \text{ for } \Lambda_{\mathrm{s},l} \subseteq \Lambda_k.
\end{equation}
Therefore, when $\Lambda_{k}$ is finer than the finest shaping lattice, i.e.
\begin{align}\label{eq_zero_dither}
	\Lambda_{\mathrm{s},l} \subseteq \Lambda_k, l\in\mathcal{I}_L,
\end{align}
$\bm v_{m,k}$ can be simplified as

\begin{subequations}\label{eq_v_m_k}
	\begin{align}
	\bm v_{m,k} &= \left(\sum_{l\in\mathcal{L}_k} a_{ml}\t_{l,k} \right) \bmod \Lambda_{k} \label{eq:component of v_m2}\\
	&= \left[\bm A(m,\mathcal{L}_k)\bm T_{k}\right]^{\mathrm{T}} \bmod \Lambda_k \label{eq:Matrix_Computed_codeword_compnent}
	\end{align}
\end{subequations}
where $l_i$ is the $i$-th element of $\mathcal{L}_k$ (ordered in an ascending manner), and $\bm A(m,\mathcal{L}_k)$ is the $m$-th row of the submatrix of $\bm A$ with the columns indexed by $\mathcal{L}_k$, and $\bm T_{k} = [\t_{l_1,k},\t_{l_2,k}, \cdots , \t_{l_{|\mathcal{L}_k|},k}]^{\mathrm{T}}$. Recall that $\{ \t_{l,k}|l\in\mathcal{I}_L,k\in\mathcal{K}_l \}$ are independent of each other. Thus, $\bm v_{m,k}$ is independent of $\{\bm v_{m',k'}|m'\in\mathcal{I}_L, k'\in\mathcal{I}_{2L-1}\backslash \{k\}\}$.
Fig. \ref{fig:illus-splitting} illustrates the codeword splitting of $\t_l$ and the relation between $\bm v_{m,k}$ and $\t_{l,k}$.



\subsection{Design of the Compression Function}

In this subsection, we describe how to design the compression functions $\{\psi_m(\cdot)\}$ based on quantization and modulo (\textit{QM}) operations. With the codeword splitting technique in Subsection A, we show that the proposed scheme is information lossless.
We also show that the one-quantization-plus-one-modulo approach proposed in \cite{tan2015compute} may be not very efficient for compression. Instead, the optimal compression function in general consists of multiple quantization and modulo pairs. 
We refer to our multiple-QM approach as generalized CCF (GCCF) to distinguish it from the single-QM approach in \cite{tan2015compute}.

To proceed, we map $\bm A \in \mathbb{R}^{L\times L}$ into $ g^{-1}(\bm A\ \bmod\ \gamma) \in \mathbb{F}_{\gamma}^{L\times L}$ following \cite{nazer2011compute}. With some abuse of notation, we replace $g^{-1}(\bm A\ \bmod\ \gamma)$ simply by $\bm A$ in circumstances without causing ambiguity.
For any $\mathcal{S}\subseteq \mathcal{I}_L,\mathcal{S}' \subseteq \mathcal{I}_L$, denote by $\bm A(\mathcal{S},\mathcal{S}')$ the submatrix of $\bm A$ with the rows indexed by $\mathcal{S}$ and the columns indexed by $\mathcal{S}'$. Denote by $\overline{\mathcal{S}}$ the complement of $\mathcal{S}$ in $\mathcal{I}_L$.
Let $\pi_{\alpha}(\cdot)$ be a permutation of $(1,2,\cdots,L)$.  Denote $ \Pi_{\alpha}(\mathcal{I}_{m})=\{\pi_{\alpha}(i)|i\in\mathcal{I}_{m}\}$ and $\Pi_{\alpha}(\emptyset)=\emptyset$. 
Define an index set $\mathcal{J}_{m}$ as 
\begin{equation}\label{eq:rank_equation}
	\begin{split}
		\mathcal{J}_{m} \triangleq \left\{ k| \textit{rank}(\bm A(\Pi_{\alpha}(\overline{\mathcal{I}_{m-1}}), \mathcal{L}_{k}))= \textit{rank}(\bm A(\Pi_{\alpha}(\overline{\mathcal{I}_{m}}), \mathcal{L}_{k})) + 1 \right\},
	\end{split}
\end{equation}
where $\textit{rank}(\bm A)$ is evaluated in $\mathbb{F}_{\gamma}^{L\times L}$. From \eqref{eq:rank_equation}, we see that $k\in\mathcal{J}_m$ implies that the row $\bm A(\pi_{\alpha}(m),\mathcal{L}_k)$ is linearly independent of the rows of $\bm A(\pi_{\alpha}(\overline{\mathcal{I}_{m}}), \mathcal{L}_{k})$. Note that $\pi_{\alpha}(\cdot)$ specifies an order of counting the row index of $\bm A$ and that $\mathcal{J}_m$ is a function of $\pi_{\alpha}(\cdot)$.
With $\eqref{eq:rank_equation}$, we design the compression function as follows.
\begin{thm}\label{thm: achive vertex}
  An information lossless compression scheme is given by
  \begin{equation}\label{eq: compression operation}
    \bm \delta_{m} = \psi_{m}(\bm v_{m}) =  \left(\sum_{k\in \mathcal{J}_{\pi_{\alpha}^{-1}(m)}} \bm v_{m,k}\right)\bmod \Lambda_{J_m^{\text{min}}}, m\in\mathcal{I}_L
  \end{equation}
  where $J_m^{\text{min}}$ is the minimal element in $\mathcal{J}_{\pi_{\alpha}^{-1}(m)}$. 
\end{thm}
To prove Theorem \ref{thm: achive vertex}, it suffices to show that the compression in \eqref{eq: compression operation} is information lossless, i.e., $\{\bm v_m\}_1^L$ can be recovered from $\{\bm \delta_{m}\}_1^L$. The detailed proof is given in Appendix \ref{app:recovery}.
\begin{rem}
	We give intuitions about how the compression functions in \eqref{eq: compression operation} work. Note that the compression aims to reduce the redundant information in $\{\bm v_m\}$. From \eqref{eq:component of v_m} and \eqref{eq:component of v_m_2}, $\bm v_{m,k}$ is a combination of $\{\t_{l,k},l\in\mathcal{L}_k\}$ by ignoring the residual dithers. Without dithers, we see from \eqref{eq:rank_equation} that $\bm v_{\pi_{\alpha}(m),k}$ is linearly independent of $\{\bm v_{\pi_{\alpha}(m'),k},m'\in\overline{\mathcal{I}_{m}}\}$ if $k\in \mathcal{J}_{m}$. Thus, the rationale of \eqref{eq: compression operation} is to choose independent codeword components of $\bm v_{m}$ in constructing $\bm \delta_m$. 
\end{rem}

\begin{thm}\label{cor:rate_of_delta}
	The achievable compression rate tuple of \eqref{eq: compression operation} is given by $(R_1,R_2,\cdots, R_L)$, with $R_m$ being the entropy rate of $\bm \delta_m$:
	\begin{equation}\label{eq_rate_delta_m}
	R_m = H(\bm \delta_m) = \sum_{k\in \mathcal{J}_{\pi_{\alpha}^{-1}(m)}} r_{\mathrm{v},k},\ \text{for}\ m \in \mathcal{I}_L.
	\end{equation}
	where $H(\cdot)$ denotes the entropy function.
	Further, the sum of the entropy rates of $\{ \bm \delta_m \}$ satisfies
	\begin{equation}\label{eq_sum_rate_delta_m}
	\sum_{m=1}^{L}H(\bm \delta_m) = \sum_{l=1}^{L} r_{l}.
	\end{equation}
\end{thm}

Eqn. \eqref{eq_sum_rate_delta_m} implies that there is no redundancy in the compressed codewords $(\bm \delta_1,\bm \delta_2,\cdots,\bm \delta_L)$. The proof of Theorem \ref{cor:rate_of_delta} is given in Appendix \ref{app:proof_of_cor_rate_of_delta}.

\subsection{Relation Between GCCF and CCF\label{subsec:GCCF&CCF}}
 The compression function in CCF \cite{tan2015compute} is given by
\begin{equation}\label{eq:CCF_compr}
\bm \delta_{m} = Q_{\Lambda_{d,m}}(\bm v_{m})\bmod\Lambda_{e,m},\ m \in \mathcal{I}_L,
\end{equation}
where the lattice pair $(\Lambda_{e,m}, \Lambda_{d,m})$ are chosen from the lattice chain $\Lambda_1\subseteq\Lambda_2\subseteq\cdots\subseteq\Lambda_{2L}$. Compared with \eqref{eq: compression operation}, the compression function in \eqref{eq:CCF_compr} only consists of a single pair of QM operations. Generally speaking, redundancy may still exists after the compression in \eqref{eq:CCF_compr}. For example, consider the case that the nested lattice chain is given in Fig. \ref{fig:illus-splitting}. Suppose $\bm A = [2,3,4;2,1,3;1,2,3]$, and $\pi_{\alpha}(m) = m$ for $m \in \mathcal{I}_L$.
 From \eqref{eq:rank_equation}, $\mathcal{K}_{\pi_{\alpha},2} = \{2,4\}$. From \eqref{eq: compression operation}, we have
 \begin{equation}
 \bm \delta_{2} = \left( \bm v_{2,2} + \bm v_{2,4} \right) \bmod \Lambda_{2}.
 \end{equation}
 In CCF, the compression function is given by 
 \begin{equation}
 \begin{split}
 \bm \delta_{2}' &= Q_{\Lambda_4}\left( \bm v_{2} \right) \bmod \Lambda_{2} = \left( \bm \delta_{2} + \bm v_{2,3} \right)\bmod \Lambda_{2}.
 \end{split}
 \end{equation}
Clearly, the rate of $ \bm \delta_{2}'$ is in general higher than the rate of $ \bm \delta_{2}$, which implies redundancy.

In the following, we show that the GCCF reduces to CCF when the coding scheme satisfies the \textit{separability condition} defined as
\begin{equation}\label{eq:separability}
\Lambda_{\mathrm{s},l} \subseteq \Lambda_{\mathrm{c},l}',\ \text{for}\ l, l' \in \mathcal{I}_L.
\end{equation}
Intuitively, the above separability condition says that, in the nested lattice chain, the finest shaping lattice is coarser than the coarsest coding lattice. The main result is presented below, with the proof given in Appendix \ref{app:compression_2}.

\begin{thm}\label{thm: achive vertex 2}
	When the separability condition in \eqref{eq:separability} is satisfied, the compression function \eqref{eq: compression operation} can be represented as
	\begin{equation}\label{eq_delta_m_2}
	\bm \delta_{m} = Q_{\Lambda_{J_m^{\text{max}}+1}}(\bm v_{m})\bmod\Lambda_{J_m^{\text{min}}},\ m \in \mathcal{I}_L
	\end{equation}
	where $J_m^{\text{max}}$ is the maximal element in $\mathcal{J}_{\pi_{\alpha}^{-1}(m)}$. 
\end{thm}

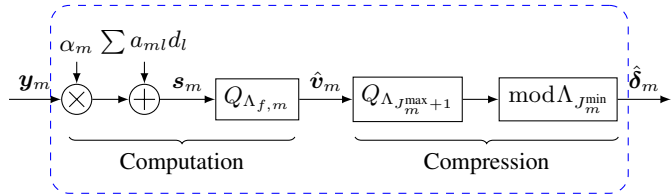
\begin{figure}[!htb]
	\centering
	\begin{tikzpicture}
	\small
	\node [draw, crosscircle,minimum width=4mm] (1) at (0.1,0) {};
	\node [draw, cross]  at (1) {};
	\node [draw, crosscircle,minimum width=4mm] (2) at (1,0) {};
	\node [draw, cross, rotate=45]  at (2) {};
	\node [draw, block,minimum width =8mm] (3) at (2.5,0) {$Q_{\Lambda_{f,m}}$};
	\node [draw, block,minimum width =10mm] (4) at (4.5,0) {$Q_{\Lambda_{J_m^{\text{max}} + 1}}$};
	\node [draw, block,minimum width =10mm] (5) at (6.5,0) {$\bmod \Lambda_{J_m^{\text{min}}}$};
	\draw [-latex] (-0.8,0) --  (1) node[midway, above, sloped]{$\bm y_m$};
	\draw [-latex] (1) --  (2) node[midway, above, sloped]{};
	\draw [-latex] (2) -- (3) node[midway, above, sloped]{$\bm s_m$};
	\draw [-latex] (0.1,0.5) -- (1.north) node[pos = 0,above]{$\alpha_m$};
	\draw [-latex] (1,0.5) -- (2.north) node[pos = 0,above]{$\sum a_{ml}d_{l}$};
	\draw [-latex] (3) -- (4) node[midway, above, sloped]{$ \hat{\bm v}_m$};
	\draw [-latex] (4) -- (5); 
	\draw [-latex] (5) -- (8.0,0) node[midway, above, sloped]{$\hat{\bm \delta}_m$};
	\draw [decoration={brace,mirror},decorate,yshift=-5mm] (0,0) -- (3.0,0) node[midway, below = 1mm of 1, sloped]{Computation};
	\draw [decoration={brace,mirror},decorate,yshift=-5mm] (3.8,0) -- (7.3,0) node[midway, below = 1mm of 1, sloped]{Compression};
	\node [draw,dottedblock, fit=(1)(2)(3)(4)(5)](dotblock){};
	\node [above= 0.5mm of dotblock]{};
	\end{tikzpicture}
	\caption{Computation and compression at relay $m$ under the separability condition.}
	\label{fig:relay operation}
\end{figure}

Under the separability condition, one \textit{QM} operation is good enough for compression at each receiver. In this case, GCCF is equivalent to CCF. The compression operation in \eqref{eq_delta_m_2} is illustrated in Fig. \ref{fig:relay operation}. 

\section{Compression Rate Region \label{sec:rate region}}

In this section, we will present an achievable compression rate region $\mathcal{R}_{\mathrm{cpr}}$ of the proposed GCCF scheme. We will also discuss the relation of the well-known Slepian-Wolf theorem and our GCCF scheme. Two examples are given to illustrate the GCCF scheme and the corresponding rate region at the end.

\subsection{Achievable Rate Region of GCCF}
We now present an achievable compression rate region $\mathcal{R}_{\mathrm{cpr}}$ of GCCF.

\begin{thm}\label{theorem: compression rate region}
		A compression rate region $\mathcal{R}_{\mathrm{cpr}}$ of GCCF is given by
		\begin{align}
		\sum_{m \in \mathcal{S}} R_m \geq f(\mathcal{S}),\ \text{for}\ \mathcal{S}\subseteq \mathcal{I}_L \label{eq:Compression rate Bound}
		\end{align}
		where
		\begin{equation}\label{eq:rank function}
			f(\mathcal{S})\! =\! \sum_{k=1}^{2L-1}\!\! \left(\textit{rank}(\!\bm A(\mathcal{I}_L, \mathcal{L}_{k})) -  \textit{rank}(\bm A(\overline{\mathcal{S}}, \mathcal{L}_{k}))\right) r_{\mathrm{v},k}.
		\end{equation}
\end{thm}
\begin{IEEEproof}
	Note that $\mathcal{R}_{\mathrm{cpr}}$ in \eqref{eq:Compression rate Bound} is a polytope.   
	Then, to prove Theorem \ref{theorem: compression rate region}, it suffices to show that GCCF can achieve all the vertices of $\mathcal{R}_{\mathrm{cpr}}$ (by noting that the other rate tuples in $\mathcal{R}_{\mathrm{cpr}}$ can be achieved by time sharing of vertices). 
	
	We first show how to determine the vertices of $\mathcal{R}_{\mathrm{cpr}}$. To this end, let 
	$\alpha_1,\cdots,\alpha_L$ be positive real numbers satisfying $\alpha_{\pi_{\alpha}(1)}\geq \alpha_{\pi_{\alpha}(2)}\geq \cdots \geq \alpha_{\pi_{\alpha}(L)}$. Then, a vertex of $\mathcal{R}_{\mathrm{cpr}}$ can be found by solving the following weighted sum-rate minimization problem:
	\begin{subequations}\label{eq:Primal Problem}
		\begin{align}
		\mathop{\text{minimize}} &\ \sum_{m\in \mathcal{I}_L}{\alpha_m R_m} \label{eq: oringinal_proble_a}\\
		\mathop{\text{subject to}} &\ 
		\begin{aligned}
		&\sum_{m\in \mathcal{S}}{R_m}\geq f\left(\mathcal{S} \right),\ \text{for}\ \mathcal{S}\subseteq \mathcal{I}_L.
		\end{aligned} \label{eq: oringinal_proble_b}
		\end{align}
	\end{subequations}
	It can be shown that $-f(\mathcal{S})$ is a submodular function by noting the rank function is submodular and the summation preserves submodularity. 
	From \cite[pp 70]{lee2004first}, the solution to \eqref{eq:Primal Problem} is given by $(R_1,R_2,\cdots,R_L)$ with
	\begin{equation}\label{eq:vertex coordinates}
	R_{\pi_{\alpha}(m)} =  f(\Pi_{\alpha}(\mathcal{I}_{m})) - f(\Pi_{\alpha}(\mathcal{I}_{m-1})),\ \text{for}\ m \in \mathcal{I}_L.
	\end{equation}
	That is, \eqref{eq:vertex coordinates} gives the coordinates of the vertex corresponding to the permutation $\pi_{\alpha}(\cdot)$. By enumerating all possible $\pi_{\alpha}(\cdot)$, we obtain all the vertices of $\mathcal{R}_{\mathrm{cpr}}$. What remains is to show that GCCF can achieve the rate tuple given by \eqref{eq:vertex coordinates}.
	By substituting $f(\mathcal{S})$ into \eqref{eq:vertex coordinates}, together with the definition of $\mathcal{J}_m$ in \eqref{eq:rank_equation}, we obtain 
	\begin{align}
	R_{\pi_{\alpha}(m)} = \sum_{k\in \mathcal{J}_{m}} r_{\mathrm{v},k},\ \text{for}\ m \in \mathcal{I}_L. \label{eq_compute_vertex_2}
	\end{align}
	From Theorem \ref{cor:rate_of_delta}, we see that GCCF achieves the vertex of $\mathcal{R}_{\mathrm{cpr}}$ in \eqref{eq:vertex coordinates}, which concludes the proof. 
\end{IEEEproof}

\begin{rem}
	Since $-f(\mathcal{S})$ is a submodular function,  
	$f(\mathcal{S})$ is a supermodular function. The rate region $\mathcal{R}_{\mathrm{cpr}}$ given by \eqref{eq:Compression rate Bound} is a contra-polymatroid\cite{piyushadvances}. 
\end{rem}
\begin{rem}
	Each $\pi_{\alpha}(\cdot)$ determines a vertex of $\mathcal{R}_{\mathrm{cpr}}$. However, the map between $\{\pi_{\alpha}(\cdot)\}$ and vertices of $\mathcal{R}_{\mathrm{cpr}}$ is not necessarily a bijection. This implies that the total number of vertices of $\mathcal{R}_{\mathrm{cpr}}$ may be less than $L!$.
\end{rem}

 \subsection{Distributed Source Coding}
  
So far, we have established the compression rate region $\mathcal{R}_{\mathrm{cpr}}$ of GCCF. A natural question is whether the compression rate region can be further enlarged or not.
From Section \ref{sec: compression at receivers}, the compression operation is information lossless, i.e., the messages $\{\bm v_m\}_1^L$ can be recovered from the compressed messages $\{\bm \delta_m\}_1^L$ without distortion. This is a distributed source coding problem with the optimal compression rate region given by the Slepian-Wolf theorem \cite{Slepian1973Wolf, cover1975proof}:
\begin{equation}\label{eq:Origin Slepian-Wolf Bound}
\sum_{m \in \mathcal{S}}\!\! R_m \!
\geq \!H\!\left(\{ \bm v_m | m \in \mathcal{S}\}|\{\bm v_m | m \in \overline{\mathcal{S}}\}\right),\! \text{ for }\! \mathcal{S}\subseteq\! \mathcal{I}_L.
\end{equation}
We henceforth refer to the rate region in \eqref{eq:Origin Slepian-Wolf Bound} as the Slepian-Wolf region  $\mathcal{R}_{\mathrm{SW}}$. 
Comparing \eqref{eq:Compression rate Bound} with \eqref{eq:Origin Slepian-Wolf Bound}, we see that $\mathcal{R}_{\mathrm{cpr}} = \mathcal{R}_{\mathrm{SW}}$ if 
\begin{equation}\label{equal_region}
f(\mathcal{S}) =  H\left(\{ \bm v_m | m \in \mathcal{S}\}|\{\bm v_m | m \in \overline{\mathcal{S}}\}\right), \text{for } \mathcal{S}\subseteq \mathcal{I}_L.
\end{equation}
The following theorem shows that \eqref{equal_region} always holds for $\mathcal{S}=\mathcal{I}_L$.
\begin{thm}\label{thm_minmize_total_rate}
	The equality in \eqref{equal_region} always holds for $\mathcal{S}=\mathcal{I}_L$, i.e., GCCF is optimal in terms of minimizing the total compression rate.  
\end{thm}
\begin{IEEEproof}
	From Theorems 2 and 4, the minimum total compression rate of GCCF is achieved at a vertex of $\mathcal{R}_{\mathrm{cpr}}$ in \eqref{eq:Compression rate Bound}, with the coordinates $(R_1,\cdots,R_L) $ given by \eqref{eq_rate_delta_m}. From \eqref{eq:vertex coordinates}, we obtain
	\begin{subequations}\label{equal_region_2_1}
		\begin{align}
			f(\mathcal{I}_L) &= f(\mathcal{I}_L) - f(\emptyset)\\
			&= \sum_{m=1}^{L}f(\Pi_{\alpha}(\mathcal{I}_{m})) - f(\Pi_{\alpha}(\mathcal{I}_{m-1}))\\
			&= \sum_{m=1}^{L} R_{\pi_{\alpha}(m)}\\
			&= \sum_{l=1}^{L} r_l \label{equal_region_2_1_1}
		\end{align}
	\end{subequations}
	where \eqref{equal_region_2_1_1} follows from \eqref{eq_sum_rate_delta_m}. For $\mathcal{S} = \mathcal{I}_L$, the right hand side (RHS) of \eqref{equal_region} is given by
	\begin{subequations}\label{equal_region_2_2}
		\begin{align}
			 H\left(\{ \bm v_m | m \in \mathcal{I}_L\}\right)
			&= H\left(\{ \bm t_l | l \in \mathcal{I}_L\}\right) \label{equal_region_2_2_1}\\
			&= \sum_{l=1}^{L} r_l.
		\end{align}
	\end{subequations}
	where \eqref{equal_region_2_2_1} follows from the fact that the map from $\{\! \t_l|l\!\in\!\mathcal{I}_L\! \}$ to $\{\bm v_m|m\in\!\mathcal{I}_L\}$ is a bijection.
	This concludes the proof.
\end{IEEEproof}

	We can further show that \eqref{equal_region} holds in the following three situations. Note that the proofs of Theorems \ref{thm:equivalent_L_2} and \ref{thm:equivalent_rate_region} are respectively given in Appendices \ref{appen:proof_equivalent_L_2} and \ref{app:proof_equivalent_rate_region}. 
	\begin{thm}\label{thm:equivalent_L_2}
		Consider the interference channel in Fig. \ref{fig:OneHop} with $L=2$. The proposed GCCF scheme is optimal, i.e. $
		\mathcal{R}_{\mathrm{cpr}} = \mathcal{R}_{\mathrm{SW}}$, where $\mathcal{R}_{\mathrm{cpr}}$ is given by \eqref{eq:Compression rate Bound} and $\mathcal{R}_{\mathrm{SW}}$ is given by \eqref{eq:Origin Slepian-Wolf Bound}.
\end{thm}
\begin{thm}\label{thm:equivalent_rate_region}
		If no dither is used (i.e., $\bm d_l = \bm 0$ for $l\in \mathcal{I}_L$ in \eqref{eq:channel input}), then the proposed GCCF scheme is optimal, i.e. $\mathcal{R}_{\mathrm{cpr}} = \mathcal{R}_{\mathrm{SW}}.$
\end{thm}
	\begin{thm}\label{thm:equivalent_common_s}
		If all the transmitters share a common shaping lattice, i.e., $\Lambda_{\mathrm{s},1} = \cdots = \Lambda_{\mathrm{s},L}$, then the proposed GCCF scheme achieves the Slepian-Wolf region in \eqref{eq:Origin Slepian-Wolf Bound}. 
	\end{thm}
	\emph{Proof of Theorem \ref{thm:equivalent_common_s}:} 
		Let $\Lambda_{\mathrm{s},1} = \cdots = \Lambda_{\mathrm{s},L} = \Lambda_{\mathrm{s}}$. Then, the coarsest lattice  is $\Lambda_1 = \Lambda_{\mathrm{s}}$. Thus, $\bm v_m$ can be represented as 
		\begin{subequations}\label{eq_cor_proof}
			\begin{align}
			\bm v_m &= \left(\sum_{l=1}^{L}a_{ml}\left(\bm t_l-Q_{\Lambda_{\mathrm{s}}}(\bm t_l-\beta_l \bm d_l)\right) \right)\bmod \Lambda_{\mathrm{s}} \label{eq_cor_proof_1} \\
			&= \left(\sum_{l=1}^{L}a_{ml}\bm t_l \right)\bmod \Lambda_{\mathrm{s}}
			\end{align}
		\end{subequations}
		where \eqref{eq_cor_proof_1} follows from \eqref{eq:computed codeword}. 
		Clearly $\bm v_m$ in \eqref{eq_cor_proof} contains no dither term. Thus, from Theorem \ref{thm:equivalent_rate_region}, we obtain $\mathcal{R}_{\mathrm{cpr}} = \mathcal{R}_{\mathrm{SW}}$.
	\hspace*{\fill} $\blacksquare$
	\begin{rem}
		It can be shown that for $L \geq 3$, the existence of dithers in general enables a compression rate region beyond $\mathcal{R}_{\mathrm{cpr}}$ in \eqref{eq:Compression rate Bound} (though the minimum total compression rate remains the same, as stated in Theorem \ref{thm_minmize_total_rate}). An example of $\mathcal{R}_{\mathrm{cpr}} \neq \mathcal{R}_{\mathrm{SW}}$ will be presented in the next subsection. It is also worth noting that to achieve a compression rate region beyond $\mathcal{R}_{\mathrm{cpr}}$, complicated distributed source coding techniques are required. Nevertheless, this is out of the scope of the paper.
	\end{rem}

\subsection{Examples}

\subsubsection{Example 1}
\begin{figure}
\centering
\includegraphics[width=3.5in]{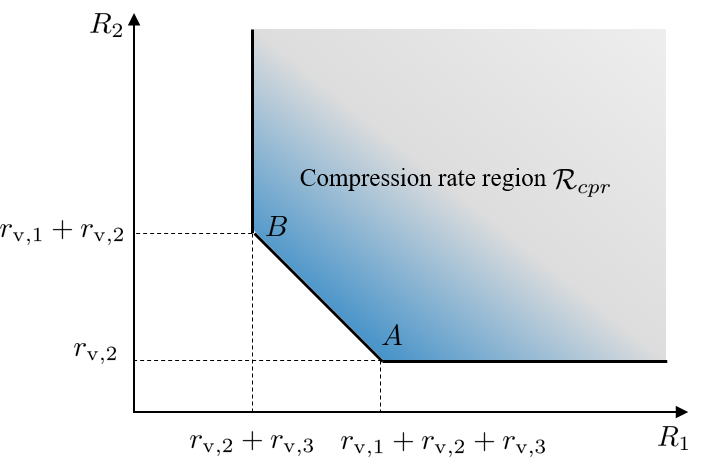}\\
\caption{The compression rate region given in \eqref{eq: SleWolf-region}, \eqref{eq: SleWolf-region1}, and \eqref{eq: SleWolf-region2}.}
\label{fig:Slepian-Wolf Region}
\end{figure}

We first present an example to illustrate the GCCF scheme and its compression rate region $\mathcal{R}_{\mathrm{cpr}}$.
Consider the channel in \eqref{eq:channel model} with two transmitters and two receivers. The nested lattice chain is set to
$\Lambda_{s,1}(=\Lambda_{1})\subseteq \Lambda_{s,2}(=\Lambda_{2})\subseteq\Lambda_{c,1}(=\Lambda_{3})\subseteq \Lambda_{c,2}(=\Lambda_{4})$. The source codewords are given by $\t_1 \in \Lambda_{3} \cap \mathcal{V}_1$ and $\t_2 \in \Lambda_{4} \cap \mathcal{V}_2$. The integer coefficient matrix is chosen as $\bm A=[1,1;1,0]$. The computed codewords are given by
\begin{align}
	\bm v_1 &= \left(\sum_{l=1}^{2}a_{1l}(\t_l - Q_{\Lambda_{\mathrm{s},l}}(\t_l - \beta_l\bm d_l))\right)\bmod \Lambda_1\\
	\bm v_2 &= \left(a_{11}(\t_1 - Q_{\Lambda_{s,1}}(\t_1 - \beta_1\bm d_1))\right)\bmod \Lambda_1.
\end{align}
From \eqref{eq:codeword splitting}, $\t_1$ and $\t_2$ can be respectively split as
$\t_1=(\t_{1,1}+\t_{1,2})\bmod\Lambda_{s,1}$ and $\t_2=(\t_{2,2} +\t_{2,3})\bmod\Lambda_{s,2}.$
The source rates are given by $r_1=r_{\mathrm{v},1}+ r_{\mathrm{v},2}$ and $r_2=r_{\mathrm{v},2}+r_{\mathrm{v},3}$, and the sum rate given by $r_{\mathrm{sum}}=r_{\mathrm{v},1} + 2r_{\mathrm{v},2} + r_{\mathrm{v},3}$.
By definitions in \eqref{eq:Index set I_k} and \eqref{eq:Index set J_l}, we have the following index sets:
\begin{equation}\label{value_set_I}
\mathcal{L}_1 = \{1\}, \mathcal{L}_2= \{1,2\}, \mathcal{L}_3 = \{2\}, \mathcal{K}_1 = \{1,2\}, \mathcal{K}_2= \{2,3\}.
\end{equation}
The prime number $\gamma$ is assumed to be large enough, so that the rank of $\bm A$ and the rank of any submatrix of $\bm A$ can be evaluated in the integer domain.
 Then, the rank functions involved in \eqref{eq:Compression rate Bound} are given by
\begin{equation}\label{rank_sub_matrix}
\begin{aligned}
&\textit{rank}(\bm A(\!\{\!1\!\},\!\mathcal{L}_1\!))\! =\! 1, \textit{rank}(\bm A(\!\{\!1\!\},\!\mathcal{L}_2\!))\! =\! 1, \textit{rank}(\bm A(\!\{\!1\!\},\!\mathcal{L}_3\!))\! =\! 1\\
&\textit{rank}(\bm A(\!\{\!2\!\},\!\mathcal{L}_1\!))\! =\! 1, \textit{rank}(\bm A(\!\{\!2\!\},\!\mathcal{L}_2\!))\! = \!1, \textit{rank}(\bm A(\!\{\!2\!\},\!\mathcal{L}_3\!))\! =\! 0\\
&\textit{rank}(\bm A(\mathcal{I}_2,\!\mathcal{L}_1\!))\! =\! 1, \textit{rank}(\bm A(\mathcal{I}_2,\!\mathcal{L}_2\!))\! =\!2, \textit{rank}(\bm A(\mathcal{I}_2,\!\mathcal{L}_3\!))\! =\! 1\\
&\textit{rank}(\bm A(\emptyset,\mathcal{L}_1)) =\! 0, \textit{rank}(\bm A(\emptyset,\mathcal{L}_2)) =\! 0, \textit{rank}(\bm A(\emptyset,\mathcal{L}_3))\! =\! 0.
\end{aligned}
\end{equation}
From Theorem \ref{theorem: compression rate region}, the compression rate region is given by the following three inequalities:
\begin{align}
\begin{split}
R_1 \geq {}&\ (\textit{rank}(\bm A(\mathcal{I}_2,\mathcal{L}_1))-\textit{rank}(\bm A(\{2\},\mathcal{L}_1))) r_{\mathrm{v},1} \\
{}&\ + (\textit{rank}(\bm A(\mathcal{I}_2,\mathcal{L}_2))-\textit{rank}(\bm A(\{2\},\mathcal{L}_2))) r_{\mathrm{v},2} \\
{}&\ + (\textit{rank}(\bm A(\mathcal{I}_2,\mathcal{L}_3))-\textit{rank}(\bm A(\{2\},\mathcal{L}_3))) r_{\mathrm{v},3}\\
={}&\ r_{\mathrm{v},2}+r_{\mathrm{v},3}
\end{split}\label{eq: SleWolf-region}\\
\begin{split}
R_2\geq {}&\ (\textit{rank}(\bm A(\mathcal{I}_2,\mathcal{L}_1))-\textit{rank}(\bm A(\{1\},\mathcal{L}_1))) r_{\mathrm{v},1}\\
{}&\ + (\textit{rank}(\bm A(\mathcal{I}_2,\mathcal{L}_2))-\textit{rank}(\bm A(\{1\},\mathcal{L}_2))) r_{\mathrm{v},2}\\
{}&\ + (\textit{rank}(\bm A(\mathcal{I}_2,\mathcal{L}_3))-\textit{rank}(\bm A(\{1\},\mathcal{L}_3))) r_{\mathrm{v},3}\\
={}&\ r_{\mathrm{v},2} 
\end{split} \label{eq: SleWolf-region1} \\
\begin{split}
R_1+R_2 \geq {}&\ (\textit{rank}(\bm A(\mathcal{I}_2,\mathcal{L}_1))-\textit{rank}(\bm A(\emptyset,\mathcal{L}_1))) r_{\mathrm{v},1}\\
{}&\ + (\textit{rank}(\bm A(\mathcal{I}_2,\mathcal{L}_2))-\textit{rank}(\bm A(\emptyset,\mathcal{L}_2))) r_{\mathrm{v},2}\\
{}&\ + (\textit{rank}(\bm A(\mathcal{I}_2,\mathcal{L}_3))-\textit{rank}(\bm A(\emptyset,\mathcal{L}_3))) r_{\mathrm{v},3}\\
={}&\ r_{sum}.
\end{split}\label{eq: SleWolf-region2}
\end{align}
Fig. \ref{fig:Slepian-Wolf Region} illustrates the compression rate region given in \eqref{eq: SleWolf-region}, \eqref{eq: SleWolf-region1}, and \eqref{eq: SleWolf-region2}. 
We now consider the compression operation in Theorem \ref{thm: achive vertex}. 
Let $\pi_{\alpha}(1) = 1$ and $\pi_{\alpha}(2) = 2$. From \eqref{eq:rank_equation}, \eqref{value_set_I}, and \eqref{rank_sub_matrix}, we obtain
\begin{equation}\label{K_index_example}
\mathcal{J}_{1} = \{2,3\}\ \text{and}\ \mathcal{J}_{2} = \{1,2\}.
\end{equation}  
From Theorem \ref{thm: achive vertex}, the compression operation at receiver $1$ and $2$ are respectively given by 
\begin{equation*}
\begin{split}
\bm \delta_1 =&\ \left(\bm v_{1,2} + \bm v_{1,3} \right) \bmod \Lambda_{2} = Q_{\Lambda_{4}}(\bm v_{1})\bmod\Lambda_{2}\\
\bm \delta_2 =&\ \left(\bm v_{2,1} + \bm v_{2,2} \right) \bmod \Lambda_{1} = Q_{\Lambda_{3}}(\bm v_{2})\bmod\Lambda_{1}.
\end{split}
\end{equation*}
The rate of $\bm \delta_1$ and $\bm \delta_2$ are respectively given by 
\begin{equation*}
\begin{split}
H(\bm \delta_1) &= r_{\mathrm{v},2} + r_{\mathrm{v},3}\\
H(\bm \delta_2) &= r_{\mathrm{v},1} + r_{\mathrm{v},2}.
\end{split}
\end{equation*}
The rate tuple $(H(\bm \delta_1), H(\bm \delta_2))$ corresponds to vertex $B$ in Fig. \ref{fig:Slepian-Wolf Region}, i.e., $(R_1,R_2) = (H(\bm \delta_1), H(\bm \delta_2))$.\footnote{Vertex A in Fig. \ref{fig:Slepian-Wolf Region} can be achieved by the permutation $\pi_{\alpha}(\cdot)$ with $\pi_{\alpha}(1) = 2$ and $\pi_{\alpha}(2) = 1$.}
Note that $\bm v_1 \in \Lambda_4\cap\mathcal{V}_1$ and $\bm v_2 \in \Lambda_3\cap\mathcal{V}_1$, the rate of $\bm v_1$ is $r_{\mathrm{v},1} + r_{\mathrm{v},2} + r_{\mathrm{v},3}$ and the rate of $\bm v_2$ is $ r_{\mathrm{v},1} + r_{\mathrm{v},2}$. Therefore, by compression, the total rate is reduced from $r_{\mathrm{v},1} + r_{sum}$ to $r_{sum}$, while $r_{sum}$ is the minimum sum rate for lossless compression.

\subsubsection{Example 2}

We now give an example of $\mathcal{R}_{\mathrm{cpr}} \neq \mathcal{R}_{\mathrm{SW}}$ for $L \geq 3$.
It suffices to show that the vertex of $\mathcal{R}_{\mathrm{cpr}}$ associated with $\pi_{\alpha}(\cdot)$ is not equal to the corresponding vertex of $\mathcal{R}_{\mathrm{SW}}$. Recall that the coordinates of the vertex of $\mathcal{R}_{\mathrm{cpr}}$ associated with $\pi_{\alpha}(\cdot)$ is given by \eqref{eq_rate_delta_m} in Theorem \ref{cor:rate_of_delta} and the coordinates of the corresponding vertex of $\mathcal{R}_{\mathrm{SW}}$ is given by \eqref{SW_vertex} in Appendix \ref{app:proof_equivalent_rate_region}. Then, we need to show
that there exists $m$ and $\pi_{\alpha}(\cdot)$ such that
\begin{equation}\label{eq_vertex_no_equal}
H(\bm \delta_{\pi_{\alpha}(m)}) \neq H\left(\bm v_{\pi_{\alpha}(m)} | \{\bm v_i, i\in \pi_{\alpha}(\overline{\mathcal{I}_{m}})\}\right).
\end{equation}
\begin{figure}
	\centering
	\includegraphics[width=2.5in]{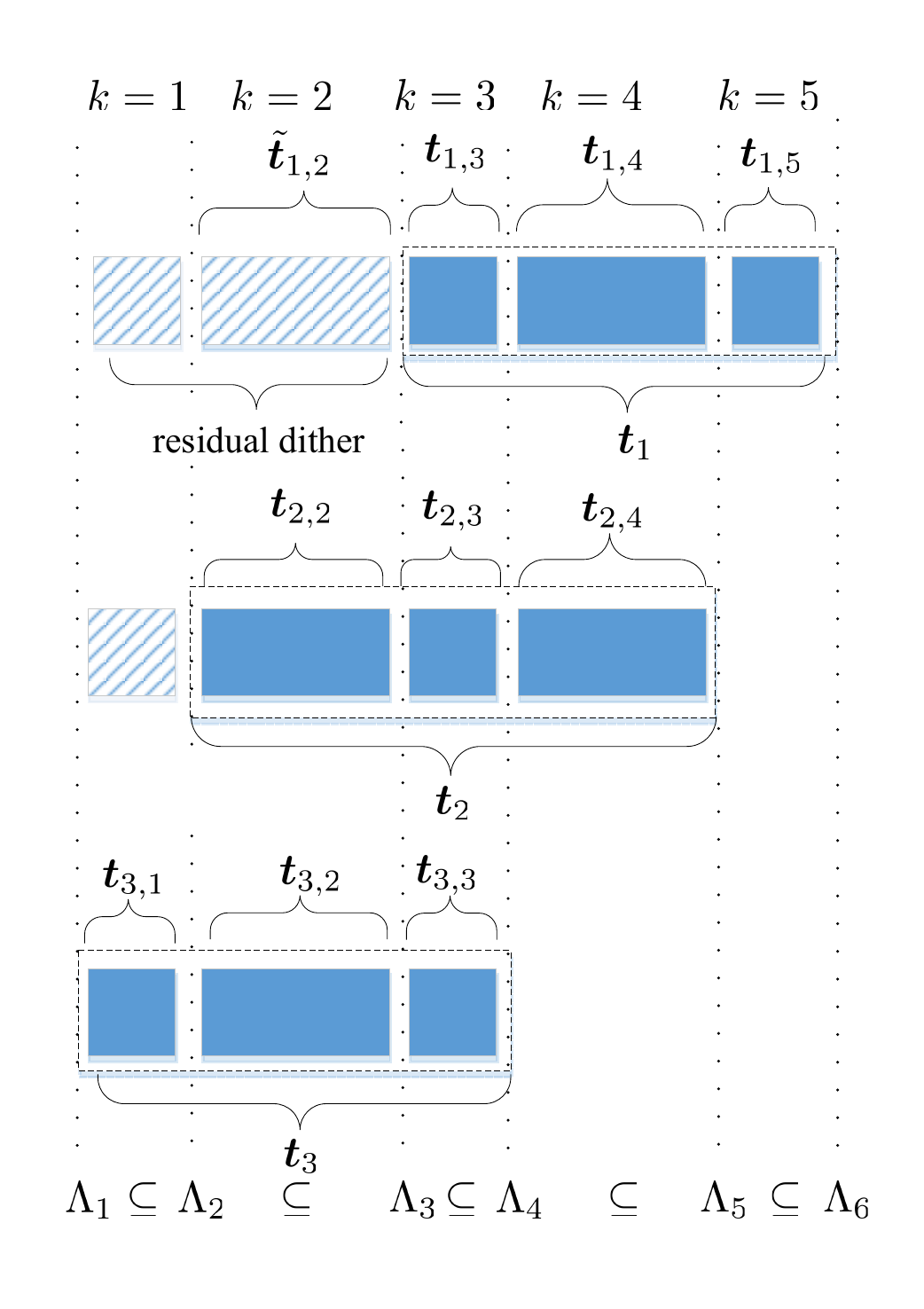}
	\caption{The codeword splitting structure of Example 2 with nested lattice chain $\Lambda_1 (\Lambda_{\mathrm{s},3}) \subseteq \Lambda_2 (\Lambda_{\mathrm{s},2}) \subseteq \Lambda_3 (\Lambda_{\mathrm{s},1}) \subseteq \Lambda_4 (\Lambda_{\mathrm{c},3}) \subseteq \Lambda_5 (\Lambda_{\mathrm{c},2}) \subseteq \Lambda_6 (\Lambda_{\mathrm{c},1})$.}
	\label{fig:app_recovery}
\end{figure}
 Consider the nested lattice chain $\Lambda_1 (\Lambda_{\mathrm{s},3}) \subseteq \Lambda_2 (\Lambda_{\mathrm{s},2}) \subseteq \Lambda_3 (\Lambda_{\mathrm{s},1}) \subseteq \Lambda_4 (\Lambda_{\mathrm{c},3}) \subseteq \Lambda_5 (\Lambda_{\mathrm{c},2}) \subseteq \Lambda_6 (\Lambda_{\mathrm{c},1})$, which is illustrated in Fig. \ref{fig:app_recovery}. Let $\pi_{\alpha}(m) = m$ for $m \in \mathcal{I}_L$. We assume
\begin{equation}\label{no_equal_1}
\begin{bmatrix}
\bm v_2\\
\bm v_3
\end{bmatrix}
=\left(
\begin{bmatrix}
1&3&3\\
2&2&3
\end{bmatrix}
\begin{bmatrix}
\tilde{\t}_1\\\tilde{\t}_2\\\tilde{\t}_3
\end{bmatrix}\right) \bmod \Lambda_1
\end{equation}
where $\bmod\ \Lambda_1$ is element-wise.
We now show
\begin{equation}\label{no_equal_2}
H(\bm \delta_2) \neq H(\bm v_2|\bm v_3).
\end{equation}
From \eqref{eq: compression operation}, we obtain
\begin{equation}\label{no_equal_3}
\bm \delta_2 = \left( \bm v_{2,2} + \bm v_{2,3} + \bm v_{2,4} \right)\bmod \Lambda_{1}.
\end{equation} 
From \eqref{eq_rate_delta_m}, the entropy rate of $\bm \delta_2$ is given by
\begin{equation}\label{no_equal_4}
H(\bm \delta_2) = \sum_{k=2}^{4}r_{\mathrm{v},k}.
\end{equation}
By the chain rule, we have
\begin{equation}\label{no_equal_5}
\begin{split}
H(\bm v_2|\bm v_3) =& H\left( \bm v_{2,1} |  \bm v_3,\{\bm v_{2,k'} \}_{k'=2}^{5}\right) \\
&\ + H\left( \bm v_{2,2} | \bm v_3,\{\bm v_{2,k'} \}_{k'=3}^{5}\right) \\
&\ + H\left( \bm v_{2,3} | \bm v_3,\{\bm v_{2,k'} \}_{k'=4}^{5}\right) \\
&\ + H\left( \bm v_{2,4} | \bm v_3,\{\bm v_{2,5} \}\right) + H\left( \bm v_{2,5} | \bm v_3\right).
\end{split}
\end{equation}
We calculate $H\left( \bm v_{2,k} |  \bm v_3,\{\bm v_{2,k'} \}_{k'=k+1}^{5}\right)$ in a descending order of $k$.

For $k=5$, 
\begin{subequations}\label{no_equal_6}
	\begin{align}
	H\left( \bm v_{2,5} | \bm v_3\right)
	&=  H\left( \bm v_{2,5} | \{\bm v_{3,k\prime}\}_{k\prime=1}^5  \right)\label{no_equal_6_2} \\
	&= H\left( \bm v_{2,5} | \bm v_{3,1},\bm v_{3,2},\bm v_{3,5} \right) \label{no_equal_6_3}
	\end{align}
\end{subequations}
where \eqref{no_equal_6_3} is from the fact that $\bm v_{2,,5}$ can be expressed by \eqref{eq:component of v_m2} and thus $\bm v_{2,5}$ is independent of $\bm v_{3,k'}$ for $k' \geq 3,k'\neq 5 $; see the discussions below \eqref{eq_v_m_k}.
From \eqref{eq:component of v_m} and \eqref{eq_t_tild_t}, we obtain 
\begin{align}
	\bm v_{2,5} &= (\t_{1,5})\bmod \Lambda_5,\label{no_equal_6_4}\\
	\bm v_{3,5} &= (\t_{1,5})\bmod \Lambda_5. \label{no_equal_6_7}
\end{align}
From \eqref{no_equal_6_4} and \eqref{no_equal_6_7}, we obtain
	\begin{align}\label{no_equal_6_8}
	H\left( \bm v_{2,5} | \bm v_{3,1},\bm v_{3,2},\bm v_{3,5} \right) = 0.
	\end{align}
 
For $k=4$,  
\begin{subequations}\label{no_equal_7}
	\begin{align}
H\left( \bm v_{2,4} | \bm v_3,\bm v_{2,5}\right)
&= H\left( \bm v_{2,4}|\{\bm v_{3,k'}\}_{k'=1}^5,\bm v_{2,5}\right)\\
&= H\left( \bm v_{2,4}| \bm v_{3,1},\bm v_{3,2},\bm v_{3,4}\right) \label{no_equal_7_1}
	\end{align}
\end{subequations}
where \eqref{no_equal_7_1} follows from the fact that $\bm v_{2,4}$ is independent of $\bm v_{2,5}$ and $\bm v_{3,3},\bm v_{3,5}$. 
From \eqref{eq:component of v_m} and \eqref{eq_t_tild_t}, we obtain
\begin{align}
\bm v_{2,4} &= \left( \t_{1,4}  + 3\t_{2,4} \right)\bmod \Lambda_4,\label{no_equal_6_9}\\
\bm v_{3,1} &= \left( 2\tilde{\t}_{1,1} + 2\tilde{\t}_{2,1} + 3\t_{3,1} \right)\bmod \Lambda_2, \label{no_equal_6_5}\\
\bm v_{3,2} &= \left( 2\tilde{\t}_{1,2} + 2\t_{2,2} + 3\t_{3,2} \right)\bmod \Lambda_2, \label{no_equal_6_6}\\
\bm v_{3,4} &= \left( 2\t_{1,4} + 2\t_{2,4} \right)\bmod \Lambda_4.\label{no_equal_6_10}
\end{align}
Since the coefficient vector $[1,3]$ is independent of $[2,2]$, we see from Lemma \ref{lem_independency} that $\bm v_{2,4}$ is independent of $\bm v_{3,4}$. 
From the Crypto lemma\cite{erez2004achieving}, we see that both $\bm v_{3,1}$ and $\bm v_{3,2}$ are independent of $\tilde{\t}_1$ and $\tilde{\t}_2$, and so are independent of $\t_1$ and $\t_2$. Thus, $\bm v_{3,1}$ and $\bm v_{3,2}$ are independent of $\bm v_{2,4}$. Thus,
\begin{subequations}\label{no_equal_7_3}
	\begin{align}
		H\left( \bm v_{2,4}| \bm v_{3,1},\bm v_{3,2},\bm v_{3,4}\right)
		&= H\left( \bm v_{2,4}\right)\\
		&= r_{\mathrm{v},4}.
	\end{align}
\end{subequations}

Similarly, for $k=3$, 
\begin{subequations}\label{no_equal_8}
	\begin{align}
	H\left( \bm v_{2,3} | \bm v_3,\bm v_{2,4},\bm v_{2,5}\right) 
	&=H\left( \bm v_{2,3}|\bm v_{3,3}\right)\\
	&=H\left( \bm v_{2,3}\right)\\
	&= r_{\mathrm{v},3},
	\end{align}
\end{subequations}
where
\begin{align}
\bm v_{2,3} &= \left(  \t_{1,3} + 3\t_{2,3} + 3\t_{3,3} \right)\bmod \Lambda_2\label{no_equal_8_1}\\
\bm v_{3,3} &= \left( 2\t_{1,3} + 2\t_{2,3} + 3\t_{3,3} \right)\bmod \Lambda_2.\label{no_equal_8_2}
\end{align}

For $k=2$, 
\begin{equation*}
	\begin{split}
		H\!\left(\! \bm v_{2,2} | \bm v_3,\{\!\bm v_{2,k'}\! \}_{k'=3}^{5}\right)
		&\! \!=\! H\!\left(\! \bm v_{2,2} |\{\! \bm v_{3,k'}\!\}_{k'=1}^{5},\!\{\bm v_{2,k'}\! \}_{k'=3}^{5}\right)\\
		& \!\!=\! H\!\left(\! \bm v_{2,2} |\bm v_{3,2},\bm v_{3,1},\!\{\!\bm v_{2,k'},\bm v_{3,k'}\! \}_{k'=3}^{5}\right)
	\end{split}
\end{equation*}
where
\begin{align}\label{no_equal_9_3}
\bm v_{2,2} &= \left(  \tilde{\t}_{1,2}  + 3\t_{2,2} + 3\t_{3,2} \right)\bmod \Lambda_2
\end{align}
Following the discussion below \eqref{no_equal_7},  $\bm v_{3,1}$ is independent of $\t_1$ and $\t_2$, and so is independent of $\bm v_{2,2}$. Also, $\bm v_{2,2}$ is independent of $\{\bm v_{2,k'},\bm v_{3,k'} \}_{k'=3}^{5}$ for given $\tilde{\t}_{1,2}$. Thus,
\begin{subequations}
	\begin{align}
		&  H\left( \bm v_{2,2} |\bm v_{3,2},\bm v_{3,1},\{\bm v_{2,k'},\bm v_{3,k'} \}_{k'=3}^{5}\right)\\
		& = H\left( \bm v_{2,2} |\bm v_{3,2}, \{\bm v_{2,k'},\bm v_{3,k'} \}_{k'=3}^{5}\right)\\
		& \geq H\left( \bm v_{2,2} |\bm v_{3,2}, \tilde{\t}_{1,2}, \{\bm v_{2,k'},\bm v_{3,k'} \}_{k'=3}^{5}\right)\\
		& = H\left( \bm v_{2,2} |\bm v_{3,2}, \tilde{\t}_{1,2}\right)\\
		&=r_{\mathrm{v},2} \label{no_equal_9_2}
	\end{align}
\end{subequations}
where \eqref{no_equal_9_2} follows from that $\bm v_{2,2}$ is independent of $\bm v_{3,2}$ for given $\tilde{\t}_{1,2}$ (by noting Lemma \ref{lem_independency} in Appendix \ref{app:proof_equivalent_rate_region}), and both $\bm v_{2,2}$ are $\bm v_{3,2}$ are uniformly distributed over $\mathcal{C}_{\mathrm{v},2}$. At the same time, we have
\begin{equation}
	\begin{split}
	H\left( \bm v_{2,2} | \bm v_3,\{\bm v_{2,k'} \}_{k'=3}^{5}\right) \leq 	H\left( \bm v_{2,2}\right) = r_{\mathrm{v},2}.
	\end{split}
\end{equation}
Thus, 
\begin{equation}\label{no_equal_9_5}
	H\left( \bm v_{2,2} | \bm v_3,\{\bm v_{2,k'} \}_{k'=3}^{5}\right) = r_{\mathrm{v},2}.
\end{equation}

For $k=1$, we have
\begin{subequations}\label{no_equal_10}
	\begin{align}
	&H\left( \bm v_{2,1} |\bm v_3,\{\bm v_{2,k'} \}_{k'=2}^{5}\right)\\
	& = H\left( \bm v_{2,1} |\{ \bm v_{3,k'}\}_{k'=1}^{5},\{\bm v_{2,k'} \}_{k'=2}^{5}\right)\\
	& = H\left( \bm v_{2,1} |\bm v_{3,1},\{\bm v_{2,k'},\bm v_{3,k'} \}_{k'=2}^{5}\right)\label{no_equal_10_1}
	\end{align}
\end{subequations}
where
\begin{align}\label{no_equal_11}
\bm v_{2,1} &= \left(  \tilde{\t}_{1,1}  + 3\tilde{\t}_{2,1} + 3\t_{3,1} \right)\bmod \Lambda_1.
\end{align}

We next show that $\bm v_{2,1}$ is deterministic for given $\bm v_{3,1}, \{\bm v_{2,k'},\bm v_{3,k'} \}_{k'=2}^{5}, \text{ and } \t_{1,3}$. From \eqref{no_equal_6_4}, we see that the value of $\t_{1,5}$ can be determined for given $\bm v_{2,5}$. Similarly, from \eqref{no_equal_6_9} and \eqref{no_equal_6_10},  $\t_{1,4}$ and $\t_{2,4}$ can be determined for given $\bm v_{2,4}$ and $\bm v_{3,4}$. Also, from \eqref{no_equal_8_1} and \eqref{no_equal_8_2}, $\t_{2,3}$ and $\t_{3,3}$ can be  determined for given $\bm v_{2,3}$, $\bm v_{3,3}$, and $\t_{1,3}$. This further implies that $\t_1 = \left(\t_{1,3} + \t_{1,4} +  \t_{1,5}\right) \bmod \Lambda_3$ is deterministic, and so are $\tilde{\t}_{1,1}$ and $\tilde{\t}_{1,2}$. 
Consequently, from \eqref{no_equal_6_6} and \eqref{no_equal_9_3},  $\t_{2,2}$ and $\t_{3,2}$ are deterministic by noting that $\bm v_{2,2}$ and $\bm v_{3,2}$ are given. Then, $\t_2 = \left(\t_{2,2} + \t_{2,3} +  \t_{2,4}\right) \bmod \Lambda_2$ is deterministic, and so is $\tilde{\t}_{2,1}$. From \eqref{no_equal_6_5}, $\t_{3,1}$ is deterministic, and so $\bm v_{2,1}$ in \eqref{no_equal_11} is also deterministic. Thus, for given $\bm v_{3,1}, \{\bm v_{2,k'},\bm v_{3,k'} \}_{k'=2}^{5}$, the randomness of $\bm v_{2,1}$ is completely determined by $\t_{1,3}$. In general, the value of $\bm v_{2,1}$ varies with the choice of $\t_{1,3}$.
Therefore, we obtain
\begin{equation}\label{no_equal_10_2}
H\left( \bm v_{2,1} |\bm v_{3,1},\{\bm v_{2,k'},\bm v_{3,k'} \}_{k'=2}^{5}\right) > 0.
\end{equation}

Combining \eqref{no_equal_5}, \eqref{no_equal_6_8}, \eqref{no_equal_7_3}, \eqref{no_equal_8}, \eqref{no_equal_9_5}, and \eqref{no_equal_10_2}, we obtain
\begin{equation}\label{eq_equal_11}
H(\bm v_2|\bm v_3) > \sum_{k=2}^{4}r_{\mathrm{v},k}.
\end{equation}
Thus, together with \eqref{no_equal_4}, we obtain \eqref{no_equal_2}. This example implies that, with random dithering, the optimal distributed source coding can achieve a rate region beyond $\mathcal{R}_{\mathrm{cpr}}$, though the minimum sum rate remains the same.

\section{Performance Optimization for Multi-hop Relay Networks \label{sec:multihop network}}
In this section, we consider a multi-hop relay network employing our GCCF relaying scheme. We first describe the GCCF scheme for a $N$-hop relay network and then present the achievable rates of the network. Then, we formulate the sum-rate maximization problem for a two hop relay network and present an algorithm to solve the problem. Finally, numerical results for the two hop relay network are provided for comparison. 

\subsection{Achievable Rates \label{sec:multihop network_A}}
\begin{figure}
	\centering
	\includegraphics[width=3.5in]{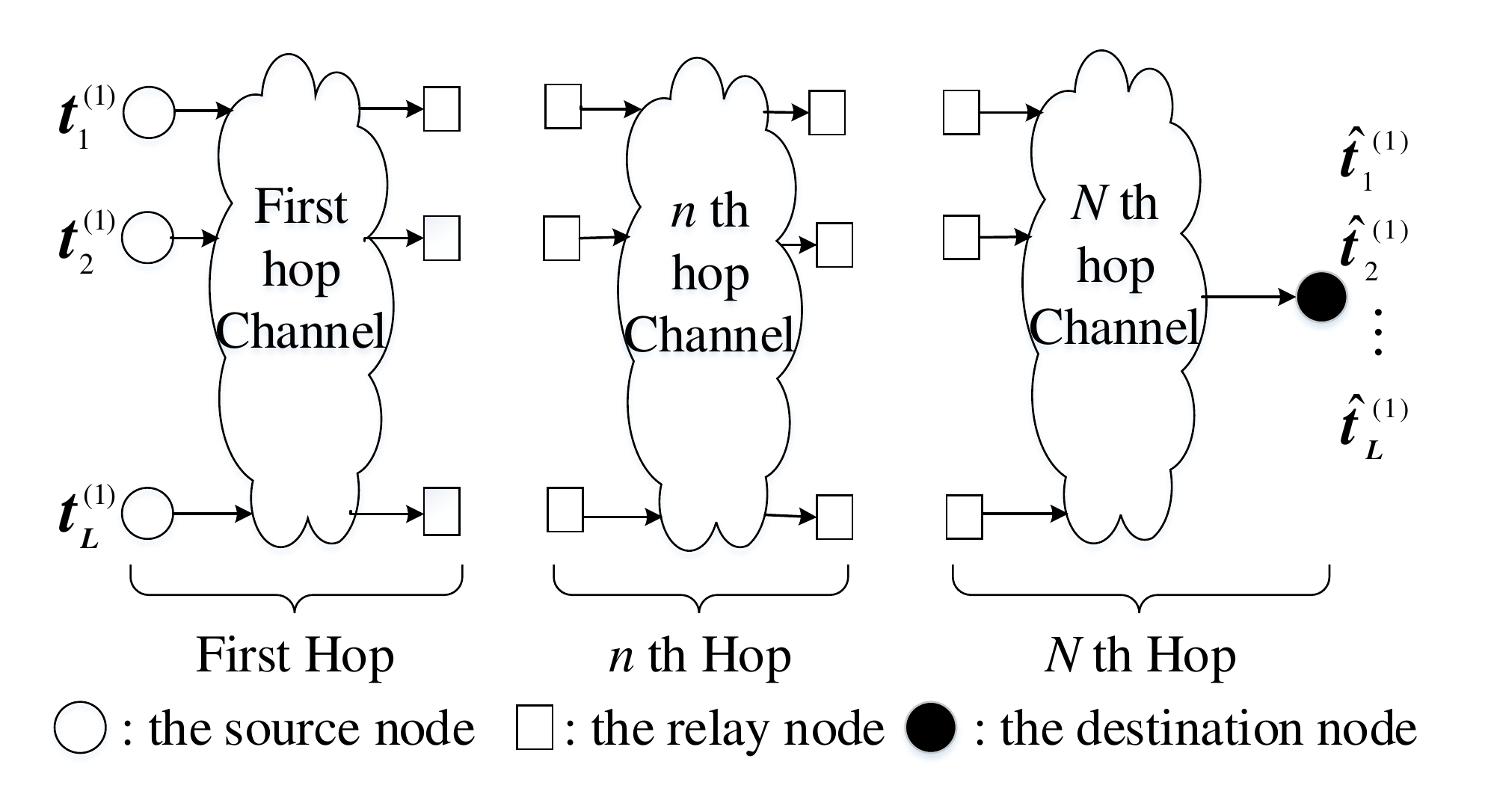}
	\caption{A multi-hop relay network with a single destination node. }
	\label{fig:MultiHop}
\end{figure}

An $N$-hop relay network is illustrated in Fig \ref{fig:MultiHop}. Each of the first $(N-1)$ hop is modelled by the interference channel in Fig. \ref{fig:OneHop}. The last hop has a unique destination node required to recovery all the source messages.  

We use superscript $ ^{(n)}$ to represent variables associated with the $n$-th hop. Specifically, denote by $\t_l^{(n)}$ the codeword of transmitter $l$ in the $n$-th hop, and by $r_l^{(n)}$ the corresponding transmission rate. Denote by  $\Lambda_1^{(n)} \subseteq \Lambda_2^{(n)} \subseteq \cdots \subseteq \Lambda_{2L}^{(n)}$ the nested lattice chain used to encode $\t_l^{(n)}$ in the $n$-th hop and by $r_{v,k}^{(n)}$ the rate of codebook $\mathcal{C}_{\mathrm{v},k}^{(n)}$ with lattice pair $(\Lambda_{k}^{(n)},\Lambda_{k+1}^{(n)})$. Denote by $\pi^{(n)}(\cdot)$ the lattice chain permutation in the $n$-th hop. 
Denote by $\bm A^{(n)}$ the corresponding coefficient matrix. 
Note that the receivers in the $n$-th hop are the transmitters in the $(n+1)$-th hop. Each receiver in the $n$-th hop needs to re-encode the compressed codeword, denoted by $\bm \delta_m^{(n)}$, into
\begin{equation}\label{eq:re-encode}
\t_m^{(n+1)} = \Psi_m^{(n)}(\bm \delta_m^{(n)}) \in \Lambda_{c,m}^{(n+1)} \cap \mathcal{V}_{s,m}^{(n+1)} 
\end{equation}
where $\Lambda_{c,m}^{(n+1)}$ is the coding lattice of $m$-th transmitter in $(n+1)$-th hop and $\mathcal{V}_{s,m}^{(n+1)}$ is the Voronoi region of the shaping lattice $\Lambda_{s,m}^{(n+1)}$. Denote by $r_m^{(n+1)}$ the rate of $\t_m^{(n+1)}$. 
The computation rate region and the compression rate region in the $n$-th hop are respectively denoted by $\mathcal{R}_{\mathrm{cpu}}^{(n)}$ and $\mathcal{R}_{\mathrm{cpr}}^{(n)}$, where $1\leq n \leq (N-1)$. Denote by $\mathcal{R}^{(N)}$ the capacity region of the $N$-th hop.
Then, an achievable transmission rate tuples of the multi-hop relay network are given in the theorem below.

\begin{thm}\label{theorem: Overall System}
  Consider the $N$-hop relay network in Fig. \ref{fig:MultiHop}. For any given lattice chain permutations $\{ \pi^{n}(\cdot)|n=1,\cdots,N-1 \}$, a transmission rate tuple $(r_1^{(1)},r_2^{(1)},\cdots,r_L^{(1)})$ is achievable when the following conditions are satisfied:
  \begin{enumerate}[leftmargin=*, labelsep=*, align=left, itemsep=0.1cm, font=\normalfont, label=(\roman*)]
    \item  $(r_1^{(n)},r_2^{(n)},\cdots,r_L^{(n)})\! \in\! \mathcal{R}_{\mathrm{cpu}}^{(n)}, n \in \{1,2,\cdots,N-1\}$,\label{enum-i}
    \item $(r_1^{(n+1)},r_2^{(n+1)},\cdots,r_L^{(n+1)})\!\in\!\mathcal{R}_{\mathrm{cpr}}^{(n)}, n\! \in\! \{1,\cdots,N\!-\!1\}$,\label{enum-ii}
    \item $ \textit{rank} (\bm A^{(n)}) = L, n \in \{1,2,\cdots,N-1\}$ \label{enum-iii},
    \item  $(r_1^{(n)},r_2^{(n)},\cdots,r_L^{(n)})\in \mathcal{R}^{(n)}, n =N$ \label{enum-iv}.
  \end{enumerate}
\end{thm}
\begin{IEEEproof}
	Condition \ref{enum-i} ensures the error-free computation at the $n$-th hop; condition \ref{enum-iii} ensures that the coefficient matrix is invertible and so the source messages can be recovered from the computed messages; condition \ref{enum-ii} ensures that the computed messages can be recovered from the compressed messages. With conditions (i)-(iii), $\{\t_l^{(n)}\}_1^L$ can be recovered from $\{\t_l^{(n+1)}\}_1^L$ (or $\{\bm \delta_m^{(n)}\}_1^L$) successfully. Besides, condition \ref{enum-iv} ensures the successful recovery of $\{\bm t_l^{(N)} \}_1^L$ at the destination node. This concludes the proof.
\end{IEEEproof}

\begin{rem}
	In Theorem \ref{theorem: Overall System},  $(r_1^{(n)},r_2^{(n)},\cdots,r_L^{(n)})$ is related to $(r_1^{(n+1)},r_2^{(n+1)},\cdots,r_L^{(n+1)})$ as follows. From Theorem \ref{theorem: compression rate region}, 
	for any given $\pi^{n}(\cdot)$, $\mathcal{R}_{\mathrm{cpr}}^{(n)}$ can be represented as a function of $\{r_{\mathrm{v},k}^{(n)}\}$. From \eqref{eq:Source rate splitting}, $r_{l}^{(n)}$ is also a function of $\{r_{\mathrm{v},k}^{(n)}\}$. Thus, condition \ref{enum-ii} gives a constraint that relates  $(r_{1}^{(n)},r_{2}^{(n)},\cdots,r_{L}^{(n)})$ to $(r_{1}^{(n+1)},r_{2}^{(n+1)},\cdots,r_{L}^{(n+1)})$.
\end{rem}

\begin{rem}
	The network configuration of $N = 2$ is illustrated in Fig. \ref{fig:Two Hop}. This configuration is of particular importance due to its connection to the cloud radio access network (C-RAN) \cite{mobile2011c}.
	In C-RAN, baseband signal processing is carried out in a central processor (CP), rather than in base stations as in a conventional cellular network. More specifically, in an uplink C-RAN, the function of a base station is reduced to receive radio signals from mobile users and then forward the signal to the CP after simple processing, while the CP collects signals from all the base stations and jointly decodes the messages of mobile users. The authors in \cite{Zhou2014CRAN}, \cite{Park2013JDD} proposed to quantize the received signal at base stations and forward the quantized signal to the CP. It was shown that with optimized quantization, C-RAN achieves a much higher sum rate than a conventional cellular network does. Interestingly, C-RAN can be modelled by the two-hop relay network described in Fig. \ref{fig:Two Hop}, where the receivers serve as the base stations in C-RAN, and the single destination node serves as the CP. With this analogy, the achievable rate region developed in Theorem \ref{theorem: Overall System} can be used to characterize the performance limits for the uplink C-RAN.
\end{rem}

\subsection{Sum-Rate Maximization}
\begin{figure}
	\centering
	\includegraphics[width = 3.5in]{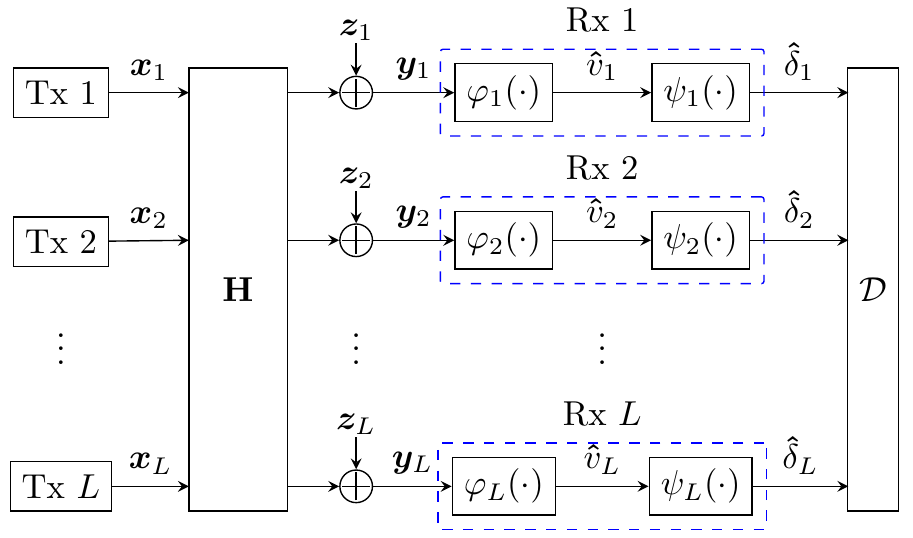}
	\caption{A two-hop relay network with $L$ transmitter nodes, $L$ receiver nodes, and one destination node.}
	\label{fig:Two Hop}
\end{figure}

In this subsection, we consider the sum-rate maximization of the network in Fig. \ref{fig:MultiHop}. We focus on the case of $N=2$ illustrated in Fig. \ref{fig:Two Hop}.
Based on Theorem \ref{theorem: Overall System} and some other encoding constraints, we can formulate the sum-rate maximization problem as
\begin{subequations}\label{eq:maximization}
	\begin{align}
	\mathop{\text{maximize}}_{
		\begin{aligned}
		\mathcal{O}
		\end{aligned}
	}\quad 
	& \sum_{l=1}^{L}r_l^{(1)} \label{eq:maxa} \\ 
	\mathop{\text{subject to}}\quad 
	& p_l \leq P_l, \beta_l > 0, \label{eq:maxb}\\
	& \pi(2l-1) < \pi(2l), \label{eq:max_pi}\\
	& r_l^{(1)}=\sum_{k\in \mathcal{K}_l}{r_{\mathrm{v},k}^{(1)}},\ \text{for}\ l\in \mathcal{I}_L,\label{eq:maxd}\\
	& \!\sum_{k \in \mathcal{P}_l}\!\! r_{\mathrm{v},k}^{(1)}\! =\! \frac{1}{2}\!\log\frac{{(\beta_{\pi_s(l)}^{(1)})}^2 p_{\pi_s(l)}^{(1)}}{{(\beta_{\pi_s(l+1)}^{(1)})}^2 p_{\pi_s(l+1)}^{(1)}},\! \text{ for }\! l\!\in\! \mathcal{I}_{L-1}, \label{eq:maxi}\\
	& r_{\mathrm{v},k}^{(1)} \geq 0, \ \text{for}\ k\in \mathcal{I}_{2L-1}, \label{eq:maxj}\\
	& (r_1^{(1)},r_2^{(1)},\cdots,r_L^{(1)}) \in \mathcal{R}_{\mathrm{cpu}}^{(1)}, \label{eq:maxe}\\
	& (r_1^{(2)},r_2^{(2)},\cdots,r_L^{(2)})\in \mathcal{R}_{\mathrm{cpr}}^{(1)}, \label{eq:maxf}\\ 
	& (r_1^{(2)},r_2^{(2)},\cdots,r_L^{(2)})\in \mathcal{R}^{(2)}, \label{eq:maxg}\\ 
	& \textit{rank}(\bm A^{(1)})=L, \label{eq:maxh}
	\end{align}
\end{subequations}
where $\mathcal{O}\! =\! \left\{\!\{\beta_l^{(1)}\}_1^L\!,\!\{p_l^{(1)}\}_1^L \!,\!\{r_{\mathrm{v},k}^{(1)}\}_1^{2L-1},\!\{r_m^{(2)}\}_1^L\!,\!\pi(\cdot)^{(1)}\!,\!\bm{A}^{(1)}\!\right\}$, and $\mathcal{P}_l = \{k|\Lambda_{s,\pi_s(l)}^{(1)} \subseteq \Lambda_{k}^{(1)} \subseteq \Lambda_{k+1}^{(1)} \subseteq \Lambda_{s,\pi_s(l+1)}^{(1)}\}$. 

In the above formulation, \eqref{eq:max_pi} is from \eqref{eq:permutation nested}; \eqref{eq:maxd} is from \eqref{eq:Source rate splitting}; \eqref{eq:maxe}-\eqref{eq:maxh} are from Theorem \ref{theorem: Overall System}. Eqn. \eqref{eq:maxi} establishes the relations between the rates of the codeword components and the source powers $\{p_l^{(1)}\}$. More specifically, consider a virtual lattice codebook with lattice pair $(\Lambda_{s,\pi_s(l)}^{(1)}, \Lambda_{s,\pi_s(l+1)}^{(1)})$.  From \eqref{volumn_shaping_lattice} and \eqref{eq:rate of virtual lattice code}, the rate of the virtual lattice codebook $\Lambda_{s,\pi_s(l+1)}^{(1)} \cap \mathcal{V}_{s,\pi_s(l)}^{(1)}$ is given by $\frac{1}{2}\log\frac{{(\beta_{\pi_s(l)}^{(1)})}^2 p_{\pi_s(l)}^{(1)}}{{(\beta_{\pi_s(l+1)}^{(1)})}^2 p_{\pi_s(l+1)}^{(1)}}$.
Note that the index set $\mathcal{P}_l$ in \eqref{eq:maxi} consists of all indices $k$ of the codeword components satisfying $\mathcal{C}_{v,k}^{(1)} \subseteq \Lambda_{s,\pi_s(l+1)}^{(1)} \cap \mathcal{V}_{s,\pi_s(l)}^{(1)}$. Therefore, the rate of the lattice code $(\Lambda_{s,\pi_s(l)}^{(1)}, \Lambda_{s,\pi_s(l+1)}^{(1)})$ can be represented by \eqref{eq:maxi}.

\subsection{Approximate Solution}  
The maximization problem in \eqref{eq:maximization} is an NP-hard mixed integer programming problem. We now present an approximate algorithm to solve \eqref{eq:maximization}. 
For any given $\{\beta_l^{(1)}\},\{p_l^{(1)}\}$, and $\pi(\cdot)^{(1)}$ satisfying \eqref{eq:max_pi} and \eqref{eq:maxb}, we can solve \eqref{eq:maximization} by the following steps:
\begin{itemize}
	\item Following \cite{tan2015compute}, find a suboptimal coefficient matrix $\bm A$ by using the LLL algorithm in \cite{osmane2011compute};
	\item Determine $\mathcal{R}_{\mathrm{cpu}}^{(1)}$ by \eqref{eq:computation rate constraint} and $\mathcal{R}_{\mathrm{cpr}}^{(1)}$ by \eqref{eq:Compression rate Bound};
	\item Optimize $\{r_{\mathrm{v},k}^{(1)}\}$ and $\{r_{\mathrm{m}}^{(2)}\}$ using linear programming.
\end{itemize}
The above procedures are summarized in Algorithm \ref{algorithm}. What remains is to optimize $\!\pi(\cdot)^{(1)}, \{\beta_l^{(1)}\}$, and $\!\{p_l^{(1)}\!\}$. Here we employ the brute-force method to optimize $\!\pi(\cdot)^{(1)}$ and the differential evolution algorithm \cite{storn1997differential} to optimize $\!\{\beta_l^{(1)}\!\}$ and $\!\{p_l^{(1)}\!\}$.

The exhaustive search over $\pi(\cdot)^{(1)}$ needs to consider $(2L)!$ different permutations, and is time-consuming even for a moderate $L$. In what follows, we describe a method to reduce the complexity when the separability condition in \eqref{eq:separability} is satisfied. With \eqref{eq:separability}, the nested lattice chain can be represented by 
$	\Lambda_{s,\pi_s(1)}^{(1)} \subseteq \Lambda_{s,\pi_s(2)}^{(1)} \subseteq \cdots \subseteq \Lambda_{s,\pi_s(L)}^{(1)} \subseteq \Lambda_{c,\pi_c(1)}^{(1)} \subseteq \Lambda_{c,\pi_c(2)}^{(1)}\subseteq \cdots \subseteq \Lambda_{c,\pi_c(L)}^{(1)}.$
From \eqref{volumn_shaping_lattice}, the permutation $\pi_s(\cdot)$ satisfies the inequality: 
\begin{equation}\label{eq:per_s}
(\beta_{\pi_{s}(1)}^{(1)})^2 p_{\pi_{s}(1)}^{(1)} \geq \cdots\geq (\beta_{\pi_{s}(L)}^{(1)})^2 p_{\pi_{s}(L)}^{(1)}. 
\end{equation}
For given $\{\beta_l^{(1)}\}_{l=1}^{L}$ and $ \{p_l^{(1)}\}_{l=1}^{L}$, $\pi_s(\cdot)^{(1)}$ is uniquely determined by \eqref{eq:per_s}. Thus, the search space of $\pi(\cdot)^{(1)}$ reduces to the set of all possible $\pi_c(\cdot)^{(1)}$ with complexity $L!$. In general, imposing the separability condition may incur a certain performance loss by reducing the search space. However, we will see from numerical results that such a performance loss is usually marginal.



\begin{algorithm}
	\begin{algorithmic}[1]
		\Require  $\!\bm{H}^{(1)}\!, \bm{H}^{(2)}\!, \pi(\cdot)^{(1)}, \{\beta_l^{(1)}\}_{l=1}^{L}, \{p_l^{(1)}\}_{l=1}^{L}, \{p_{l}^{(2)}\}_{m=1}^{M}$.
		\Ensure $\sum_{l=1}^{L}r_l^{(1)}$.
		
		\State Reorder $\{\beta_l^{(1)}\}_{l=1}^{L}, \{p_l^{(1)}\}_{l=1}^{L}$ to satisfy \eqref{eq:per_s}.
		\State Apply LLL algorithm \cite{osmane2011compute} to find a full rank $\bm A$. \label{alg:main_fun2}
		\State With $\{p_l^{(1)}\}_{l=1}^{L},\{\beta_l^{(1)}\}_{l=1}^{L}, \bm{H}^{(1)}$, and $\bm{A}^{(1)}$, compute $\mathcal{R}_{\mathrm{cpu}}^{(1)}$ in \eqref{eq:computation rate constraint} and \eqref{computation rate region}. \label{alg:main_fun3}
		\State With $\bm{A}^{(1)}$ and $\pi(\cdot)^{(1)}$, compute $\mathcal{R}_{\mathrm{cpr}}^{(1)}$ in \eqref{eq:Compression rate Bound}. \label{alg:main_fun4}
		\State With $\bm{H}^{(2)}$ and $\{p_{l}^{(2)}\}_{m=1}^{M}$, compute $\mathcal{R}^{(2)}$. \label{alg:main_fun5}
		\State Solve \eqref{eq:maximization} by linear programming. \label{alg:main_fun6}
	\end{algorithmic}
	\caption{Approximate Algorithm}
	\label{algorithm}
\end{algorithm}

\subsection{Numerical Results \label{subsec: numerical result}}

In simulation, we assume that the second hop channel of Fig. \ref{fig:Two Hop} is a parallel channel. That is, the destination observes $\bm{y}_m' = h_m \bm{x}_m' + \bm z_m'$ from relay $m$, where $h_m$ is the channel gain, $\bm{x}_m'\in\mathbb{R}^{n\times 1}$ is the signal forwarded by relay $m$ with power $P_{R,m} = \frac{1}{n}\lVert\bm{x}_m'\rVert^2$, and $z_m'$ is independently drawn from $\mathcal{N}(0,1)$. Therefore, the capacity region of the second hop is given by 
\begin{equation*}
	\mathcal{R}^{(2)}\!=\!\left\{\!(r_1^{(2)},r_2^{(2)},\cdots,r_L^{(2)})|r_m^{(2)}\! <\! \frac{1}{2}\log(1+p_{m}^{(2)}), m \!\in\!\mathcal{I}_L\!\right\}\!.
\end{equation*}
In our simulation, we use the toolbox Scipy\cite{Scipy} to realize the differential evolution and the linear programming algorithm. We average the numerical results over $500$ channel realizations. The following settings are employed: $P_l = P, 0.1 \leq \beta_{l} \leq 4, l\in\mathcal{I}_L$, and $ p_{m}^{(2)} = 0.25P, m\in\mathcal{I}_L$.


%

\subsubsection{Comparison of Various Relaying Schemes}
We compare the following four relaying schemes in a $3\times 3$ network:
\begin{itemize}
	\item AF: amplify-and-forward;
	\item DF: decode-and-forward;
	\item GCCF: generalized compute-compress-and-forward;
	\item GCCF-S: generalized compute-compress-and-forward under the separability condition.
\end{itemize}

\begin{figure}[h]
	\centering
	\includegraphics[width=3.3in]{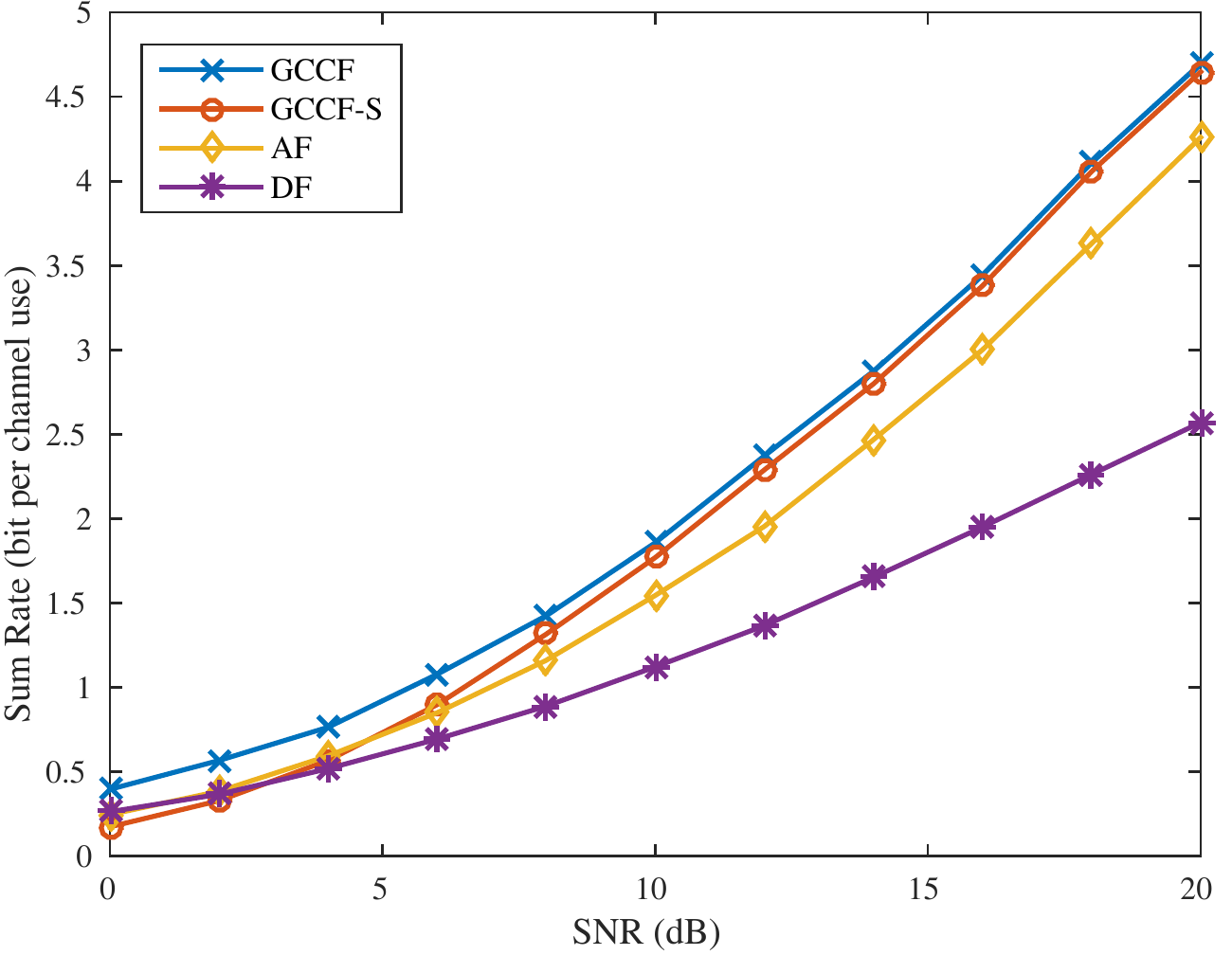}
	\caption{The performance comparison of various relaying schemes in the two-hop relay network in Fig. \ref{fig:Two Hop}.}
	\label{fig:2}
\end{figure}

The numerical results are presented in Fig. \ref{fig:2}.
From Fig. \ref{fig:2}, we see that the GCCF scheme performs much better than the conventional DF scheme, and also outperforms AF by about $1.5$ dB at relatively high SNR. GCCF and GCCF-S perform very close to each other, especially in the high SNR regime. This implies that GCCF-S is an attractive low-complexity alternative to GCCF. 

\subsubsection{Comparison in Various Network Sizes}
\begin{figure}[h]
	\centering
	\includegraphics[width=3.3in]{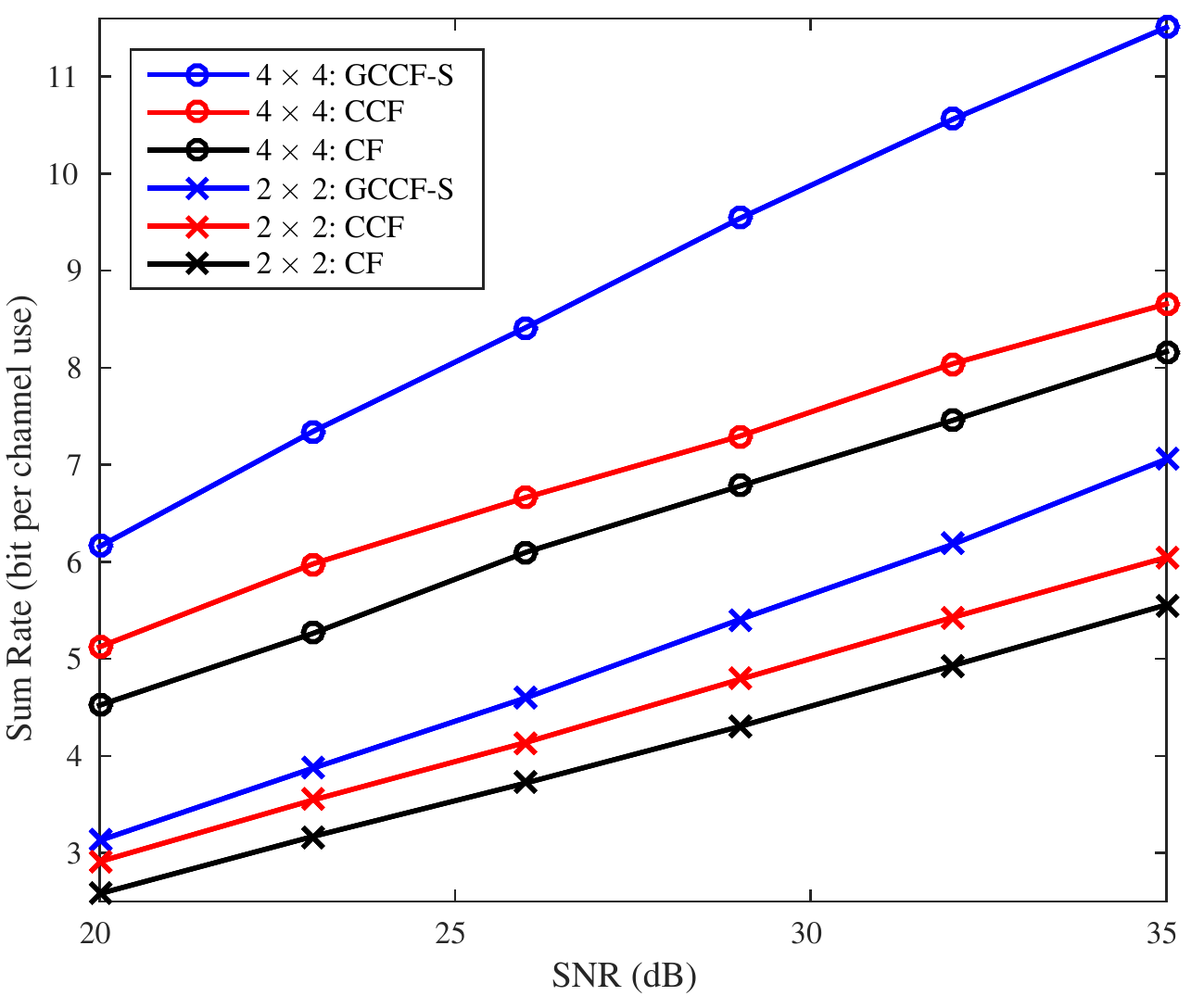}
	\caption{The performance comparison of various schemes in the two-hop relay network in Fig. \ref{fig:Two Hop} with different network sizes.}
	\label{fig:1}
\end{figure}

The numerical results for the considered relay network with sizes $2\times 2$ and $4\times 4$ are presented in Fig. \ref{fig:1}. Note that CCF refers to the original CCF scheme \cite{tan2015compute}; ACF refers to the asymmetric CF scheme \cite{ntranos2013asymmetric}. Also note that the performance of GCCF is not included due to high complexity. From Fig. \ref{fig:1}, we see that GCCF-S achieves a much higher sum rate than CCF and ACF does, especially as the increase of the network size.

%
%



\section{Conclusion \label{sec:conclusion}}

In this paper, we developed a general compression framework, termed GCCF, for CF-based relay networks. In contrast to the one-pair QM operation in the original CCF, our proposed GCCF scheme allows each relay to take multiple pairs of QM operations, so as to reduce the information redundancy more efficiently. We showed that the compression rate region of GCCF is a contra-polymatroid and is broader than that of CCF. We also showed that GCCF is optimal in the sense of minimizing the total compression rate, and established sufficient conditions for GCCF to achieve the optimal SW region. Based on that, we studied the sum-rate maximization of the GCCF-based two-hop relay network, and demonstrated the superior performance of GCCF over the other relaying schemes.

With GCCF, there are a number of future research directions worth of pursuing. First, recall that a connection between the two-hop relay network and C-RAN was established in Section \ref{sec:multihop network_A}. This inspires us to utilize the analytical results in this paper to characterize the fundamental performance limits of C-RAN. Some initial results on the incorporation of CF techniques into C-RAN can be found, e.g., in \cite{TaoSSA}. Second, the GCCF scheme can be potentially combined with other relaying strategies, such as decode-and-forward and amplify-and-forward, to enhance the network performance. How to analyze and optimize these hybrid-relaying schemes is an interesting research topic. Third, the GCCF scheme considered in this paper assumes single-antenna transmitters and single-antenna receivers. It is known that multi-antenna techniques can be employed to dramatically increase the system capacity. As such, how to extend the results in this paper to multi-antenna relay networks is an interesting topic worth of future research effort.


\appendices

\section{Derivation of \eqref{eq:component of v_m} \label{app:derivation}}

From \eqref{eq_v_splitting}, we obtain
\begin{subequations}\label{app_deriv}
\begin{align}
\bm v_{m,k}\!&{} = Q_{\Lambda_{k+1}}(\bm v_{m})\bmod\Lambda_{k} \label{app_deriv_1}\\
&{}= \left( \bm v_{m} - \bm v_{m} \bmod \Lambda_{k+1} \right) \bmod \Lambda_{k} \label{app_deriv_2} \\
&{}= \left( \bm v_{m}\bmod \Lambda_{k} - \bm v_{m} \bmod \Lambda_{k+1} \right) \bmod \Lambda_{k} \label{app_deriv_3}\\
&{}= \!\left(\! \sum_{l=1}^{L}\!a_{ml}(\tilde{\t}_{l})\!\bmod\! \Lambda_{k}\! -\!\! \sum_{l=1}^{L}\!a_{ml}(\tilde{\t}_{l})\!\bmod\! \Lambda_{k+1}\!\! \right)\!\!\bmod\!\! \Lambda_{k} \label{app_deriv_4}\\
\begin{split}\label{app_deriv_5}
&{}=\! \left(\! \sum_{l=1}^{L}\!a_{ml}\!\left(Q_{\Lambda_{k+1}}\!(\tilde{\t}_{l})\! \bmod\! \Lambda_{k}\! +\! \tilde{\t}_{l}\!\bmod\! \Lambda_{k+1} \right.\right.\\
&{}\qquad \left.\left.- \tilde{\t}_{l} \bmod \Lambda_{k+1} \vphantom{Q}\right)\vphantom{\sum_{l=1}^{L}} \right)\bmod \Lambda_{k}
\end{split}\\
&{}= \left( \sum_{l=1}^{L}a_{ml} \left(Q_{\Lambda_{k+1}}(\tilde{\t}_{l})\bmod \Lambda_{k}\right) \right)\bmod \Lambda_{k} \label{app_deriv_6}\\
&{}= \left( \sum_{l=1}^{L}a_{ml} \left(\tilde{\t}_{l,k}\right) \right)\bmod \Lambda_{k}\label{app_deriv_8}
\end{align}
\end{subequations}
where \eqref{app_deriv_2} follows from \eqref{eq:modulo}, \eqref{app_deriv_4} follows by substituting \eqref{eq:computed codeword} and letting $\tilde{\t}_{l} = \t_l - Q_{\Lambda_{\mathrm{s},l}}(\t_l - \beta_l \bm d_l)$, \eqref{app_deriv_5} follows by utilizing \eqref{eq:modulo}, and
\eqref{app_deriv_8} follows by letting $\tilde{\t}_{l,k} = Q_{\Lambda_{k+1}}(\tilde{\t}_{l})\bmod \Lambda_{k}$.

\section{Proof of Theorem \ref{thm: achive vertex} \label{app:recovery}}

	Since $\bm A$ is invertible, there is a bijection between $\{\bm v_m\}_1^L$ and $ \{\bm t_l\}_1^L$. Thus, to prove that $\{\bm v_m\}_1^L$ can be recovered from $\{\bm \delta_{m}\}_1^L$, it suffices to show that $\{\bm t_l\}_1^L$ can be recovered from $\{\bm \delta_{m}\}_1^L$. From \eqref{eq:codeword splitting}, for any given $l$, $\t_l$ can be recovered from $\{\t_{l,k}|k\in \mathcal{K}_l\}$. Thus, we only need to show that $\{ \t_{l,k} | l\in \mathcal{I}_L, k\in \mathcal{K}_{l} \}$ can be recovered from $\{ \bm \delta \}_1^L$, or equivalently,  $\{\t_{l,k}|l\in \mathcal{L}_k\}$ for each $k \in \mathcal{I}_{2L-1}$ can be recovered for from $\{ \bm \delta \}_1^L$. In the following, we show that $\{\bm \t_{l,k}|l\in \mathcal{L}_k\}$ can be recovered recursively in a descending order of $k$. 
	
	To start with, we establish a relation between $\{ \bm \delta_m \}$ and $\{\bm v_{m,k}\}$ as follows.
    From \eqref{eq: compression operation} in Theorem \ref{thm: achive vertex}, we see that $\bm \delta_{m}$ contains $\bm v_{m,k}$ for $k\in \mathcal{J}_{\pi_{\alpha}^{-1}(m)}$. 
    For any given $k$, define 
    \begin{equation}\label{eq_recover_M_index}
    \mathcal{M}_{k} = \left\{ \text{all}\ m\ \text{satisfying } k \in \mathcal{J}_{\pi_{\alpha}^{-1}(m)}\right\}.
    \end{equation} 
	Following the codeword splitting in Section \ref{sec:codeword splitting}, we define
    \begin{equation}
    	\bm \delta_{m,k} = Q_{\Lambda_{k+1}}(\bm \delta_{m})\bmod\Lambda_{k} \label{recover_eq_1}.
    \end{equation}
    For any $m \in \mathcal{M}_{k}$, we have 
    	\begin{subequations}\label{recover_eq_2}
    		\begin{align}
    		\bm \delta_{m,k} 
    		 &= Q_{\Lambda_{k+1}}\left(\sum_{k'\in \mathcal{J}_{\pi_{\alpha}^{-1}(m)}} \bm v_{m,k'}\right)\bmod \Lambda_{J_m^{\text{min}}}\bmod\Lambda_{k} \label{recover_eq_2_1}\\
    		&=  Q_{\Lambda_{k+1}}\left(\sum_{k'\in \mathcal{J}_{\pi_{\alpha}^{-1}(m)}}
    		\bm v_{m,k'}\right)\bmod\Lambda_{k} \label{recover_eq_2_2} \\
    		&= \bm v_{m,k} \label{recover_eq_2_5}
    		\end{align}
    	\end{subequations}
    	where \eqref{recover_eq_2_1} follows by substituting \eqref{eq: compression operation} into \eqref{recover_eq_1}, \eqref{recover_eq_2_2} is from $\Lambda_{J_m^{\text{min}}} \subseteq \Lambda_k$ for $k \in \mathcal{J}_{\pi_{\alpha}^{-1}(m)}$, \eqref{recover_eq_2_5} is from the definition in \eqref{eq_v_splitting}.
    	If $m \notin \mathcal{M}_{k}$, we have
    		\begin{subequations}\label{recover_eq_22}
    			\begin{align}
    			\bm \delta_{m,k} 
    			&=  Q_{\Lambda_{k+1}}\left(\sum_{k'\in \mathcal{J}_{\pi_{\alpha}^{-1}(m)}}
    			\bm v_{m,k'}\right)\bmod\Lambda_{k}\\
    			&= \bm 0.
    			\end{align}
    		\end{subequations}
    	
   		We now consider the recovery of $\{\t_{l,k}|l\in \mathcal{L}_k\}$ for $k$ satisfying \eqref{eq_zero_dither} (implying no residual dither in $\bm v_{m,k}$). 
   		Without loss of generality, we henceforth assume $\Lambda_{s,L} \subseteq \Lambda_{s,L-1} \subseteq \cdots \subseteq \Lambda_{s,1}$. Note that $\Lambda_{\mathrm{s},l} = \Lambda_{\pi(2l-1)}$ in \eqref{eq:WhichLatticePair}, we see that, the finest shaping lattice is $\Lambda_{\pi(1)}$. Therefore, the range of $k$ satisfying \eqref{eq_zero_dither} is given by $\pi(1) \leq k \leq 2L-1$. \footnote{For the example in Fig. \ref{fig:app_recovery}, the finest shaping lattice is $\Lambda_3$. Thus, we have $\pi(1) = 3$, and so the considered range of $k$ is $3\leq k \leq 5$. Clearly, there is no dithering signal in $\bm v_{m,k}$ for $k=3,4,$ and $5$.}
   		Thus, for any $k$ satisfying \eqref{eq_zero_dither} and  $m \in \mathcal{M}_{k}$, $\bm \delta_{m,k}$ can be represented as
   		\begin{subequations}\label{recover_eq_2_6}
   			\begin{align}
   			\bm \delta_{m,k} &= \bm v_{m,k} \label{recover_eq_2_6_1} \\
   			&=  \left(\sum_{l\in \mathcal{L}_k}a_{ml}\t_{l,k}\right)\bmod \Lambda_{k} \label{recover_eq_2_6_2}
   			\end{align}
   		\end{subequations}
   		where \eqref{recover_eq_2_6_1} is from \eqref{recover_eq_2_5}, and \eqref{recover_eq_2_6_2} follows from the fact that \eqref{eq:component of v_m2} holds for $k$ satisfying \eqref{eq_zero_dither}.

   		Following \eqref{eq:encoding mapping}, we map $\bm \delta_{m,k}$ from $\mathbb{R}^n$ to $\mathbb{F}_{\gamma}^{K}$ by defining 
   		\begin{equation}\label{recover_eq_2_7}
   		\phi_{\mathrm{v},k}(\bm x) = (\bm{B}\gamma^{-1} g(\bm{G}\bm x)) \bmod \Lambda_{k}
   		\end{equation}
   		where $\bm x \in \mathbb{F}^{K\times 1}$ is a zero-padded vector with $i_k$ initial zeros and $K-i_{k+1}$ ending zeros. 
   		Note that, $\phi_{\mathrm{v},k}(\bm x)$ is a bijection between $\mathbb{F}_{\gamma}^{i_{k+1} - i_{k}}$ and $\mathcal{C}_{\mathrm{v},k}$.
   	 Denote by $\phi_{\mathrm{v},k}^{-1}$ the inverse of $\phi_{\mathrm{v},k}(\cdot)$. Let $\bm w_{l,k} = \phi_{\mathrm{v},k}^{-1}(\t_{l,k})$ be the finite field message of $\t_{l,k}$. 
	Following \cite[Lemma 6]{nazer2011compute}, we can map $\bm \delta_{m,k}$ in \eqref{recover_eq_2_5} from $\mathbb{R}^n$ to $\mathbb{F}_{\gamma}^{K}$, yielding
	\begin{equation}\label{recover_eq_3}
	\phi_{\mathrm{v},k}^{-1}\left(\bm \delta_{m,k}\right) = \sum_{l\in \mathcal{L}_k} g^{-1}(a_{ml}\bmod\gamma) \bm w_{l,k}.
	\end{equation}
	By taking transpose on the both sides of \eqref{recover_eq_3} and then stacking row by row for $m\in \mathcal{M}_{k}$, we obtain
	\begin{equation}\label{recover_eq_4}
		\bm \Delta_{k} = \bm A(\mathcal{M}_{k},\mathcal{L}_k) \bm W_{k}
	\end{equation}
	where the $i$-th rows of $\bm \Delta_{k}$ is given by the transpose of $\phi_{v,k}^{-1}\left(\bm \delta_{m_i}\bmod \Lambda_{k}\right)$ with $m_i$ being the $i$-th element of $\mathcal{M}_{k}$ (ordered in an ascending manner), and the $i$-th row of $\bm W_{k}$ is given by the transpose of $\bm w_{l_i,k}$ with $l_i$ being the $i$-th element of $ \mathcal{L}_k$ (ordered in an ascending manner). The following lemma states that $\bm A(\mathcal{M}_{\pi_{\alpha}}^k,\mathcal{L}_k)$ is invertible.
	\begin{lem}\label{lem:invertible_submatrix}
		The two sets defined in \eqref{eq:Index set I_k} and \eqref{eq_recover_M_index} have the same cardinality, i.e., $|\mathcal{M}_{k}| = |\mathcal{L}_k|$. Further, submatrix $\bm A(\mathcal{M}_{k},\mathcal{L}_k)$ is invertible over $\mathbb{F}^{|\mathcal{L}_k|\times |\mathcal{L}_k|}$.
	\end{lem}
\begin{IEEEproof}
	From the definition of $\mathcal{M}_k$ in \eqref{eq_recover_M_index}, we can construct $\mathcal{M}_k$ as follows. For each $m$, we consider the $\pi_{\alpha}(m)$-th row of the submatrix $\bm A(:,\mathcal{L}_k)$. If the $\pi_{\alpha}(m)$-th row is independent of the $\pi_{\alpha}(m+1)$-th, $\cdots,\pi_{\alpha}(L)$-th rows of $\bm A(:,\mathcal{L}_k)$, we have $m \in \mathcal{M}_k$. As $\bm A$ is invertible, the columns of $\bm A$ are independent. Thus, $\text{rank}(\bm A(:,\mathcal{L}_k)) = |\mathcal{L}_k|$, and so there are $|\mathcal{L}_k|$ independent rows in $\bm A(:,\mathcal{L}_k)$. Thus, $|\mathcal{M}_k| = |\mathcal{L}_k|$. Also, as the $|\mathcal{M}_k|$ selected rows are linearly independent, we obtain that  $\bm A(\mathcal{M}_{k},\mathcal{L}_k)$ is invertible.
\end{IEEEproof}

	 From Lemma \ref{lem:invertible_submatrix}, we recover $\bm W_{k}$ by
	\begin{equation}\label{recover_eq_4_1}
		\bm W_{k} = \left(\bm A(\mathcal{M}_{k},\mathcal{L}_k)\right)^{-1} \bm \Delta_{k}
	\end{equation}
	Thus, $\{\bm \t_{l,k}|l\in \mathcal{L}_k\}$ can be recovered by $\t_{l,k} = \phi_{\mathrm{v},k}(\bm w_{l,k})$ for $\pi(1) \leq k \leq 2L-1$. Recall from \eqref{eq:WhichLatticePair} that $\Lambda_{\mathrm{c},l} = \Lambda_{\pi(2l)}$ for $l\in \mathcal{I}_L$. Thus, from \eqref{eq:Index set J_l}, we obtain $\mathcal{K}_1 = \{ k| \pi(1) \leq k \leq \pi(2)\} \subseteq \{ k| \pi(1) \leq k \leq 2L-1 \}$. Therefore, $\{\bm \t_{1,k}|k\in \mathcal{K}_1\}$ are all recoverable and so $\t_1$ can be recovered by using \eqref{eq:codeword splitting}.
	

	We now consider the recovery of $\{\t_{l,k}\}$ on the $k$-th virtual codebook with $k$ not satisfying \eqref{eq_zero_dither}, i.e., $1\leq k \leq \pi(1) - 1$. We first show the recovery of $\{\t_{l,k}\}$ for $ k = \pi(1)-1$.\footnote{ For the example in Fig. \ref{fig:app_recovery}, $\pi(1)-1 = 2$.} From $\Lambda_{s,2} \subseteq \Lambda_{s,1} = \Lambda_{\pi(1)}$ and the fact that $\Lambda_{k}$ is the finest lattice which is coarser than $ \Lambda_{\pi(1)}$, we have 
	\begin{equation}\label{eq_recover_neste_1}
		\Lambda_{s,2} \subset \Lambda_{k}.
	\end{equation}
	In this case, the residual dithers $Q_{\Lambda_{\mathrm{s},l}}(\t_l - \beta_l \bm d_l)\bmod\Lambda_k = 0$ for $L\leq l \leq 2$.  Also, the remaining residual dither $Q_{\Lambda_{s,1}}(\t_1 - \beta_1 \bm d_1)$ is known since $\beta_1$ and  $\bm d_1$ are known, and $\t_1$ is just recovered. Thus, $Q_{\Lambda_{s,1}}(\t_1 - \beta_1 \bm d_1)$ can be pre-cancelled from $\bm \delta_{m,k}$ as follows . Let 
	\begin{equation}\label{recover_eq_4_2}
		\tilde{\t}_{1,k} = Q_{\Lambda_{k+1}}(\bm t_1 - Q_{\Lambda_{s,1}}(\bm t_1 - \beta_1 \bm d_1))\bmod \Lambda_{k}.
	\end{equation}
	 Then, for $m \in \mathcal{M}_k$,
	 \begin{subequations}\label{recover_eq_5}
	 	\begin{align}
	 	&\left(\bm \delta_{m,k} - a_{m1} \tilde{\t}_{1,k}\right)\bmod \Lambda_{k}\\
	 	&=\left(\bm v_{m,k} - a_{m1} \tilde{\t}_{1,k}\right)\bmod \Lambda_{k} \label{recover_eq_5_2}\\
	 	&=\left(\sum_{l=2}^{L}a_{ml}\t_{l,k} + a_{m1}\tilde{\t}_{1,k} - a_{m1} \tilde{\t}_{1,k} \right) \bmod \Lambda_{k} \label{recover_eq_5_1}\\
	 	& = \left(\sum_{l\in \mathcal{L}_k}a_{ml}\t_{l,k}\right)\bmod\Lambda_{k} \label{recover_eq_5_3}
	 	\end{align}
	 \end{subequations}
	where \eqref{recover_eq_5_2} is from \eqref{recover_eq_2}, \eqref{recover_eq_5_1} is from \eqref{eq_v_splitting} and \eqref{recover_eq_4_2}, \eqref{recover_eq_5_3} is from the fact that $\t_{1,k} = \bm 0$ for $ k = \pi(1)-1$. 
	 Then, following the approach in \eqref{recover_eq_2_7}-\eqref{recover_eq_4_1}, we can recover $\{ \t_{l,k} | l\in \mathcal{L}_k \}$ for $k=\pi(1)-1$.
	
	Note that $\Lambda_{\mathrm{s},2} = \Lambda_{\pi(3)}$. Then, \eqref{eq_recover_neste_1} still holds for $\pi(3) \leq k \leq \pi(1)-2$. Then, following the approach in \eqref{recover_eq_4_2}-\eqref{recover_eq_5}, we can recover $\{\bm \t_{l,k}|l\in \mathcal{L}_k\}$ for $ \pi(3) \leq k \leq \pi(1)-2$ in the same way. Recall from \eqref{eq:Index set J_l} that $\mathcal{K}_2 = \{ k| \pi(3) \leq k \leq \pi(3)\} \subseteq \{ k| \pi(3) \leq k \leq 2L-1\}$. Therefore, we can also recover $\t_2$ by \eqref{eq:codeword splitting}.
	
	By induction, we can recover $\{\t_l\}_{1}^{L}$ recursively. This complete the proof.
	
	
\section{Proof of Theorem \ref{cor:rate_of_delta} \label{app:proof_of_cor_rate_of_delta}}

	We first show \eqref{eq_rate_delta_m}. 
	The compression function in \eqref{eq: compression operation} gives a bijection between $\bm \delta_m$ and $\{ \bm v_{m,k}|k\in \mathcal{J}_{\pi_{\alpha}^{-1}(m)} \}$. Thus,
	\begin{equation}
	H(\bm \delta_m) = H(  \bm v_{m,k}|k\in \mathcal{J}_{\pi_{\alpha}^{-1}(m)} ).
	\end{equation}
	By definition, the entropy rate of $\bm v_{m,k}$ is given by 
	\begin{equation}
	H(\bm v_{m,k}) = r_{\mathrm{v},k}.
	\end{equation}
	Thus, to prove \eqref{eq_rate_delta_m}, it suffices to show that $\bm v_{m,k}$ is independent of $\bm v_{m,k'}$ for any $k\neq k'$ and $k,k' \in \mathcal{J}_{\pi_{\alpha}^{-1}(m)}$. 
	
	From \eqref{eq:component of v_m} and \eqref{eq_t_tild_t}, we obtain
	\begin{equation}
	\begin{split}
		\bm v_{m,k} &= \left( \sum_{l=1}^{L} a_{ml} \tilde{\bm{t}}_{l,k} \right)\bmod \Lambda_k\\
		&= \left( \sum_{l\in \mathcal{L}_k} a_{ml} \bm{t}_{l,k} + \sum_{l\notin \mathcal{L}_k} a_{ml}\tilde{\bm{t}}_{l,k} \right)\bmod \Lambda_k.
	\end{split}
	\end{equation}
	Note that $\{ a_{m,l}|l\in\mathcal{L}_k \}$ defines $\bm A(m,\mathcal{L}_k)$. Also note that the submatrix $\bm A(\overline{\mathcal{I}_{m-1}}, \mathcal{L}_k)$ has one more row (i.e. $\bm A(m,\mathcal{L}_k)$) than the submatrix $\bm A(\overline{\mathcal{I}_{m}}, \mathcal{L}_k)$. Then, from the definition in \eqref{eq:rank_equation}, for $k \in \mathcal{J}_{\pi_{\alpha}^{-1}(m)}$, 
	$\bm A(m,\mathcal{L}_k)$ can not be a zero vector, and so $\{ a_{m,l}|l\in\mathcal{L}_k \}$ are not all zeros.
	Thus, from the Crypto lemma\cite{erez2004achieving}, $\bm v_{m,k}$ is uniformly distributed over $\mathcal{V}_{k}$ and is independent of $\{ \tilde{\t}_{l,k} \}$.  Also note that $\t_{l,k}$ is independent of $\t_{l',k'}$ for $k\neq k'$. Thus, $\bm v_{m,k}$ is independent of $\bm v_{m,k'}$ for $k\neq k'$ and $k,k' \in \mathcal{J}_{\pi_{\alpha}^{-1}(m)}$.
	Therefore, the entropy rate of $\bm \delta_{m}$ is given by 
	\begin{subequations}\label{cor_entrpy_eq0}
		\begin{align}
		H(\bm \delta_{m}) &=  H(  \bm v_{m,k}|k\in \mathcal{J}_{\pi_{\alpha}^{-1}(m)} )\\
		&= \sum_{k\in \mathcal{J}_{\pi_{\alpha}^{-1}(m)}}H(\bm v_{m,k}) \label{cor_entrpy_eq0_1} \\
		& = \sum_{k\in \mathcal{J}_{\pi_{\alpha}^{-1}(m)}} r_{\mathrm{v},k} \label{cor_entrpy_eq0_2}.
		\end{align}
	\end{subequations}

	We then show \eqref{eq_sum_rate_delta_m}. The left hand side (LHS) of \eqref{eq_sum_rate_delta_m} can be represented as
	\begin{subequations}\label{cor_entrpy_eq2}
		\begin{align}
		\sum_{m=1}^{L}	H(\bm \delta_{m}) &= \sum_{m=1}^{L} \sum_{k\in\mathcal{J}_{\pi_{\alpha}^{-1}(m)}} H(\bm \delta_{m,k}) \label{cor_entrpy_eq2_2_1}\\
		&= \sum_{k=1}^{2L-1} \sum_{m\in\mathcal{M}_k} H(\bm \delta_{m,k})\label{cor_entrpy_eq2_2_2}\\
		&=  \sum_{k=1}^{2L-1} |\mathcal{M}_k| r_{\mathrm{v},k}\\
		&= \sum_{k=1}^{2L-1} |\mathcal{L}_k| r_{\mathrm{v},k}
		\end{align}
	\end{subequations}
	where \eqref{cor_entrpy_eq2_2_1} is from \eqref{cor_entrpy_eq0_1} together with $\bm \delta_{\pi_{\alpha}(m),k} = \bm v_{\pi_{\alpha}(m),k}$ for $k\in \mathcal{J}_{\pi_{\alpha}(m)}$ and $\bm \delta_{\pi_{\alpha}(m),k} = \bm 0$ for $k\notin \mathcal{J}_{\pi_{\alpha}(m)}$, \eqref{cor_entrpy_eq2_2_2} is from the definition of $\mathcal{M}_k$ in \eqref{eq_recover_M_index}.
	The RHS of \eqref{eq_sum_rate_delta_m} can be represented as
	\begin{subequations}\label{cor_entrpy_eq1}
		\begin{align}
		\sum_{l=1}^{L} r_l &= H(\{\bm t_l|l\in \mathcal{I}_L\}) \\
		&= H(\{\bm t_{l,k}|l\in \mathcal{I}_L,k\in\mathcal{K}_l\})\\
		& = H(\{\bm t_{l,k}|l\in \mathcal{L}_k,k\in\mathcal{I}_{2L-1}\}) \\
		&= \sum_{k=1}^{2L-1}|\mathcal{L}_k|r_{\mathrm{v},k}.
		\end{align}
	\end{subequations}
	By combining \eqref{cor_entrpy_eq2} and \eqref{cor_entrpy_eq1}, we obtain \eqref{eq_sum_rate_delta_m}.

\section{Proof of Theorem \ref{thm: achive vertex 2} \label{app:compression_2}}
	
	We first assume that  $\mathcal{J}_{m}$ can be represented by 
	\begin{equation}\label{eq:index_K}
	\mathcal{J}_{m} = \{J_{m}^{\text{min}},J_{m}^{\text{min}}+1,\cdots, J_{m}^{\text{max}}\}.
	\end{equation}
	We now show \eqref{eq_delta_m_2}. Note that
	\begin{subequations}\label{eq_reduce}
		\begin{align}
			\begin{split}\label{eq_reduce_1}
				Q_{\Lambda_{k+2}}(\bm v_m)\bmod \Lambda_{k}= &{} \left(Q_{\Lambda_{k+2}}(\bm v_m)\bmod \Lambda_{k}\right)\bmod \Lambda_{k+1} + Q_{\Lambda_{k+ 1}} \left(Q_{\Lambda_{k+2}}(\bm v_m)\bmod \Lambda_{k}\right) 
			\end{split}\\
			\begin{split}\label{eq_reduce_2}
				= &{} Q_{\Lambda_{k+2}}(\bm v_m)\bmod \Lambda_{k+1} + Q_{\Lambda_{k+1}}(\bm v_m)\bmod \Lambda_{k}
			\end{split}\\
			= &{} \bm v_{m,k+1} + \bm v_{m,k}
		\end{align}
	\end{subequations}
	where \eqref{eq_reduce_1} follows from \eqref{eq:modulo}, and \eqref{eq_reduce_2} from $\Lambda_{k}\subseteq \Lambda_{k+1}\subseteq \Lambda_{k+2}$. By induction, we obtain from \eqref{eq_reduce} that
	\begin{equation*}
		Q_{\Lambda_{J_m^{\text{max}}}+1}(\bm v_{m})\bmod\Lambda_{J_m^{\text{min}}} = \left(\sum_{k\in \mathcal{J}_{\pi_{\alpha}^{-1}(m)}} \bm v_{m,k}\right)\bmod \Lambda_{J_m^{\text{min}}},
	\end{equation*}
	which implies that \eqref{eq: compression operation} simplifies to \eqref{eq_delta_m_2}.
	 Thus, to prove Theorem \ref{thm: achive vertex 2}, it suffices to show that the separability condition \eqref{eq:separability} implies the continuity of the index set in \eqref{eq:index_K}.

	With the left $L$ lattices in the nested lattice chain $\Lambda_1 \subseteq \cdots \subseteq \Lambda_L\subseteq \Lambda_{L+1} \subseteq \cdots \subseteq \Lambda_{2L}$ are all shaping lattices. That is, $\Lambda_{\mathrm{s},l} \subseteq \Lambda_{L} \subseteq \Lambda_{L+1} \subseteq \Lambda_{\mathrm{c},l}$ for $l\in\mathcal{I}_L$. Then, by the definition in \eqref{eq:Index set I_k},  $\mathcal{L}_{L} = \mathcal{I}_L$. From \eqref{eq:rank_equation} and the fact that $\bm{A}$ is full rank, we have $ L \in \mathcal{J}_{m}, m \in \mathcal{I}_L$.  
	Note that $|\mathcal{L}_{k}|$ is monotonically increasing in $k$ when $k < L $, and monotonically decreasing in $k$ when $k > L$. Also note that, for any given $m$ and $k \in \mathcal{J}_{m}$, the rank equality in \eqref{eq:rank_equation} holds for $k'$ satisfying $ \mathcal{L}_{k} \subset \mathcal{L}_{k'}$, and so $k' \in \mathcal{J}_{m}$. Thus, we have the following two results:
	\begin{enumerate}
			\item For any $k \leq L$, $ k \in \mathcal{J}_{m}$ implies $ k' \in \mathcal{J}_{m}$.
			\item For $k \geq L$, $ k \in \mathcal{J}_{m}$ implies $ k' \in \mathcal{J}_{m}$.
	\end{enumerate}
	Therefore, the continuity of $\mathcal{J}_{m}$ holds with the separability condition.

\section{Proof of Theorem \ref{thm:equivalent_L_2} \label{appen:proof_equivalent_L_2}}  

	To prove Theorem \ref{thm:equivalent_L_2}, we need to show that \eqref{equal_region} holds for $L=2$.
	The RHS of \eqref{equal_region} can be represented as 
	\begin{equation}\label{eq_compression_bound2}
	\begin{split}
	&H\!\left(\!\{\bm v_m | m \!\in\! \mathcal{S}\}|\{\bm v_m | m \!\in\! \overline{\mathcal{S}}\}\!\right)\!=\!H\!\left(\!\bm v_m|m \!\in\! \mathcal{I}_L\!\right)\!-\!H\!\left(\!\bm v_m | m \!\in\! \overline{\mathcal{S}}\right).
	\end{split}
	\end{equation}
	Thus, we need to show that
	\begin{equation}\label{eq_the6ToBeProved}
	H\left(\{\bm v_m | m \in \mathcal{S} \}\right)=\sum_{k=1}^{2L-1}{\textit{rank}(\bm A(\mathcal{S}, \mathcal{L}_{k}))} r_{\mathrm{v},k}, \text{ for } \mathcal{S} \in \mathcal{I}_2.
	\end{equation}
	The non-empty subsets of $\mathcal{I}_2$ are given by $\{1\},\{2\},\mathcal{I}_2.$ For $\mathcal{S} = \mathcal{I}_2$, we have
	\begin{subequations}\label{eq_thm6_1}
		\begin{align}
			H\left(\{\bm v_m | m \in \mathcal{S} \}\right) 
			&= H(\t_1,\t_2) \label{eq_thm6_1_1}\\
			&= \sum_{k=1}^{2L-1} |\mathcal{L}_k| r_{\mathrm{v},k} \label{eq_thm6_1_2}
		\end{align}
	\end{subequations}
	where \eqref{eq_thm6_1_1} follows from the fact that $\{\t_1,\t_2\}$ can be recovered from $\{\bm v_1,\bm v_2\}$, and \eqref{eq_thm6_1_2} from \eqref{cor_entrpy_eq1}. Note that for $\mathcal{S} = \mathcal{I}_2$, $\textit{rank}(\bm A(\mathcal{S}, \mathcal{L}_{k})) = |\mathcal{L}_{k}|$. Thus, \eqref{eq_the6ToBeProved} holds for $\mathcal{S} = \mathcal{I}_2$.
	
	We now prove \eqref{eq_the6ToBeProved} for $\mathcal{S} = \{1\}$. The proof for $\mathcal{S} = \{2\}$ is similar and thus omitted. By the chain rule of the entropy, 
	\begin{equation}\label{eq_the7ToBeProved}
	\begin{split}
	H(\bm v_1) &= H(\bm v_{1,1},\bm v_{1,2},\bm v_{1,3})\\
	&=H(\bm v_{1,3}) + H(\bm v_{1,2}|\bm v_{1,3}) + H(\bm v_{1,1}|\bm v_{1,2},\bm v_{1,3}).
	\end{split}
	\end{equation}
	With \eqref{eq_the7ToBeProved}, to prove \eqref{eq_the6ToBeProved}, it suffices to show 
	\begin{equation}\label{eq_thm6_2}
		H(\bm v_{1,k}|\{\bm v_{1,k'} \}_{k'=k+1}^{3}) = \textit{rank}(\bm A(1, \mathcal{L}_{k})) r_{\mathrm{v},k}, \text{ for } k\in\mathcal{I}_3.
	\end{equation}
	where $\{\bm v_{1,k'} \}_{k'=k+1}^{3} = \emptyset$ if $k+1 > 3$.
	We assume $\Lambda_{\mathrm{s},2} \subseteq \Lambda_{\mathrm{s},1}$ without loss of generality. Then the finest shaping lattice is $ \Lambda_{\mathrm{s},1} = \Lambda_{\pi(1)}$. Note from \eqref{eq_v_m_k} that $\bm v_{m,k}$ contains no dither term. We next prove \eqref{eq_thm6_2} for two cases: $\pi(1) \leq k \leq 3$ and $1 \leq k \leq \pi(1) - 1$.
	
	For $\pi(1) \leq k \leq 3$, $\bm v_{1,k}$ can be expressed in the form of \eqref{eq:Matrix_Computed_codeword_compnent}, and so $\bm v_{1,k}$ is independent of $\bm v_{1,k'}$ for $k'>k$ from the discussion below \eqref{eq_v_m_k}. 
	Thus, we have 
	\begin{equation}\label{eq_thm6_3}
			H(\bm v_{1,k}|\{\bm v_{1,k'} \}_{k'=k+1}^{3}) = H(\bm v_{1,k}).
	\end{equation}
	Furthermore, if there exist non-zero elements in $\bm A(1, \mathcal{L}_{k})$, $\textit{rank}(\bm A(1, \mathcal{L}_{k})) = 1$ and $H(\bm v_{1,k}) = r_{\mathrm{v},k} = \textit{rank}(\bm A(1, \mathcal{L}_{k})) r_{\mathrm{v},k}$, where the second equality follows from the definition of $\bm v_{1,k}$ in \eqref{eq_v_splitting}. If $\bm A(1, \mathcal{L}_{k})$ is an all-zero vector, then $\textit{rank}(\bm A(1, \mathcal{L}_{k})) = 0$ and $H(\bm v_{1,k}) = H(\bm 0) = 0$. Therefore, \eqref{eq_thm6_2} holds for $\pi(1) \leq k \leq 3$.

\begin{figure}[!t]
		\captionsetup[subfigure]{justification=centering}
	\centering
	\subfloat[$\pi(1)=2$]{\includegraphics[height=2.3in]{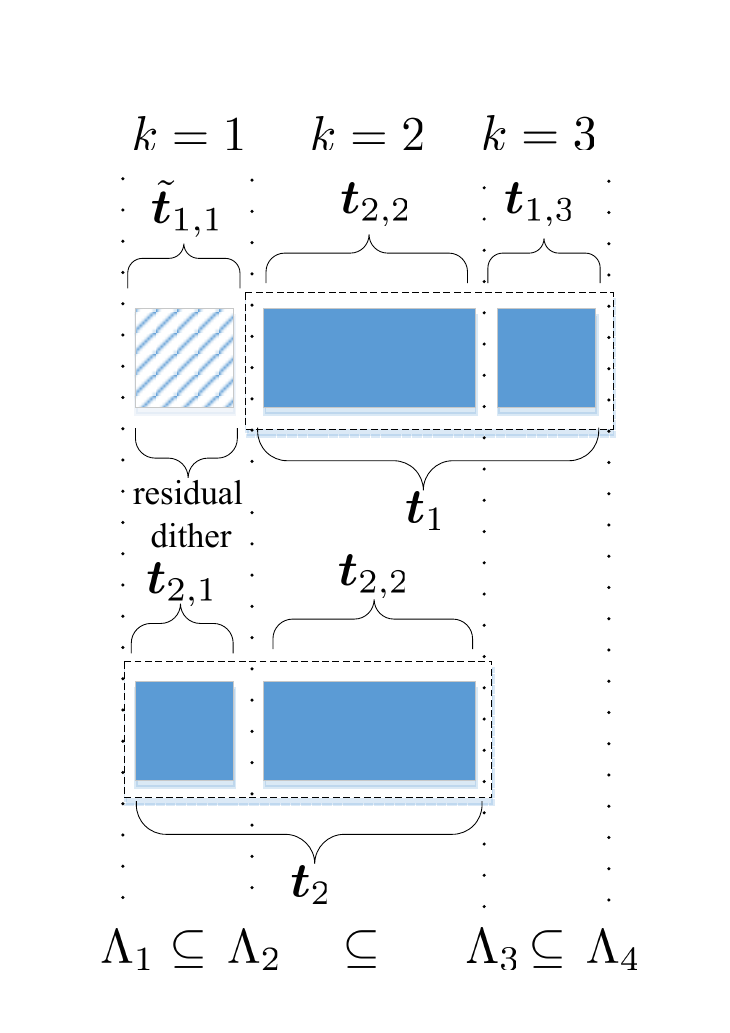}%
	\label{fig_app_L_2_1}}
	\qquad
	\subfloat[$\pi(1)=3$]{\includegraphics[height=2.3in]{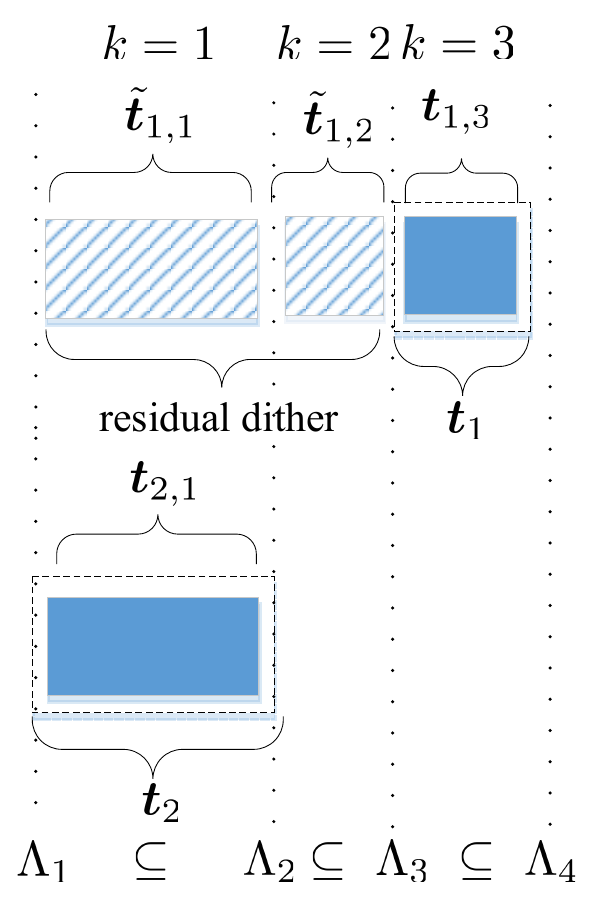}%
	\label{fig_app_L_2_2}}
	\caption{ The nested lattice chains discussed in Appendix \ref{appen:proof_equivalent_L_2}. }
	\label{fig_app_L_2}
	\end{figure}

	What remains is to show that \eqref{eq_thm6_2} holds for  $1 \leq k \leq \pi(1)-1$. Recall that $\Lambda_{\pi(1)}$ is the finest shaping lattice. Then, $\Lambda_{\pi(1)}\neq \Lambda_1$ since $\Lambda_1$ being the coarsest lattice, and $\Lambda_{\pi(1)}\neq \Lambda_4$ since $\Lambda_4$ being the finest lattice (which must be a coding lattice). Thus, the possible choices of $\pi(1)$ are $\pi(1)=2$ or $\pi(1)=3$, corresponding to Fig. \ref{fig_app_L_2}\protect\subref{fig_app_L_2_1} and Fig. \ref{fig_app_L_2}\protect\subref{fig_app_L_2_2}, respectively.
	
	We first consider the choice of $\pi(1) = 2$. 
	Then, the nested lattice chain can be $\Lambda_{1}(\Lambda_{\mathrm{s},2}) \subseteq \Lambda_{2}(\Lambda_{\mathrm{s},1}) \subseteq \Lambda_{3}(\Lambda_{\mathrm{c},2}) \subseteq \Lambda_{4}(\Lambda_{\mathrm{c},1})$ or $\Lambda_{1}(\Lambda_{\mathrm{s},2}) \subseteq \Lambda_{2}(\Lambda_{\mathrm{s},1}) \subseteq \Lambda_{3}(\Lambda_{\mathrm{c},1}) \subseteq \Lambda_{4}(\Lambda_{\mathrm{c},2})$. We next focus on the first lattice chain, which is illustrated in 
	Fig. \ref{fig_app_L_2}\protect\subref{fig_app_L_2_1}. The discussion for the second lattice chain is similar and thus omitted for the simplicity.  
	For $\pi(1) = 2$, $1 \leq k \leq \pi(1)-1$ implies $k=1$. Thus, from \eqref{eq_thm6_2}, we need to show
	\begin{equation}\label{eq_thm6_4}
	H(\bm v_{1,1}|\bm v_{1,2},\bm v_{1,3}) = \textit{rank}(\bm A(1, \mathcal{L}_{1})) r_{\mathrm{v},1}.
	\end{equation}  
	To this end, we first give more details about the variables involved in \eqref{eq_thm6_4}.
	From  \eqref{eq:computed codeword} and \eqref{eq_tilde_t_l}, $\bm v_{1}$ is given by $\bm v_{1} = (a_{11} \tilde{\t}_1 + a_{12} \tilde{\t}_2)\bmod \Lambda_1$. 
	Note that $\Lambda_{\mathrm{s},2} = \Lambda_{1} \subseteq \Lambda_{1} \subseteq \Lambda_{2} \subseteq \Lambda_{3}$. Thus, by letting $l=2$ in \eqref{eq_t_tild_t}, we obtain that $\t_{2,k} = \tilde{\t}_{2,k}$ for $k=1,2,3$. Similarly, by noting $\Lambda_{\mathrm{s},1} = \Lambda_{2} \subseteq \Lambda_{2} \subseteq \Lambda_{3}$, we obtain from \eqref{eq_t_tild_t} that $\t_{1,k} = \tilde{\t}_{1,k}$ for $k=2,3$.
	Thus, from \eqref{eq:component of v_m}, we obtain
	\begin{subequations}\label{eq_thm6_5}
		\begin{align}
		\bm v_{1,1} &= (a_{11} \tilde{\t}_{1,1} + a_{12}\t_{2,1})\bmod \Lambda_1 \label{eq_thm6_5_1}\\
		\bm v_{1,2} &= (a_{11} \t_{1,2} + a_{12}\t_{2,2})\bmod \Lambda_2 \label{eq_thm6_5_2}\\
		\bm v_{1,3} &= (a_{11} \t_{1,3} + a_{12}\t_{2,3})\bmod \Lambda_3 \label{eq_thm6_5_3}.
		\end{align}
	\end{subequations}
	In addition, we obtain $\mathcal{L}_{1} = \{2\}$ from \eqref{eq:Index set I_k}, and so $\textit{rank}(\bm A(1, \mathcal{L}_{1})) = \textit{rank}([a_{12}])$. We are now ready to determine the value of $H(\bm v_{1,1}|\bm v_{1,2},\bm v_{1,3})$ for $a_{12} = 0$ and $a_{12} \neq 0$, respectively. 
	\begin{itemize}
		\item 
		For $a_{12} = 0$, we obtain $ a_{11} \neq 0$  since $\bm A$ is invertible. Then, \eqref{eq_thm6_5} can be rewritten as
		\begin{subequations}\label{eq_thm6_6}
			\begin{align}
				\bm v_{1,1} &= (a_{11} \tilde{\t}_{1,1})\bmod \Lambda_1 \label{eq_thm6_6_1}\\
				\bm v_{1,2} &= (a_{11} \t_{1,2})\bmod \Lambda_2 \label{eq_thm6_6_2}\\
				\bm v_{1,3} &= (a_{11} \t_{1,3})\bmod \Lambda_3 \label{eq_thm6_6_3}.
			\end{align}
		\end{subequations}
		Note that $\t_{1,2}$ can be recovered from $\bm v_{1,2}$ and $\t_{1,3}$ can be recovered from $\bm v_{1,3}$. Also note from \eqref{eq:codeword splitting} that $\t_1 = \t_{1,2} + \t_{1,3}$. Thus, for given $\bm v_{1,2}$ and $\bm v_{1,3}$, $\t_1$ can be recovered and so $\tilde{\t}_1$ can be recovered. Therefore, $\bm v_{1,1}$ is deterministic for given $\bm v_{1,2}$ and $\bm v_{1,3}$ and so $H(\bm v_{1,1}|\{\bm v_{1,2},\bm v_{1,3}\}) =0$.
		On the other hand, $\textit{rank}(\bm A(1, \mathcal{L}_{1})) = \textit{rank}([a_{12}]) =0$. Therefore, \eqref{eq_thm6_4} holds.
		
		\item 
		For $a_{12} \neq 0$,  since $\t_{2,1}$  is independent of $\t_{1,2},\t_{2,2},\t_{1,3}$, and $\t_{2,3}$, 
		from the Crypto lemma\cite{erez2004achieving}, $\bm v_{1,1}$ in \eqref{eq_thm6_5_1}
		is independent of $\tilde{\t}_{1,1}$ and thus is independent of $\bm v_{1,2}$ and $\bm v_{1,3}$. 
		Therefore, $H(\bm v_{1,1}|\bm v_{1,2},\bm v_{1,3}) = H(\bm v_{1,1}) = r_{\mathrm{v,k}}$.
	 	Noting $ \textit{rank}(\bm A(1, \mathcal{L}_{1}) = \textit{rank}([a_{12}]) = 1$, we see that \eqref{eq_thm6_4} holds.
	\end{itemize}
	The above discussion concludes the proof for $\pi(1) = 2$.
	
	We now consider the choice of $\pi(1) = 3$. With this choice, the nested lattice chain is given by
	$\Lambda_{1}(\Lambda_{\mathrm{s},2}) \subseteq \Lambda_{2}(\Lambda_{\mathrm{c},2}) \subseteq \Lambda_{3}(\Lambda_{\mathrm{s},1}) \subseteq \Lambda_{4}(\Lambda_{\mathrm{c},1})$, 
	 which is illustrate in Fig. \ref{fig_app_L_2}\protect\subref{fig_app_L_2_2}. 
	 For $\pi(1) = 3$, $1 \leq k \leq \pi(1)-1$ implies $k=1,2$. 
	 For $ k = 2$, from \eqref{eq_thm6_2}, we need show 
	 \begin{equation}\label{eq_thm6_9}
	 	H(\bm v_{1,2}|\bm v_{1,3}) = \textit{rank}(\bm A(1, \mathcal{L}_{2}))  r_{\mathrm{v,2}}.
	 \end{equation}
	 From \eqref{eq:Index set J_l}, $\t_1 = \t_{1,3}$ and $\t_2 = \t_{2,1}$($\t_{2,2} = \t_{2,3} =0$). 
	 Note that $\Lambda_{\mathrm{s},2} = \Lambda_{1} \subseteq \Lambda_{1}$. Thus, by letting $l=2$ and $k=1$, we obtain from \eqref{eq_t_tild_t} that $\t_{2,k} = \tilde{\t}_{2,k}$ for $k=1,2,3$. Also, $\Lambda_{\mathrm{s},1} = \Lambda_{3} \subseteq \Lambda_{3}$, we obtain from \eqref{eq_t_tild_t} that $\t_{1,3} = \tilde{\t}_{1,3}$.
	 From \eqref{eq:component of v_m}, we have
	\begin{subequations}
		\begin{align}
			\bm v_{1,2} &= (a_{11}\tilde{\t}_{1,2})\bmod \Lambda_{2}\\
			\bm v_{1,3} &= (a_{11}\t_{1})\bmod \Lambda_{3}.
		\end{align}
	\end{subequations}
	We now calculate $H(\bm v_{1,2}|\bm v_{1,3})$ by discussing the value of $a_{11}$. 
	\begin{itemize}
		\item
		If $a_{11} = 0$, $H(\bm v_{1,2}|\bm v_{1,3}) = H(\bm 0|\bm 0) = 0$.
		\item
		If $a_{11} \neq 0$, $ \t_{1}$ is deterministic for given $\bm v_{1,3}$. Thus, $\bm v_{1,2}$ is deterministic for given $\bm v_{1,3}$, and so $H(\bm v_{1,2}|\bm v_{1,3}) =  0$.
	\end{itemize}
	Since from \eqref{eq:Index set I_k} $\mathcal{L}_{2} = \{0\}$, $\textit{rank}(\bm A(1, \mathcal{L}_{2})) = 0$. Thus, \eqref{eq_thm6_9} holds for $k=2$. 
	For $k=1$, following the approach in the choice of $\pi(1) = 2$, we see that \eqref{eq_thm6_9} also holds, 
	which concludes the proof of Theorem \ref{thm:equivalent_L_2}. 

\section{Proof of Theorem \ref{thm:equivalent_rate_region} \label{app:proof_equivalent_rate_region}}

		We prove Theorem \ref{thm:equivalent_rate_region} by showing that when $\bm d_l = \bm 0$ for $l\in\mathcal{I}_L$, GCCF achieves the vertices of $\mathcal{R}_{SW}$. The following lemma gives the vertices of $\mathcal{R}_{\mathrm{SW}}$.
				\begin{lem}\label{cor_SW_vertex}
					The vertex of $\mathcal{R}_{SW}$ specified by $\pi_{\alpha}(\cdot)$ is given by $(R_1,R_2,\cdots,R_L)$ with
					\begin{equation}\label{SW_vertex}
					R_{\pi_{\alpha}(m)} = H\left(\bm v_{\pi_{\alpha}(m)} |\{ \bm v_i, i\in \Pi_{\alpha}(\overline{\mathcal{I}_m})\}\right), \text{ for } m \in \mathcal{I}_L.
					\end{equation}
				\end{lem}
				\begin{IEEEproof}
					We follow the proof of Theorem \ref{theorem: compression rate region}. 
					Note that the vertices of $\mathcal{R}_{SW}$ is also given by the weighted sum-rate minimization problem given in \eqref{eq:Primal Problem} with $f(\mathcal{S})$ replaced by 
					\begin{equation}\label{SW_vertex_1}
					h(\mathcal{S}) \triangleq  H\left(\{ \bm v_m | m \in \mathcal{S}\}|\{\bm v_m | m \in \overline{\mathcal{S}}\}\right),\ \text{for}\ \mathcal{S}\subseteq \mathcal{I}_L.
					\end{equation}
					Also note that $-h(\mathcal{S})$ is a submodular function since the entropy function is submodular \cite[pp. 31]{fujishige2005submodular}.
					Thus, similar to \eqref{eq:vertex coordinates}, we have
					\begin{subequations}\label{SW_vertex_2}
						\begin{align}
						R_{\pi_{\alpha}(m)}\! &=\! h(\Pi_{\alpha}(\mathcal{I}_{m})) - h(\Pi_{\alpha}(\mathcal{I}_{m-1})) \label{SW_vertex_2_1}\\
						&=\! H\!\left(\! \{\bm v_m | m \in \Pi_{\alpha}(\overline{\mathcal{I}_{m-1}})\} \!\right)\! -\! H\!\left(\! \{\bm v_m | m \in \Pi_{\alpha}(\overline{\mathcal{I}_{m}})\}\! \right) \label{SW_vertex_2_2} \\
						&=H\left(\bm v_{\pi_{\alpha}(m)} |\{ \bm v_i, i\in \Pi_{\alpha}(\overline{\mathcal{I}_m})\}\right)
						\end{align}
					\end{subequations}
					where \eqref{SW_vertex_2_2} follows from \eqref{SW_vertex_1} and the chain rule of the entropy.
		\end{IEEEproof}
		
		From Theorem \ref{thm: achive vertex}, the achievable rate tuples of GCCF are given by \eqref{eq_rate_delta_m}.  Thus, together with Lemma \ref{cor_SW_vertex}, to prove Theorem \ref{thm:equivalent_rate_region}, it suffices to show that
		\begin{equation}\label{prov_entrpy_eq}
		H(\bm \delta_{\pi_{\alpha}(m)}) = H\left(\bm v_{\pi_{\alpha}(m)} | \{\bm v_i, i\in \Pi_{\alpha}(\overline{\mathcal{I}_{m}})\}\right),\ \text{for}\ m \in \mathcal{I}_L.
		\end{equation}

		From \eqref{cor_entrpy_eq0_1}, the LHS of \eqref{prov_entrpy_eq} can be represented as
		\begin{equation}\label{prov_entrpy_eq2}
		H(\bm \delta_{\pi_{\alpha}(m)}) = \sum_{k\in\mathcal{J}_m}H(\bm v_{\pi_{\alpha}(m),k}).
		\end{equation} 
		The RHS of \eqref{prov_entrpy_eq} can be written as
		\begin{subequations}\label{prov_entrpy_eq11}
			\begin{align}
			&{}H\left(\bm v_{\pi_{\alpha}(m)} | \{ \bm v_i, i\in \Pi_{\alpha}(\overline{\mathcal{I}_{m}})\}\right)\\
			&{}= H\left( \{\bm v_{\pi_{\alpha}(m),k} \}_{k=1}^{2L-1} | \{ \bm v_i, i\in \Pi_{\alpha}(\overline{\mathcal{I}_{m}})\}\right) \label{prov_entrpy_eq11_1}\\
			\begin{split}
			&{}= H\left( \bm v_{\pi_{\alpha}(m),1} | \{ \bm v_i, i\in \Pi_{\alpha}(\overline{\mathcal{I}_{m}})\},\{\bm v_{\pi_{\alpha}(m),k} \}_{k=2}^{2L-1}\right) \\
			&{}\quad + H\left( \bm v_{\pi_{\alpha}(m),2} | \{ \bm v_i, i\in \Pi_{\alpha}(\overline{\mathcal{I}_{m}})\},\{\bm v_{\pi_{\alpha}(m),k} \}_{k=3}^{2L-1}\right) \\
			&{}\quad+ \cdots + H\left( \bm v_{\pi_{\alpha}(m),2L-1} | \{ \bm v_i, i\in \Pi_{\alpha}(\overline{\mathcal{I}_{m}})\}\right)
			\end{split}
			\label{prov_entrpy_eq11_2}
			\end{align}
		\end{subequations}
		where \eqref{prov_entrpy_eq11_1} is from the bijection between $\bm v_{m}$ and $\{\bm v_{m,k} \}_{k=1}^{2L-1}$, and \eqref{prov_entrpy_eq11_2} is from the chain rule of the entropy.
		To ensure \eqref{prov_entrpy_eq}, it suffices to show 
		\begin{subequations}\label{prov_entrpy_eq12_1}
			\begin{align}
				\begin{split}\label{prov_entrpy_eq13}
					&{}H\left( \bm v_{\pi_{\alpha}(m),k} | \{ \bm v_i, i\in \Pi_{\alpha}(\overline{\mathcal{I}_{m}})\},\{\bm v_{\pi_{\alpha}(m),k'} \}_{k' = k+1}^{2L-1}\right) = 0,\text{ for } k\notin \mathcal{J}_{m} 
				\end{split}\\
				\begin{split}\label{prov_entrpy_eq12}
					&{}H\left( \bm v_{\pi_{\alpha}(m),k} | \{ \bm v_i, i\in \Pi_{\alpha}(\overline{\mathcal{I}_{m}})\},\{\bm v_{\pi_{\alpha}(m),k'} \}_{k' = k+1}^{2L-1}\right) = H(\bm v_{\pi_{\alpha}(m),k}), \text{ for } k\in \mathcal{J}_{m}
				\end{split}
			\end{align}
		\end{subequations}
		where $\{\bm v_{\pi_{\alpha}(m),k'}\}_{k'=k+1}^{2L-1} = \emptyset$ for $k+1 > 2L-1$. By assumption, we have $\bm d_l = \bm 0$ for $l\in\mathcal{I}_L$. Thus, $\bm v_m$ in \eqref{eq:computed codeword} and $\bm v_{m,k}$ in \eqref{eq:component of v_m_2}  reduce to
		\begin{align}
			\bm v_m &= \left(\sum_{l=1}^{L} a_{ml} \t_l\right)\bmod \Lambda_{1} \label{prov_entrpy_eq3_1}\\
			\bm v_{m,k} &= \left(\sum_{l=1}^{L} a_{ml} \t_{l,k}\right)\bmod \Lambda_{k} \label{prov_entrpy_eq3_2}.
		\end{align}
		Recall from the discussion below \eqref{eq_v_m_k} that $\bm v_{m,k}$ is independent of $\{\bm v_{m',k'}|m'\in\mathcal{I}_L, k'\in\mathcal{I}_{2L-1}\backslash \{k\}\}$. Then, the LHS of \eqref{prov_entrpy_eq13} can be represented as
			\begin{align*}
			&H\left( \bm v_{\pi_{\alpha}(m),k} | \{ \bm v_i, i\in \Pi_{\alpha}(\overline{\mathcal{I}_{m}})\},\{\bm v_{\pi_{\alpha}(m),k'} \}_{k' = k+1}^{2L-1}\right) \\
			&=\!H\!\left(\! \bm v_{\pi_{\alpha}(m),k} | \{\!\{ \bm v_{i,k'}, i\!\in\! \Pi_{\alpha}(\overline{\mathcal{I}_{m}})\}\!\}_{k'=1}^{2L-1},\!\{\!\bm v_{\pi_{\alpha}(m),k'}\!\}_{k' = k+1}^{2L-1}\!\right)\\
			& = H\left( \bm v_{\pi_{\alpha}(m),k} | \{ \bm v_{i,k}, i\in \Pi_{\alpha}(\overline{\mathcal{I}_{m}})\}\right).
			\end{align*}
		Thus, \eqref{prov_entrpy_eq12_1} reduces to
		\begin{subequations}\label{prov_entrpy_eq14_1}
			\begin{align}
			H\!\left(\! \bm v_{\pi_{\alpha}(m),k} | \{ \bm v_{i,k}, i\!\in\! \Pi_{\alpha}(\overline{\mathcal{I}_{m}})\}\right)  &\!=\! 0, \text{ for } k\notin \mathcal{J}_{m},\label{prov_entrpy_eq14}\\
			H\!\left(\! \bm v_{\pi_{\alpha}(m),k} | \{ \bm v_{i,k}, i\!\in\! \Pi_{\alpha}(\overline{\mathcal{I}_{m}})\}\right)  &\!=\! H(\!\bm v_{\pi_{\alpha}(m),k}\!), \text{ for } k\in \mathcal{J}_{m}. \label{prov_entrpy_eq15}
			\end{align}
		\end{subequations}
	
		To show \eqref{prov_entrpy_eq14_1}, it suffices to show that $\bm v_{\pi_{\alpha}(m),k}$ is deterministic given $\{ \bm v_{i,k}, i\in \Pi_{\alpha}(\overline{\mathcal{I}_{m}})\}$ for $k\notin \mathcal{J}_{m}$ and $\bm v_{\pi_{\alpha}(m),k}$ is independent of $\{ \bm v_{i,k}, i\in \Pi_{\alpha}(\overline{\mathcal{I}_{m}})\}$ for $k \in \mathcal{J}_{m}$.
			
		Note that $k \notin \mathcal{J}_{m}$ can be interpreted as \textcolor{blue}{:} $\bm A(\pi_{\alpha}(m), \mathcal{L}_k)$ is linear dependent of the rows of  $\bm A(\pi_{\alpha}(\overline{\mathcal{I}_{m}}), \mathcal{L}_{k})$ and $k \in \mathcal{J}_{m}$ can be interpreted as \textcolor{blue}{:} $\bm A(\pi_{\alpha}(m), \mathcal{L}_k)$ is linear independent of the rows of $\bm A(\pi_{\alpha}(\overline{\mathcal{I}_{m}}), \mathcal{L}_{k})$. 
		Therefore, the proof of Theorem \ref{thm:equivalent_rate_region} concludes by the following Lemma.
		\begin{lem}\label{lem_independency}
			The codeword component $\bm v_{m,k}$ in \eqref{prov_entrpy_eq3_2}, 
			$\bm v_{m,k}$ is deterministic given $\{\bm v_{m',k}|m'\in \mathcal{S}\}$ if $\bm A(m, \mathcal{L}_k)$ is linearly dependent of the row vectors of $\bm A(\mathcal{S}, \mathcal{L}_k)$; otherwise, $\bm v_{m,k}$ is independent of $\{\bm v_{m',k}|m'\in \mathcal{S}\}$ if $\bm A(m, \mathcal{L}_k)$ is linearly independent of the row vectors of $\bm A(\mathcal{S}, \mathcal{L}_k)$.
		\end{lem}
	\begin{IEEEproof}
		Following the approach in Appendix \ref{app:recovery}, we map $\bm v_{m',k}$ from $\mathbb{R}^{n}$ into $\mathbb{F}_{\gamma}^{K}$, yielding
		\begin{equation}\label{entropy_eq_3}
		\phi_{\mathrm{v},k}^{-1}\left(\bm v_{m',k}\right) = \sum_{l\in \mathcal{L}_k} g^{-1}(a_{m'l}\bmod\gamma) \bm w_{l,k}.
		\end{equation}
		where $\phi_{\mathrm{v},k}(\cdot)$ is defined in \eqref{recover_eq_3} and $\bm w_{l,k} = \phi_{l,k}^{-1}(\t_{l,k})$. 
		By taking transpose on the both sides of \eqref{entropy_eq_3} and then stacking the result row by row for $m'\in \Pi_{\alpha}(\overline{\mathcal{I}_{m-1}})$, we obtain
		\begin{equation}\label{prov_entrpy_eq5}
		\begin{bmatrix}
		\bar{\bm v}_{\pi_{\alpha}(m),k}^\mathrm{T}\\
		\bar{\bm v}_{\pi_{\alpha}(m+1),k}^\mathrm{T}\\
		\vdots\\
		\bar{\bm v}_{\pi_{\alpha}(L),k}^\mathrm{T}\\
		\end{bmatrix}
		=
		\begin{bmatrix}
		\bm A(\pi_{\alpha}(m), \mathcal{L}_k)\\
		\\
		\bm A(\Pi_{\alpha}(\overline{\mathcal{I}_{m}}), \mathcal{L}_k)\\
		\\
		\end{bmatrix} 
		\begin{bmatrix}
		\bm w_{l_1,k}^\mathrm{T}\\ \bm w_{l_2,k}^\mathrm{T}\\ \vdots\\ \bm w_{l_{|\mathcal{L}_k|},k}^\mathrm{T}.
		\end{bmatrix}
		\end{equation}
		where $\bar{\bm v}_{\pi_{\alpha}(m'),k} \triangleq (\phi_{\mathrm{v},k}^{-1}\left(\bm v_{m',k}\right))$.
		
		We first consider the case that  $\bm A(\pi_{\alpha}(m), \mathcal{L}_k)$ is linearly dependent of $\bm A(\pi_{\alpha}(\overline{\mathcal{I}_{m}}), \mathcal{L}_{k})$. 
		Then, $\bm A(\pi_{\alpha}(m), \mathcal{L}_k)$ can be represented by a linear combination of $\bm A(\pi_{\alpha}(\overline{\mathcal{I}_{m}}), \mathcal{L}_{k})$. Thus,  $\bar{\bm v}_{\pi_{\alpha}(m),k}$ can also be represented by a linear combination of $\{ \bar{\bm v}_{i,k}, i\in \pi_{\alpha}(\overline{\mathcal{I}_{m}})\}$.
		Therefore, $\bar{\bm v}_{\pi_{\alpha}(m),k}$ is deterministic for given $\{ \bar{\bm v}_{i,k}, i\in \pi_{\alpha}(\overline{\mathcal{I}_{m}})\}$ and so $\bm v_{\pi_{\alpha}(m),k}$ is deterministic for given $\{ \bm v_{i,k}, i\in \pi_{\alpha}(\overline{\mathcal{I}_{m}})\}$.
		
		We next consider the case that $\bm A(\pi_{\alpha}(m), \mathcal{L}_k)$ is linearly independent of $\bm A(\pi_{\alpha}(\overline{\mathcal{I}_{m}}), \mathcal{L}_{k})$.
		 By reducing $\bm A(\Pi_{\alpha}(\overline{\mathcal{I}_{m}}), \mathcal{L}_k)$ into row reducing echelon form, we obtain
		\begin{equation}\label{prov_entrpy_eq6}
		\begin{bmatrix}
		\bar{\bm v}_{\pi_{\alpha}(m+1),k}^{\prime\mathrm{T}}\\
		\bar{\bm v}_{\pi_{\alpha}(m+2),k}^{\prime\mathrm{T}}\\
		\vdots\\
		\bar{\bm v}_{\pi_{\alpha}(L),k}^{\prime\mathrm{T}}\\
		\end{bmatrix}
		=
		\begin{bmatrix}
		\bm I_{\lambda}& \bm F\\
		\bm 0& \bm 0\\
		\end{bmatrix} 
		\begin{bmatrix}
		\bm w_{l_1,k}^\mathrm{T}\\ \bm w_{l_2,k}^\mathrm{T}\\ \vdots\\ \bm w_{l_{|\mathcal{L}_k|},k}^\mathrm{T}
		\end{bmatrix}.
		\end{equation}
		where $\lambda$ is the rank of $\bm A(\Pi_{\alpha}(\overline{\mathcal{I}_{m}}), \mathcal{L}_k)$, $\bm F$ is the free matrix, and $\bar{\bm v}_{\pi_{\alpha}(m'),k}^{\prime\mathrm{T}}$ are obtained from $\bar{\bm v}_{\pi_{\alpha}(m'),k}^{\mathrm{T}}$ by the row operations that transforms $\bm A(\Pi_{\alpha}(\overline{\mathcal{I}_{m}}), \mathcal{L}_k)$ into $[\bm I_{\lambda}, \bm F;\bm 0, \bm 0]$. Note that the linear transform from  $\{\bar{\bm v}_{\pi_{\alpha}(m'),k}^{\mathrm{T}}\}$ to $\{\bar{\bm v}_{\pi_{\alpha}(m'),k}^{\prime\mathrm{T}}\}$ is invertible.
		 From \eqref{prov_entrpy_eq6}, we have
		\begin{equation}
			\begin{bmatrix}
			\bm w_{l_1,k}^\mathrm{T}\\ \bm w_{l_2,k}^\mathrm{T}\\ \vdots\\ \bm w_{l_{\lambda},k}^\mathrm{T}.
			\end{bmatrix}
			= 
			\begin{bmatrix}
			\bar{\bm v}_{\pi_{\alpha}(m+1),k}^{\prime\mathrm{T}}\\
			\bar{\bm v}_{\pi_{\alpha}(m+2),k}^{\prime\mathrm{T}}\\
			\vdots\\
			\bar{\bm v}_{\pi_{\alpha}(m+\lambda),k}^{\prime\mathrm{T}}\\
			\end{bmatrix} -
			\bm F \begin{bmatrix}
			\bm w_{l_{\lambda + 1},k}^\mathrm{T}\\ \bm w_{l_2,k}^\mathrm{T}\\ \vdots\\ \bm w_{l_{|\mathcal{L}_k|},k}^\mathrm{T}
			\end{bmatrix}.
		\end{equation}
		Thus, we can represent $\bar{\bm v}_{\pi_{\alpha}(m),k}$ as a linear combination of $\{\bar{\bm v}_{\pi_{\alpha}(m'),k}, m'\in\Pi_{\alpha}(\overline{\mathcal{I}_{m}})\}$ and $\{\bm w_{l_i,k},l_i\in\{ l_{\lambda+1},\cdots,l_{|\mathcal{L}_k|} \}  \}$.

		From the assumption that $\bm A(\pi_{\alpha}(m), \mathcal{L}_k)$ is linear independent of the rows of $\bm A(\Pi_{\alpha}(\overline{\mathcal{I}_{m}}), \mathcal{L}_{k})$, the coefficients of $\{\bm w_{l_i,k},l_i\in\{ l_{\lambda+1},\cdots,l_{|\mathcal{L}_k|} \}  \}$ are not all-zero; otherwise, $\bar{\bm v}_{\pi_{\alpha}(m),k}$ is a linear combination of $\{\bar{\bm v}_{\pi_{\alpha}(m'),k}', m'\in\Pi_{\alpha}(\overline{\mathcal{I}_{m}})\}$, which contradicts to the assumption.
		Thus, from the Crypto lemma\cite{erez2004achieving}, we see that $\bar{\bm v}_{\pi_{\alpha}(m),k}$ is 
		independent of $\{ \bar{\bm v}_{i,k}, i\in \Pi_{\alpha}(\overline{\mathcal{I}_{m}})\}$.
		By mapping $\bar{\bm v}_{\pi_{\alpha}(m'),k}$ into $\bm v_{\pi_{\alpha}(m'),k}$, we see that $\bm v_{\pi_{\alpha}(m),k}$ is independent of $\{ \bm v_{i,k}, i\in \pi_{\alpha}(\overline{\mathcal{I}_{m}})\}$, which concludes the proof.
	\end{IEEEproof}

\ifCLASSOPTIONcaptionsoff
  \newpage
\fi

%

\bibliographystyle{IEEEtran}
\bibliography{referencebib}

%








\end{document}